# X-rays from Solar System Objects

Anil Bhardwaj
Space Physics laboratory, Vikram Sarabhai Space Centre, Trivandrum 695022, India;
anil_bhardwaj@vssc.gov.in     bhardwaj_spl@yahoo.com

Ronald F. Elsner
NASA Marshall Space Flight Center, NSSTC/XD12, Space Science Branch, 320
Sparkman Drive, Huntsville, AL 35805, USA; ron.elsner@msfc.nasa.gov

G. Randall Gladstone
Southwest Research Institute, 6220 Culebra Road, San Antonio, TX 78228-0510, USA;
randy.gladstone@swri.org

Thomas E. Cravens
Department of Physics and Astronomy, University of Kansas, Lawrence, KS 66045,
USA; cravens@ku.edu

Carey M. Lisse
Department of Astronomy, University of Maryland, College Park, MD 20742, USA;
lisse@astro.umd.edu

Konrad Dennerl
MPI fur extraterrestrische Physik, Giessenbachstrasse, Garching, D-85748 Germany;
kod@mpe.mpg.de

Graziella Branduardi-Raymont
Mullard Space Science Laboratory, University College London, Holmbury St Mary,
Dorking, Surrey RH5 6NT, UK; gbr@mssl.ucl.ac.uk

Bradford J. Wargelin
Harvard-Smithsonian Center for Astrophysics, 60 Garden St., Cambridge, MA 02138,
USA; bwargelin@cfa.harvard.edu

J. Hunter Waite, Jr.
Department of Atmospheric, Oceanic, and Space Sciences, University of Michigan, 2455
Hayward Street, Ann Arbor, MI 48109-2143, USA; hunterw@umich.edu

Ina Robertson
Department of Physics and Astronomy, University of Kansas, Lawrence, KS 66045, USA
robertin@ku.edu

Nikolai Østgaard

Department of Physics and Technology, University of Bergen, Bergen N-5007, Norway; nikost@ift.uib.no

Peter Beiersdorfer
Department of Physics, Lawrence Livermore National Laboratory, Livermore, CA 94550, USA;   Also at : Space Sciences Laboratory, 7 Gauss Way, University of California, Berkeley, CA 94720, USA.
beiersdorfer@llnl.gov

Steven L. Snowden
NASA Goddard Space Flight center, Code 662, Greenbelt, MD 20771, USA
snowden@milkyway.gsfc.nasa.gov

Vasili Kharchenko
Harvard-Smithsonian Center for Astrophysics, 60 Garden St., Cambridge, MA 02138, USA; vkharchenko@cfa.harvard.edu

**Abstract**

During the last few years our knowledge about the X-ray emission from bodies within the solar system has significantly improved. Several new solar system objects are now known to shine in X-rays at energies generally below 1 keV. Apart from the Sun, the known X-ray emitters now include planets (Venus, Earth, Mars, Jupiter, and Saturn), planetary satellites (Moon, Io, Europa, and Ganymede), all active comets, the Io plasma torus, the rings of Saturn, the coronae (exospheres) of Earth and Mars, and the heliosphere. The advent of higher-resolution X-ray spectroscopy with the Chandra and XMM-Newton X-ray observatories has been of great benefit in advancing the field of planetary X-ray astronomy. Progress in modeling X-ray emission, laboratory studies of X-ray production, and theoretical calculations of cross sections, have all contributed to our understanding of processes that produce X-rays from the solar system bodies.

At Jupiter and Earth, both auroral and non-auroral disk X-ray emissions have been observed. X-rays have been detected from Saturn's disk, but no convincing evidence of an X-ray aurora has been observed. The first soft X-ray observation of Earth's aurora by Chandra shows that it is highly variable. The non-auroral X-ray emissions from Jupiter, Saturn, and Earth, and those from the disk of Mars, Venus, and Moon are mainly produced by scattering of solar X-rays. The spectral characteristics of X-ray emission from comets, the heliosphere, the geocorona, and the Martian halo are quite similar, but they appear to be quite different from those of Jovian auroral X-rays.

This paper reviews studies of X-ray emission from the solar system bodies. Processes of production of solar system X-rays are discussed and an overview is provided of the main source mechanisms of X-ray production at each object. A brief account on recent development in the area of laboratory studies of X-ray production is also provided.

# 1. Introduction

With 9 planets, more than 100 satellites, and a large number of small bodies (comets, asteroids, and planetoids), our solar system is a fantastic natural laboratory with a plethora of phenomena that challenge our understanding of the underlying physics and provides valuable insights into similar processes occurring elsewhere in the universe. X-ray emission from the solar system bodies is one such phenomenon. X-ray emission is generally associated with hot plasmas in which collisions with energetic electrons both ionize and excite levels in atomic species (Griem, 1997; Mewe, 1990). The million degree solar corona is a prime example of a hot astrophysical plasma producing X-rays. However, most solar system bodies do not contain hot gas, yet they do emit X-rays, as will be described in this review. That is, planetary and cometary atmospheres are much colder than the solar corona with neutral temperatures of only ≈10 K for comets (Krankowsky et al., 1986) up to about 1000 K for the upper atmospheres of most planets (Schunk and Nagy, 2000). Thus, study of solar system X-ray emission is an interesting discipline, with X-ray emission from a wide variety of objects under a broad range of conditions. This paper provides a review of X-ray emission from solar system bodies, other than the Sun, in the soft X-ray energy range. This paper will focus on soft X-ray emission (~0.1-2.0 keV) from throughout the solar system, although emission at lower and higher energies will also be discussed.

Terrestrial X-rays from auroral region were discovered in the 1950s. In 1962, the first attempt to detect X-rays from the Moon discovered the first extrasolar X-ray source, Scorpius X-1 (Giacconi et al., 1962), leading to the birth of X-ray astronomy. In the early 1970s, the Apollo 15 and 16 missions studied fluorescently scattered X-rays from the Moon. The launch of the first X-ray satellite, Uhuru, in 1970 marked the beginning of satellite-based X-ray astronomy. Observations with the Einstein X-ray Observatory in 1979 discoved X-rays from Jupiter (Metzger et al., 1983; see Bhardwaj and Gladstone, 2000, for a history of earlier searches for X-ray emission from this planet). Up to 1990, the 3 solar system objects, other than the Sun, known to emit X-rays were the Earth, the Moon and Jupiter. In 1996, observations with the Roentgensatllit (ROSAT) made the surprising discovery of X-ray emission from comets (Lisse et al., 1996; Dennerl et al. 1997). This discovery revolutionized the field of solar system X-ray emission and demonstrated the importance of the solar wind charge exchange (SWCX) mechanism (Cravens, 1997, 2002) in the production of X-rays in the solar system. This process will be discussed in detail in this paper in various sections.

The advent of the sophisticated X-ray observatories Chandra and XMM-Newton is advancing study of solar system X-ray emission at a fast pace. Several more solar system objects are now known to shine in X-rays at energies generally below 2 keV. These include Venus, Mars and its halo, Saturn and its rings, the Galilean satellites Io and Europa, the Io plasma torus. The superb spatial and spectral resolution of these new X-ray observatories is improving our understanding of the physics of X-ray production in the solar system. In Table 1 we have tabulated all solar system objects known to emit X-rays and classified them in different categories.

The goal of this paper is to summarize the recent results from X-ray studies of solar system bodies and to present a comparative overview of X-ray emission from different objects. Although this paper extensively covers various topics in solar system X-ray emission, it is not an exahaustive review of the vast literature in this field. Readers are referred to recent reviews for more details and different perspectives, e.g., on Jupiter (Bhardwaj and Gladstone, 2000; Bhardwaj, 2003), on comets (Cravens, 2002; Lisse et al. 2004; Krasnopolsky et al. 2004), and on solar system objects in general (Waite and Lummerzheim, 2002; Cravens, 2000a; Bhardwaj et al., 2002, Bhardwaj, 2006).

**Table 1.  Classes of Solar System Objects Detected in X-rays**

| Class | Object(s) |
|---|---|
| Star | Sun |
| Planets | Venus, Earth, Mars, Jupiter, Saturn |
| Satellites | Moon, Io, Europa, Ganymede, [Titan]* |
| Minor Bodies | Comets, Asteroids |
| Planetary Coronae (exospheres) | Geocorona, Martian Halo (corona) |
| Extended Objects (around planets) | Io Plasma Torus, Rings of Saturn |
| Large Cavity | Heliosphere |

*X-rays from Titan have not been observed, but in a rare celestial event captured by the Chandra X-ray Observatory on January 5, 2003, Titan passed in front of the Crab Nebula. The X-ray shadow cast by Titan allowed astronomers to make the first X-ray measurement of the extent of its atmosphere (Mori et al., 2004).

**2. Production Mechanisms for Solar System X-Ray Emission**

Electromagnetic radiation with wavelengths between about 0.01 nm and 10 nm constitutes the X-ray part of the spectrum.  The corresponding photon energies range from about 100 eV up to tens of keV, with soft X-rays between about 100 eV and 1 keV and extreme ultraviolet (EUV) emission between roughly 10 and 100 eV  (wavelengths (λ) from 10 nm to about 100 nm). The distinctive feature of both X-ray and EUV radiation is that such radiation can ionize neutral atoms and molecules, and thus plays an important role in the formation of planetary ionospheres (Schunk and Nagy, 2000).

A variety of physical mechanisms responsible for solar system X-ray emission will be briefly described here, although reviews covering some aspects of this topic can also be consulted (Krasnopolsky et al., 2004; Lisse et al., 2004; Cravens, 2000a, 2002; Bhardwaj et al., 2002). The main mechanisms that produce X-rays in solar system environments include: (1) collisional excitation of neutral species and ions by charged particle impact (particularly electrons) followed by line emission, (2) electron collisions with neutrals and ions producing continuum bremsstrahlung emission, (3) solar photon scattering from neutrals in planetary atmospheres – both elastic scattering and K-shell fluorescent scattering, (4) charge exchange of solar wind ions with neutrals, followed by X-ray emission, and (5) X-ray production from the charge exchange of energetic heavy ions with neutrals or by direct collisional excitation of ions.

X-ray emission can be understood by considering a highly-excited atom or ion. Fig. 1 is a schematic showing the electron energy levels for a hydrogenic atomic species. The energies for a species with nuclear charge, Z, and just one electron are given by the Bohr energy expression: $E_n = - Z^2\ 13.6\ eV\ /\ n^2$, where n is the principal quantum number. The ground-state energy for atomic hydrogen (Z=1) is – 13.6 eV (this is also the ionization potential), whereas for $O^{7+}$ ions (Z=8), the ground-state energy (n=1 level) is 64 times this, or - 870 eV. The energy levels for a multi-electron atom/ion are not so easily described but the X-ray emission processes are the same in their essential features. The types of transitions include bound-bound (denoted bb), bound-free (bf), and free-free (ff). For example, collisions with fast electrons can excite an atom from the ground-state to an excited state. This would then be followed by the emission of one, or more, photons in the form of line emission. For example, the n=2 to n=1 transition in H produces Lyman alpha photons with energies of 10.2 eV ($\lambda$ = 121.6 nm – ultraviolet radiation), whereas the same transition for $O^{7+}$ ions produces 653 eV photons ($\lambda$ = 1.9 nm – soft X-ray radiation).

Now we consider in more detail each of the five X-ray production mechanisms listed above.

2.1. *Electron collisions – line emission*

Consider first a coronal-type plasma, such as the Sun's. X-ray line radiation is produced when the particle kinetic energies (that is, the temperature for a thermal plasma) are high enough such that: (1) atoms are multiply-ionized and high-charge state species (high Z) are created, (2) collisional excitation of high-lying levels can take place. Typically in a plasma, electron-ion recombination balances ionization for any particular species and charge-state. Radiative recombination is simply the inverse of photoionization, and involves the transition from an energy level in the continuum to a bound state; this is a free-bound transition which contributes to continuum emission (Fig. 1). If the bound state is not the ground state, then the resulting radiative cascade will also produce line emission. In high-temperature plasmas, dielectronic recombination may be more important than radiative recombination for high-Z ion species with several orbital electrons (cf., Mewe, 1990). In this case, the capture of a free-electron into an excited state is accompanied by a transition in which an electron in the K-shell or L-shell is excited to a higher level. The resulting excited atom sometimes stabilizes by emitting an X-ray photon in which case, recombination has occurred.

*2.2. Electron collisions-- bremsstrahlung emission*

Continuum X-ray emission can also be generated by free-free transitions (Fig. 1), if the electron energies are sufficiently large. This is the bremsstrahlung (German for braking radiation) process (cf., Griem, 1997), and in the classical description is due to the electromagnetic radiation emitted by the electron acceleration associated with its Coulomb interaction with the electrically-charged atomic nucleus (shielded or unshielded by atomic electrons). Bremsstrahlung is an important process for the formation of the X-ray continuum of the Sun, and it also explains the hard X-ray emission in the terrestrial aurora (Vij et al., 1975; Stadsnes et al., 1997; Østgaard et al., 2001), and perhaps the Jovian aurora also (Barbosa, 1990; Waite, 1991; Singhal et al. 1992; Branduardi-Raymont et al., 2006a). For the auroral cases, the fast electrons responsible are produced externally to the planetary atmosphere in the magnetosphere and then precipitate along magnetic field lines into the atmosphere. This topic will be discussed in more detail later. In a "thermal plasma", such as the solar corona (Zirin, 1988), in which the electron energy distribution is Maxwellian, "thermal bremsstrahlung" radiation is produced. Thermal bremsstrahlung is important in the Sun and it might also be important for explaining some of the emission from hot plasma in the Io Plasma Torus (Elsner et al., 2002).

X-ray emission from the solar corona mainly involves mechanisms 1 and 2, and is not surprising given the temperatures between $\approx 10^6$ K in coronal holes and $3 \times 10^6$ K in active regions. The heavy ions in the solar corona are highly-ionized ($O^{6+}$, $O^{7+}$, $Fe^{14+}$, $Si^{12+}$, etc.) and copious X-ray line emission is produced via the processes described above. This hot solar corona is also the source of the solar wind.

X-ray emission from planetary atmospheres, and other solar system environments, requires somewhat different mechanisms (cf., Cravens, 2000a; Bhardwaj et al., 2002), as we will briefly describe below and in the individual sections of this paper. As already alluded to, auroral bremsstrahlung radiation can be produced by energetic electrons interacting with neutrals and ions, but these electrons are not produced in the cold atmospheres, but are produced externally in the more electrodynamically active magnetospheres (see sections on Earth and Jupiter of this paper), and then travel to the atmospheres. In all the mechanisms below, the energy needed to produce the X-rays from a cold neutral gas comes externally from either the Sun (solar wind plasma or X-ray photons) or from planetary magnetospheres.

Consider X-ray production by electron collisions with neutrals and ions in planetary or cometary environments. Fast electrons can collide with the target species and excite them to either bound states or to the continuum (resulting in ionization). The former can lead to line emission, including X-ray emission as occurs at the solar corona, but this is probably not an important source of X-rays elsewhere in the solar system (as Kransnopolsky 1997 demonstrated for comets). However, free-free collisions producing X-ray photons in the continuum (i.e., the bremsstrahlung process) are known to produce auroral X-rays at the Earth. The electron energies must exceed the energies of the photons produced; hence, electrons with energies at least about a keV are needed to generate X-rays this way. Recent observations (Branduardi-Raymont et al., 2006a) have shown that bremsstrahlung X-rays are being produced in the Jovian aurora with photon energies in excess of a keV, although the total power is much less than the soft X-ray power produced in line emission.

2.3. *Solar photon scattering and fluorescence from planetary atmospheres*

X-rays can be both absorbed and elastically scattered (both incoherently and coherently – Chantler et al., 1995) by atoms or molecules in an atmosphere. In particular, solar X-rays can be scattered from planetary atmospheres, which act, in effect, as diffuse mirrors. In the soft X-ray part of the spectrum scattering cross sections are much samller than absorption cross sections (Chantler, 1995), indicating that the bulk of solar X-rays deposit their energy in the target atmosphere rather than reflect. Nonetheless, as will be discussed later in this review, planetary X-rays produced by this process have been observed. On the Earth scattered X-rays have been observed both during major solar flares (Petrenic et al., 2000b) as well as during non-flaring conditions (McKenzie et al., 1982; Fink et al., 1988). Maurellis et al. (2000) first calculated the intensity of solar X-rays elastically scattered from the disk of Jupiter. Cravens et al. (2006) showed that the scattering albedo for this process at the outer planets is quite small (e.g., $\approx 10^{-3}$ at 3 nm wavelength). The X-ray spectrum associated with this scattering process is essentially the solar soft X-ray spectrum, with its many emission lines from highly ionized coronal species, but modified by the smooth albedo function associated with the atmospheric scattering. Its detailed application to planetary atmospheres will be reviewed later in this paper.

The absorption of X-rays, usually beyond the K-shell edge, can also result in X-ray emission. In the K-shell fluorescence process, ionization from the K-shell leaves a vacancy and an X-ray photon is emitted when a valence electron makes a transition to fill this vacancy. However, the excess energy is usually taken up by the emission of an Auger electron rather than a photon. For oxygen, the photon yield is only about 0.2 % in the soft X-ray part of the spectrum. K-shell fluorescence from carbon (found in atmospheric methane for Jupiter or in carbon dioxide for Venus or Mars) makes a minor contribution to the disk emission of the outer planets (Cravens et al., 2006), but is the dominant disk X-ray source at Venus and Mars, for which carbon dioxide is the major neutral species (Cravens and Maurelllis, 2001; Dennerl, 2002; Dennerl et al., 2002, 2005a,b; Bhardwaj et al., 2005a,b). L-shell fluorescence can also occur, but the radiative yields are much lower than for the K shell.

Theoretical calculations have demonstrated that solar scattering and K-shell fluorescence are not important as X-ray sources for comets (Krasnopolsky et al., 2004; Krasnopolsky, 1997). The reason is that to obtain a unit optical depth for X-ray absorption requires a neutral column density of $\approx 10^{20}$ cm$^{-2}$ (the inverse of the total cross section) that is easy to obtain in a planetary atmosphere, but not in the more tenuous cometary atmospheres.

The fluorescence mechanism can also operate when solar X-ray photons are absorbed by solid surfaces, such as the surface of the Moon (Schmitt et al., 1991; Wargelin et al.

2004), the surfaces of the Galilean satellites (Elsner et al., 2002) or ring particles at Saturn (Bhardwaj et al., 2005c). The ultimate energy source for this emission derives from the solar photons (and, hence, the solar corona) rather than the relatively low-temperature surfaces themselves.

## 2.4. *Solar Wind Charge Exchange*

A very important X-ray production mechanism for many solar system environments is the solar wind charge exchange (SWCX) mechanism (some review material can be found in Cravens, 2002; Krasnopolsky et al., 2004; Lisse et al., 2004). X-rays are generated by ions left in excited states after charge transfer collisions with target neutrals. As will be discussed later, this mechanism was first proposed (Cravens, 1997) to explain the surprising ROSAT observations of soft X-ray and extreme ultraviolet emission from comet Hyakutake in 1996 (Lisse et al., 1996). Since then, the SWCX mechanism has also been shown to operate in the heliosphere, in the terrestrial magnetosheath (geocoronal emission), and at Mars (halo emission; see later sections of this paper).

An example of a charge transfer reaction that leads to X-ray production is the reaction of fully-stripped oxygen with atomic hydrogen leading to an excited $O^{7+}$ ion:

$$O^{8+} + H \rightarrow O^{7+*} + H^+ \qquad (1)$$

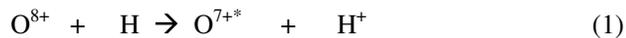

The product ion in this reaction typically has a principal quantum number of $n \approx 5$ and the radiative cascade that follows the collision must include an X-ray photon. $O^{7+}$ produces a hydrogen-like spectrum (see Fig. 1) in which the n=2 to n=1 transition has an energy of $(3/4)\ 13.6\ eV\ Z^2 = 653\ eV$ (given that Z=8). The energy derives from whatever originally ionized the oxygen. The source of the solar wind is the million degree solar corona, which is the ultimate source of power for the X-ray emission in the SWCX mechanism. While multiple-electron transfer can occur with multi-electron neutral speices, our discussion of charge exchange focuses on single-electron transfer, which dominates in all cases of relevance here.

More generally, the charge transfer reaction can be represented by:

$$A^{q+} + M \rightarrow A^{(q-1)+*} + M^+ \qquad (2)$$

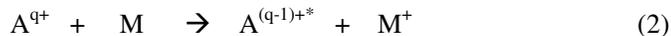

where A denotes the projectile (e.g., O, C, Fe….), q is the projectile charge (e.g., q = 5, 6, or 7) and M denotes the target neutral species (e.g., $H_2O$, O, H….). The product ion de-excites by emitting one or more photon ($A^{(q-1)+*} \rightarrow A^{(q-1)+} + h\nu$, where $h\nu$ represents a photon. For species and charge states relevant to the operation of SWCX at comets or planets, the principal quantum number of the ion $A^{(q-1)+}$ is usually about n = 4, 5, or 6. X-ray photons are emitted by the excited product ions. Collision cross sections for charge exchange of high-charge state ions are quite large. Charge exchange cross sections at solar wind ion energies are quite large, typically a few times $10^{-15}\ cm^2$, which is several orders of magnitude larger than corresponding cross sections for electron collisional excitation (Cravens, 2002; Krasnopolsky et al., 2004; Gilbody, 1986; Hoekstra et al., 1989; Greenwood et al., 2000, 2001; Stancil, 2002).

After highly excited ions are created by the charge transfer process, a radiative cascade occurs. Much work on the detailed X-ray spectrum associated with this cascade has been carried out by Kharchenko and Dalgarno (2001), Kharchenko et al. (2003), and Rigazio et al. (2002). The most obvious lines in measured cometary spectra, as will be discussed later in the paper, are the helium-like transitions of $O^{6+}$, produced by the charge exchange of solar wind $O^{7+}$ ions with neutral targets. In particular, the lines at 561, 568, and 574 eV are very strong in cometary X-ray spectra (cf., Krasnopolsky et al., 2004).

The charge-exchange X-ray production requires highly-charged ion species. Such ions are found as minor ions in the solar wind. Minor, heavy ions account for about 0.1% of the solar wind and exist in highly charged states such as: $O^{7+}$, $O^{6+}$, $C^{6+}$, $C^{5+}$, $N^{6+}$, $Ne^{8+}$, $Si^{9+}$, and $Fe^{12+}$ (Schwadron and Cravens, 2000; Steiger et al., 2000). The composition of the solar wind is variable and depends on where in the solar corona it originated. The slow solar wind originates in a hotter corona than does the fast solar wind and tends to have a greater abundance of the most highly-charged ions (Schwadron and Cravens, 2000; Steiger et al., 2000).

For the SWCX mechanism a relatively simple expression can be written (cf., Cravens, 1997, 2002) for the photon volume emission rate (denoted by $P_{sqjn}(\mathbf{r})$ with units of photons cm$^{-3}$ s$^{-1}$) at a particular location ($\mathbf{r}$) in a solar system environment, originating with a particular solar wind ion species (denoted by index, s) and charge-state (denoted, q), and for a particular transition (denoted, j) of this species:

$$P_{sqjn}(\mathbf{r}) = n_{sq}(\mathbf{r})\, n_n(\mathbf{r})\, <g>\, f_{sqj}\, \sigma_{sqn}(g) \qquad (3)$$

$n_n(\mathbf{r})$ is the number density of "target" neutral species, n, at a position vector, r. Examples of species in the cometary environment include $H_2O$ and its dissociation products OH, O, and H. $n_{sq}(\mathbf{r})$ is the number density of a solar wind ion species, s (e.g., O, C, N, Fe) with charge-state q (e.g., q = 8 for fully-stripped oxygen). In a collisionally-thin regime where the neutral density is low, $n_{sq}$ is just the unperturbed solar wind density, but close to the object (i.e., cometary nucleus) the original solar wind flux of a particular ion species will be attenuated due to charge-transfer collisions or the density can be altered by dynamical processes (e.g., bow shock in the supersonic solar wind flow around an obstacle). $<g>$ is the average collision velocity and is almost equal to the solar wind speed, $u_{sw}$, at most locations. $\sigma_{sqn}(g)$ is the total charge transfer cross section at collision speed, g, for the designated ion species and for neutral species, n. $f_{sqj}$ is the probability, given a charge-transfer collision, of the occurrence of a specific transition, j, (e.g., the n=2 to n=1 transition in $O^{7+}$ which originates from $O^{8+}$, q = 8, solar wind ions).

The complete spectrum requires the application of equation (3) to all target species, solar wind species and charge-states, and to all relevant X-ray transitions. Determining the spectral intensity in a particular direction from a particular observational platform requires integrating the volume emission rate from equation (3) over the relevant optical path.

2.5. *Charge exchange and direct collisional excitation of energetic heavy ions*

X-rays have been observed from the high-latitude auroral regions of Jupiter, as will be discussed later. Some small part of this emission consists of harder X-rays produced by electron bremsstrahlung (mechanism 2) but most of the observed X-ray power is in softer X-rays with energies less than 1 keV (e.g., Bhardwaj and Gladstone, 2000; Bhardwaj et al. 2002; Waite et al., 1994; Gladstone et al., 2002; Elsner et al., 2005a; Braduardi-Raymont et al., 2004). The two most probable explanations for the bulk of the Jovian auroral X-ray emission are: (1) the SWCX mechanism applied to solar wind ions precipitating along open cusp field lines, (2) the precipitation of energetic magnetospheric heavy ions (i.e., sulfur and oxygen ions) in the outer magnetosphere (see Cravens et al., 2003).

The first mechanism has already been discussed in the last section, although for this mechanism to successfully produce sufficient X-ray power the solar wind ions must be accelerated to about 200 keV in order to allow a high enough flux of ions to reach the atmosphere of Jupiter along the cusp field lines (Cravens et al., 2003).

For the second mechanism, ambient ions in Jupiter's outer magnetosphere are accelerated to high energies and precipitate into the upper atmosphere. The Jovian magnetosphere is known to contain sulfur and oxygen ions with a wide range of energies (cf., Mauk et al., 2004). Heavy ions in the magnetosphere are mostly $S^+$, $S^{2+}$, $S^{3+}$, $O^+$, and $O^{2+}$ but the collision of these ion species with neutral $H_2$ or H in Jupiter's atmosphere will not produce X-rays because their charge-states are too low (see the beginning of this section) and the transitions are in the ultraviolet part of the spectrum rather than the X-ray part. But at energies of a few hundred keV/nucleon or more (Cravens et al., 1995), the ions become more highly charged due to electron removal collisions when they encounter the atmosphere. X-rays emitted from these highly charged ions either by direct collisional excitation or by charge transfer collisions.

Originally, energetic ions already present in the middle magnetosphere (Metzger et al., 1983; Cravens et al., 1995; Kharchenko et al., 1998; Liu and Schultz, 2000; Waite et al., 1994) were thought to be responsible for the X-ray aurora. However, Chandra observations (Gladstone et al, 2002; Elsner et al., 2005a) demonstrated that the Jovian X-ray emission originated from very high-latitudes well above the main auroral oval in the polar cap. The X-ray auroral region magnetically connects either to open field lines and/or to the outer magnetosphere where the population of heavy ions at higher energies is insufficient to produce the required X-ray luminosity. Cravens et al. (2003) suggested that the precipitating ions obtain their required high energies by acceleration associated with field-aligned electrical potentials located a few Jovian radii above the planet.

The collisional processes that take place for energetic heavy ions in Jupiter's atmosphere were discussed by Cravens et al. (1995), Kharchenko et al. (1998), and Liu and Schultz (2000) and will be briefly summarized here. The key process that creates high charge-state ions in the atmosphere is electron removal (or stripping), which can be represented by the reaction:

$$O^{q+} + H_2 \rightarrow O^{(q+1)+} + H_2 + e \qquad (4)$$

The cross section for this process remains high for ion energies of a few hundred keV per nucleon. A similar reaction can be written for other ion species such as sulfur. The inverse process, charge exchange, has already been discussed and is represented by the reaction:

$$O^{q+} + H_2 \rightarrow O^{(q-1)+*} + H_2^+ \qquad (5)$$

The cross sections for this process are quite high at lower energies, as was discussed earlier, but for energies more than about 100 keV/nucleon the cross sections become quite small. Note that the product ions are left in highly-excited states, as has already been discussed, and thus emit X-ray photons when they de-excite.

Cravens et al. (1995) used an "equilibrium fraction" approach to the energy deposition of energetic oxygen ions with Jupiter's atmosphere. The population distribution of the different ion charge states in a monoenergetic beam reaches equilibrium at a given energy if the mean free path is less than the characteristic length scale over which the target species (molecular hydrogen in this case) changes. Figure 2 shows the equilibrium fraction as a function of energy for the charge states of a precipitating oxygen beam in $H_2$. Note that for X-rays to be produced requires that q be 7 or 8, which occurs if the beam energy is about 1 MeV/nucleon or higher. Kharchenko et al. (1998) performed Monte Carlo calculations that simulate the charge state histories of energetic oxygen ions as they precipitate into Jupiter's atmosphere. They find that the ions are slow to reach the charge state equilibrium, which can significantly alter the X-ray spectrum.

After introducing main mechanisms of X-ray production in the previous subsections, in the following sections we will review the soft X-rays studies on each object in the solar system.

## 3. EARTH

### 3.1. Auroral Emissions

As a result of dynamic processes in the Earth's magnetosphere and the coupling between its magnetosphere and ionosphere, charged particles (electrons and protons) from the magnetosphere precipitate into the ionosphere and deposit their energy by ionization, excitation, dissociation, and heating of the neutral gas. The interaction between precipitating particles and the upper atmosphere give rise to emissions in visible, UV and X-ray wavelength ranges, which constitute the auroral oval. All these emissions can be used to derive information about the incoming particles using different techniques. On the dayside in a small region defined as the cusp, solar wind particles have direct access to the ionosphere and will also deposit their energy in a similar way. The produced emissions create the cusp aurora.

Particles precipitating into the Earth's upper atmosphere give rise to discrete emission lines in the X-ray range when interacting with the electrons of the atoms in the atmosphere. The characteristic inner-shell line emissions for the main species of the Earth's atmosphere, Nitrogen (K$\alpha$ at 0.393 keV), Oxygen (K$\alpha$ at 0.524 keV) and Argon (K$\alpha$ at 2.958 keV, K$\beta$ at 3.191 keV) are all in the low energy range. Very few X-ray observations of the Earth's ionosphere have been made at energies where these lines emit. However, for low energy measurements of X-rays these emission lines contribute significantly to the total X-ray production, as will be shown below. In addition precipitating particles interact with the nucleus of atoms or molecules, de-celerate the incoming particles and give rise to a continuous spectrum of X-ray photons, which are called bremsstrahlung (see Section 2.2). The main X-ray production mechanism in the Earth's auroral zones, for energies above 3-4 keV, is electron bremsstrahlung, and therefore the X-ray spectrum of the aurora has been found to be very useful in studying the characteristics of energetic electron precipitation (Vij et al, 1975; Stadsnes et al., 1997; Østgaard et al., 2001). Since the X-ray measurements are not contaminated by sunlight, the remote sensing of X-rays can be used to study energetic electron precipitation on the dayside as well as on the nightside of the Earth (Petrinec et al., 2000a).

According to the thick-target bremsstrahlung theory the energy of the X-ray photon produced in this process have any energy up to the energy of the incoming particle, but the probability distribution increases exponentially to lower energies (Kulenkampff, 1922). The X-ray bremsstrahlung production efficiency depends on the square of the de-celeration and thus is proportional to $1/m^2$ where m is the mass of the precipitating particle. This implies that electrons are $10^6$ times more efficient than protons at producing X-ray bremsstrahlung. The production efficiency is a non-linear function of energy, with increasing efficiency for increasing incident energies. For example, for a 200 keV electron the probability of producing an X-ray photon at any energy below 200 keV is 0.5%, while the probability for a 20 keV electron to produce an X-ray photon below 20 keV is only 0.0057% (Berger and Seltzer, 1972).

The angular distribution of bremsstrahlung X-rays will have a peak normal to the de-acceleration direction of the incoming electrons. However, for relativistic electrons the angular distribution will be increasingly peaked forward. Consequently, the majority of the X-ray photons in Earth's aurora are directed normal to the Earth's magnetic field, but for higher energies with a preferential direction toward the Earth. Downward propagating X-rays cause additional ionization and excitation in the atmosphere below the altitude where the precipitating particles have their peak energy deposition (e.g., Sharber et al., 1993; Winningham et al., 1993). The fraction of the X-ray emission that is moving away from the ground can be studied using satellite-based imagers, e.g, AXIS (Chenette et al., 1993) on UARS and PIXIE (Imhof et el., 1995) on Polar.

Figure 3 shows the modeled X-ray energy spectrum for eight different zenith angle intervals produced by an exponential electron distribution with an e-folding energy of 22 keV. These spectra are generated on the basis of the "general electron-photon transport code" of Lorence (1992), where both the scattering of electrons, production of secondary electrons, angular dependent X-ray production, photoelectric absorption of X rays and Compton scattering of X rays have been taken into account.

Auroral X-ray bremsstrahlung has been observed from balloons and rockets since the 1960s and from spacecraft since the 1970s (Anderson, K., 1965; Vij et al., 1975; Imhof, 1981; Parks et al., 1993; Stadsnes et al., 1997; Van Allen, 1995). Because of absorption of the low energy X-rays propagating from the production altitude (~100 km) down to balloon altitudes (35-40 km) such measurements were limited to >20 keV X-rays. Nevertheless, these early omni-directional measurements of X-rays revealed detailed information of temporal structures from slowly varying bay events (e.g., Sletten et al., 1971) to fast pulsations and microbursts (Trefall et al., 1966; Barcus and Rosenberg, 1966; Parks et al., 1968). Figure 4 shows the temporal variations at all local times of electron distributions with different e-folding energies deduced from X-ray bremsstrahlung (Barcus and Rosenberg, 1966).

When going to space (rockets and satellites) it is not the absorption of low energy X-rays that sets limitations but rather the detector techniques that have been used, and thus most observations of X-rays from the Earth's ionosphere have been limited to energies above ~3 keV. Although imaging from space has been performed from the 1970s, the PIXIE (Polar Ionospheric X-Ray Imaging Experiment) instrument aboard Polar was the first X-ray detector that provides true 2-D global X-ray image at energies >~3 keV. Because of the high apogee of the Polar satellite (~9 $R_E$), PIXIE was able to image the entire auroral oval with a spatial resolution of ~700 km for long duration when the satellite is around apogee. This has helped to study the morphology of the X-ray aurora and its spatial and temporal variation, and consequently the evolution of energetic electron precipitation during magnetic storms (days) and substorms (1-2 hours).

Figure 5 shows two images taken by PIXIE on two different days from the northern and southern hemispheres, respectively. Both energy range and time resolution are very different, but the auroral X-ray zone can be clearly seen. Data from the PIXIE camera have shown that the X-ray bremsstrahlung intensity statistically peaks in the midnight substorm onset, is significant in the morning sector, and has a minimum in the early dusk sector (Petrinec et al., 2000b). During substorms X-ray imaging shows that the energetic electron precipitation brightens up in the midnight sector and has a prolonged and delayed maximum in the morning sector due to the scattering of magnetic-drifting electrons (Østgaard et al., 1999; Asnes et al., 2005) with an evolution significantly different than viewing, e.g., in the UV emissions (Østgaard et al., 1999). Anderson et al. (2002) showed that the peak intensities during stormtime substorms were highly modified by the solar wind convective electric field. During the onset/expansion phase of a typical substorm the global electron energy deposition power is 60-90 GW, which produces 10-30 MW of bremsstrahlung X-rays (Østgaard et al., 2002).

By combining the results of PIXIE with the UV imager aboard Polar, it has been possible to derive the energy distribution of precipitating electrons in the 0.1-100 keV range with a time resolution of about 5 min (e.g., Østgaard et al., 2001), as shown in Figure 6. As these energy spectra cover the entire energy range important for the electrodynamics of the ionosphere, important parameters like Hall and Pedersen conductivity and Joule heating can be determined on a global scale with smaller uncertainties than parameterized models can do (Aksnes et al., 2002; 2004). Aksnes et al. (2004) showed how the inclusion of the high energy component of the energy spectra, which can only be derived globally from X-ray imaging, influences the estimated Hall conductance and Joule heating significantly, whereas Pedersen conductance can be estimated from UV imaging only. Electron energy deposition estimated from global X-ray imaging also gives valuable information on how the constituents of the upper atmosphere, like NOX, is modified by energetic electron precipitation (Sætre et al., 2004).

X-ray measurements of the Earth aurora below 2-3 keV are very rare, as there has been no dedicated search for auroral X-ray emissions at energies <2 keV. A few limb scans of the nighttime Earth at low latitude by the X-ray astronomy satellite, HEAO-1, in the energy range 0.15 keV to 3 keV, showed clear evidence of the K$\alpha$ lines for Nitrogen and Oxygen sitting on top of the bremsstrahlung spectrum (Luhmann et al., 1979). Recently, the High-Resolution Camera (HRC-I) aboard the Chandra X-ray Observatory imaged the northern auroral regions of the Earth in the 0.1-10 keV X-ray range using 10 epochs (each ~20 min duration) between mid-December 2003 and mid-April 2004 (Bhardwaj et al., 2005d). These observations aimed at searching for Earth's soft (<2 keV) X-ray aurora in a comparative study with Jupiter's X-ray aurora, where Chandra has previously observed a pulsating X-ray "hot-spot" (see section on Jupiter). These first soft X-ray observations of Earth's aurora showed that it is highly variable (intense arcs, multiple arcs, diffuse patches, at times absent). In at least one of the observations an isolated blob of emission was observed near the expected cusp location. A fortuitous overflight of DMSP satellite F13 provided SSJ/4 energetic particle measurements above a bright arc seen by Chandra on 24 January 2004, 20:01–20:22 UT (cf., Figure 7). A model of the emissions expected strongly suggests that the observed soft X-ray signal is a combination of bremsstrahlung and characteristic K-shell line emissions of Nitrogen and Oxygen in the atmosphere produced by electrons. In the soft X-ray energy range of 0.1-2 keV these line emissions are 4-6 times more intense than the X-ray bremsstrahlung.

## 3.2. Non-Auroral Emissions

The non-auroral X-ray background above 2 keV from the Earth is almost completely negligible except for brief periods during major solar flares (Petrinec et al., 2000b). However, at energies below 2 keV soft X-rays from the sunlit Earth's atmosphere have been observed even during quiet (non-flaring) Sun conditions (e.g., McKenzie et al., 1982; Fink et al., 1988; Snowden and Freyberg, 1993). The two primary mechanisms for the production of X-rays from the sunlit atmosphere are: 1) the Thomson (coherent) scattering of solar X-rays from the electrons in the atomic and molecular constituents of the atmosphere, and 2) the absorption of incident solar X-rays followed by the emission of characteristic K lines of Nitrogen, Oxygen, and Argon. Figure 8 shows the PIXIE image of Earth demonstrating X-ray (2.9-10 keV) production in the sunlit atmosphere during a solar flare of August 17, 1998. During flares, solar X-rays light up the sunlit side of the Earth by Thomson scattering, as well as by fluorescence of atmospheric Ar to produce characteristic X-rays at 3 keV, which can be observed by the PIXIE camera. The X-ray brightness can be comparable to that of a moderate aurora. Petrinec et al. (2000b) examined the X-ray spectra from PIXIE for two solar flare events during 1998. They showed that the shape of the measured X-ray spectra was in fairly good agreement with modeled spectra of solar X-rays subject to Thomson scattering and Argon fluorescence in the Earth's atmosphere.

Within the past decade, a different and new type of X-ray source from the Earth has been discovered. Possibly related to sprites and/or lightening discharges, very short-lived (1 ms) X-ray and γ-ray bursts (~25 keV – 1 MeV) from the atmosphere above thunderstorms, were fortuitously recorded from the Compton Gamma Ray Observatory (CGRO) satellite (Fishman et al., 1994; Nemiroff et al., 1997) and further supported by the recent Reuvan Ramaty High Energy Solar Spectroscopic Imager (RHESSI) results (Smith et al, 2005). It has been suggested that these emissions are bremsstrahlung from upward-propagating, relativistic (MeV) electrons generated in a runaway electron discharge process above thunderclouds by the transient electric field following a positive cloud-to-ground lightning event (Lehtinen et al., 1996), and that "Terrestrial Gamma-ray Flashes" (TGFs) are associated with Sprites (Roussel-Dupré and Gurevich, 1996). However, no conclusive evidence of production altitude or mechanism has yet been found, and several missions from space are planned to study this phenomena (e.g., the French small satellite TARANIS and the ESA ASIM on the ISS).

X-ray emissions are also produced in Earth's exosphere (corona), which is described in Section 4 on Geocoronal X-rays.

**4. Lunar X-rays**

The Moon's X-ray emissions have been studied in two ways: close up from lunar orbiters (e.g., Apollo 15 and 16, SMART-1, and from planned missions such as SELENE, Chandrayaan-1, and Chang'e), and more distantly from Earth-orbiting X-ray telescopes (e.g., ROSAT and Chandra). In addition to a low level of scattered solar radiation and perhaps a very low level of bremsstrahlung from solar-wind electrons impacting the surface, lunar X-rays result from fluorescence of sunlight by the surface. Thus, X-ray fluorescence studies provide an excellent way to determine the elemental composition of the lunar surface by remote sensing, since at X-ray wavelengths the optical properties of the surface are dominated by elemental abundances (rather than mineral abundances, which determine the optical properties at visible and longer wavelengths). Elemental abundance maps produced by the X-ray spectrometers on the Apollo 15 and 16 orbiters were necessarily limited to the equatorial regions but succeeded in finding geochemically interesting variations in the relative abundances of Al, Mg, and Si (Adler et al., 1972a; 1972b; 1973). Although the energy resolution of the Apollo proportional counters was low, important results were obtained, such as the enhancement of Al/Si in the lunar highlands relative to the mare. These derived Al/Si ratios were consistent with the dichotomy of plagioclase-rich highland and basaltic mare in returned samples, providing confirmation that the Moon's highland crust is differentiated and of feldspathic composition. Currently, the D-CIXS instrument on SMART-1 is obtaining abundances of Al, Si, Fe, and even Ca at 50-km resolution from a 300-km altitude orbit about the Moon (Grande, 2005). Upcoming missions planned for launch in 2007-2008 by Japan (SELENE), India (Chandrayaan-1), and China (Chang'e), will each carry X-ray spectrometers to obtain further improved maps of the Moon's elemental composition, at ~20-km resolution from ~100-200 km altitude polar orbits. It is important to recall that the solar X-ray irradiance and spectrum is highly variable, so that the X-ray spectrometers in lunar orbit must also monitor the solar spectrum in order to accurately derive elemental abundances.

Early observations from Earth orbit were made by Schmitt et al. (1991) using the ROSAT PSPC proportional counter and a marginal detection by ASCA using CCDs was recorded by Kamata et al. (1999). Figure 9 shows the Schmitt et al. (1991) data, while the right image is unpublished ROSAT data from a lunar occultation of the bright X-ray source GX5-1 (the higher energy X-rays from GX5-1 have been suppressed in this figure, but a faint trail to the upper left of the Moon remains). The power of the reflected and fluoresced X-rays observed by ROSAT in the 0.1-2 keV range coming from the sunlit surface was determined by Schmitt et al. (1991) to be only 73 kW, corresponding to $7.8 \times 10^{-15}$ erg/s/cm$^2$/arcsec$^2$. The faint but distinct lunar nightside emissions (100 times less bright than the dayside) were until recently a matter of controversy. Schmitt et al. (1991) suggested that solar wind electrons of several hundred eV might be able to impact the nightside of the Moon on the leading hemisphere of the Earth-Moon orbit around the Sun. However, this was before the GX5-1 data were acquired, which clearly show lunar night side X-rays from the trailing hemisphere as well. A much better explanation is the accepted mechanism for cometary X-rays, heavy ion solar wind charge exchange (SWCX) (e.g., Cravens, 2002). In this case, however, the heavy ions in the solar wind would be charge exchanging with geocoronal H atoms that lie between the Earth and Moon but outside the Earth's magnetosphere, thus resulting in foreground X-ray emissions between ROSAT and the Moon's dark side. This has now been confirmed by Wargelin et al. (2004) with Chandra ACIS CCD data, as discussed in the Section on geocoronal X-rays.

The first remote observations to clearly resolve discrete K-shell fluorescence lines of O, Mg, Al, and Si on the sunlit side of the Moon were also made by Chandra (Wargelin et al., 2004). This set of six 3-ks observations was made in July 2001 near the peak of the solar cycle (as was also the case for the ROSAT lunar observations) and thus with a relatively high solar irradiance. The Moon was 28% illuminated and drifted through the field of view during each observation. There was some leakage of optical photons through the ACIS UV-ion shield and one of the observations had to be discarded; the net exposure time of the spectrum in Fig. 10 corresponds to approximately 700 seconds if the Moon were fully illuminated. 1300 counts were observed in the O-K line, corresponding to a flux of $3.8 \times 10^{-5}$ photons/s/cm$^2$/arcmin$^2$ ($3.2 \times 10^{-14}$ erg/s/cm$^2$/arcmin$^2$). The Mg-K, Al-K, and Si-K lines each had roughly 10% as many counts and 3% as much flux as the O-K line, and statistics were inadequate to draw any conclusions regarding differences in element abundance ratios between highlands and maria.

The most recent Chandra observations of the Moon used the HRC-I imager to look for albedo variations due to elemental composition differences between highlands and maria, as described above. As shown in Fig. 11, the albedo contrast is noticeable, but very slight. Future observations with, for example, the RGS on XMM-Newton could provide even higher spectral resolution observations, and perhaps lead to the identification of additional elements, but bright-object constraints currently prohibit observations of the Moon.

## 5. Geocoronal X-rays

During the ROSAT all-sky survey, Snowden et al. (1994) observed unexplained long term enhancements (LTE) in the soft X-ray background (see Section 16 for more details), which were not of galactic origin. These enhancements had a noticeable time varying component, which Snowden et al. removed when constructing the "cosmic" soft X-ray background. The discovery of cometary X-rays has offered an alternative explanation already in 1997 (Dennerl et al. 1997). Cox (1998) suggested that the solar wind charge exchange mechanism applied to interstellar neutrals, and neutrals in the Earth's geocorona, could explain the LTE and might be part of the observed soft X-ray background. Freyberg (1998) also attributed the LTE to variations in the solar wind and speculated that the SWCX mechanism applied to the vicinity of Earth might be responsible.

Cravens et al. (2001) developed a simple time-dependent model of the soft X-ray emission from the SWCX mechanism applied to interstellar neutrals (atomic hydrogen and atomic helium) and geocoronal atomic hydrogen. The X-ray production rate was calculated using the following production rate expression (also see section 2 of this paper, equation 3):

$$P_{\text{X-ray}} = \alpha \, n_{sw} \, n_n \, u_{sw} \; (\text{eV cm}^{-3} \text{ s}^{-1}) \qquad (4)$$

where $\alpha$ is an efficiency factor that contains all the atomic cross sections, the transition information, and relative solar wind heavy ion composition; $n_n$ is the neutral species the solar wind charge exchanges with, $u_{sw}$ is the solar wind speed and $n_{sw}$ is the solar wind proton density. The parameter $\alpha$ should be different for each neutral target as well as for different composition states of the solar wind (Schwadron and Cravens, 2000). A reasonable value for the "overall" efficiency factor is $6 \times 10^{-16}$ eV cm$^2$ (Wargelin et al., 2004). The neutral hydrogen density was determined using the following equation: $n_H \sim= n_{H0} \, (10 \, R_E/r)^3$ ; with $n_{H0} = 25$ cm$^{-3}$. To determine the X-ray intensity, the production rates are integrated along lines of sight.

Next, Robertson et al. (2003) modified the SWCX model to look at geocoronal X-ray emissions only. The solar wind speed, density and temperature distributions in the magnetosheath were from a numerical hydrodynamic model (Spreiter et al., 1966). The cusps were not included in this model. Instead of the simple equation for atomic hydrogen used previously, Robertson and Cravens (2003) took the neutral hydrogen densities from the Hodges (1994) Monte Carlo model. Robertson and Cravens modeled SWCX emission as seen from an observation point 50 $R_E$ away from Earth, well outside the magnetosheath. Figure 12 shows the resulting X-ray intensities as observed from this observation point. The magnetopause and bow shock are clearly visible in this simulation.

Robertson and Cravens (2003) also studied the effect of an increase of solar wind flux on geocoronal SWCX X-ray intensities. If the solar wind flux increases, the distance from Earth to the magnetopause reduces, which exposes the solar wind to higher hydrogen densities. They noted that an increase in solar wind density and/or speed will drastically increase the X-ray intensities, not only because of the increase in solar wind flux, but also because of the exposure to higher hydrogen densities (which go as $1/r^3$). Since the integration volume is small, a variation in solar wind flux can immediately be seen in the X-ray intensities. Due to this time variation Cravens and Robertson (2003) concluded that it should be possible to remove the observed time varying component of geocoronal X-ray emission from the more steady state back ground emission due to charge exchange with interstellar neutrals or any other form of X-ray production. This time-varying component shows us the solar wind flow in the magnetosheath.

A final version of the geocoronal SWCX model includes the cusps (Robertson et al., 2005). X-ray emission due to the March 31, 2001, coronal mass ejection which pushed the magnetosheath inside the geosynchronous orbit, was modeled, using simulated solar wind data generated by the BATSRUS MHD model developed at the university of Michigan, but run at Goddard Space Flight Center. Figure 13 shows the modeled X-ray emission under these conditions as observed from the same observation point as was used in Figure 12. The largest X-ray production rates are in the cusps, where the solar wind flux remains high and the neutral hydrogen density is also high. In addition to temporal evidence for SWCX from ROSAT observations, there is also spectral proof from Chandra observations of the Moon that revealed line emission from hydrogenic and He-like oxygen lines (Wargelin et al., 2004) with intensities that match theoretical predictions quite well.

The Chandra Moon observations are particularly interesting because they cleanly separate geocoronal charge exchange (CX) emission from heliospheric CX emission and all other contributors to the Soft X-Ray Background (SXRB) as it appears from Earth orbit. Two sets of Moon observations were made in July and September 2001 in order to calibrate the background of the ACIS CCD detectors on Chandra. The intention was to use the Moon to block out the cosmic SXRB, revealing the intrinsic detector background that arises from cosmic rays. Later calibration measurements placed the detector midway between the telescope focus and an off-axis position illuminated by a radioactive calibration source. In this "stowed" position the detector was unilluminated by any photons while still being exposed to the same high-energy cosmic rays that produce its intrinsic background.

Spectra from the July 2001 Moon observations were statistically identical to the ACIS-stowed detector background apart from the first of the six 3000-second exposures, which showed a roughly three-sigma excess around 600 eV. The September 2001 observations, with a total exposure of 14 ks, showed much stronger emission, primarily in the "triplet" 2→1 lines of He-like O (~565 eV) and in Lyman-alpha (654 eV) and the high-n lines (775-~850 eV) of H-like O (see Figure 14). There was also a three-sigma detection of the He-like Mg 2→1 triplet.

As explained above (see Eq. 4), the intensity of CX emission from a given ion is proportional to that ion's density and velocity in the solar wind (approximately, because the cross sections vary with the relative velocity). Based on data from the ACE solar-wind monitoring satellite stationed at the L1 point (1.5 million km from Earth in the direction of the Sun), the intensity of geocoronal CX emission from O between 500 and 900 eV was predicted to be 60 times larger during the September Chandra Moon observations than during the 5 quiescent July observations. Most of the difference was in the wind density and the fraction of fully stripped and H-like ions. This expectation was consistent with the measured ratio of greater than 8, the determination of which was limited by the statistics of the null detection in July.

Predictions of the absolute flux of CX emission require models of the size of the geomagnetosphere and the density of neutral H surrounding the Earth. Using the geocoronal neutral-gas density model described above, and calculating the emission from each CX line separately (rather than using the global $\alpha$ term of Eq. 4), the predicted flux for the September Chandra Moon observations was $140 \times 10^{-6}$ counts/s/arcmin$^2$, in relatively good agreement with the measured flux of $(287+/-39) \times 10^{-6}$ counts/s/arcmin$^2$. It is historically interesting, as noted by Wargelin et al. (2004), that geocoronal X-rays were in fact detected by ROSAT in a much earlier observation of the dark side of the Moon (Schmitt et al., 1991), although their true nature was not understood at the time.

Another observation of note is that described by Snowden et al. (2004). During one of four XMM-Newton observations of the Hubble Deep Field-North, a significant enhancement of the X-ray background was seen, with a differential spectrum that clearly showed CX emission lines from O, Ne, and Mg. Unlike with the Chandra Moon observation, no obvious correlation between the solar wind flux and the CX emission strength was found. Instead, it may be that the varying CX emission was the result of XMM's changing viewing geometry with respect to the geocorona, perhaps looking through or near the cusp (see Figure 13) that caused the relatively sudden jump in the X-ray background.

**6. Venus**

Orbiting the Sun at heliocentric distances of 0.718–0.728 AU, the angular separation of Venus from the Sun, as seen from the Earth, never exceeds 47.8°. This distance is too small for most imaging X-ray astronomy satellites, because they can only observe objects at solar elongations of at least 60–70°. A remarkable exception is Chandra, the first such satellite that is able to observe as close as 45° from the limb of the Sun. Thus, with Chandra, an observation of Venus became possible for the first time.

It was expected that Venus would be an X-ray source, due to the presence of two processes: (i) charge exchange interactions between highly charged ions in the solar wind and the Venusian atmosphere (Cravens, 2000a; Krasnopolsky, 2000; Holmström et al., 2001), and (ii) scattering of solar X-rays in the Venusian atmosphere (Cravens and Maurellis, 2001). The predicted X-ray luminosities were ~0.1–1.5 MW for the first process, and ~35 MW for the second one, with an uncertainty factor of about two.

The first X-ray observations of Venus took place on January 10 and 13, 2001 (Dennerl et al., 2002). They were performed with Chandra and consisted of two parts: grating spectroscopy with LETG/ACIS–S and direct imaging with ACIS–I. This combination yielded data of high spatial, spectral, and temporal resolution. With ACIS–I, Venus was clearly detected as a half–lit crescent, exhibiting considerable brightening on the sunward limb (Fig. 15); the LETG/ACIS–S data showed that the spectrum was dominated by O–K$\alpha$ and C–K$\alpha$ emission (Fig. 16), and both instruments indicated temporal variability of the X-ray flux. An average luminosity of $55 \pm 14$ MW was found, which agreed well with that predicted by Cravens and Maurellis (2001) for scattering of solar X-rays. The LETG/ACIS–S spectrum showed, in addition to the C–K$\alpha$ and O–K$\alpha$ emission at 0.28 and 0.53 keV, also evidence for N–K$\alpha$ emission at 0.40 keV. An additional emission line was indicated at 0.29 keV, which might be the signature of the C 1s $\to \pi*$ transition in $CO_2$.

Dennerl et al. (2002) performed detailed computer simulations of fluorescent scattering of solar X-rays on Venus. The ingredients to the model were the composition and density structure of the Venusian atmosphere, the photoabsorption cross-sections and fluorescence efficiencies of the major atmospheric constituents, and the incident solar spectrum at the time of the observation. These simulations showed that fluorescence is most efficient in the Venusian thermosphere, at heights of ~120 km, where an optical depth of one is reached for incident X-rays with energy 0.2–0.9 keV (Fig. 17a). Images derived from the simulations (Fig. 18a–c) show a pronounced brightening of the sunward limb, in agreement with the observed X-ray image (Fig. 18d), while the optical image (Fig. 18e) is characterized by a different brightness distribution.

The reason for the different appearance of Venus in the optical and X-ray band is that the optical light is reflected from clouds at a height of 50–70 km, while scattering of X-rays takes place at higher regions extending into the tenuous, optically thin parts of the thermosphere and exosphere (Fig.17b). From there, the volume emissivities are accumulated along the line of sight without considerable absorption, so that the observed brightness is mainly determined by the extent of the atmospheric column along the line of sight. As a result, Venus' sun–lit hemisphere appears surrounded by an almost transparent luminous shell in X-rays, and Venus looks brightest at the limb since more luminous material is there.

Detailed comparison of the simulated images (Fig. 18a–18c) shows that the amount of limb brightening is different for the three energies. The computer simulations by Dennerl et al. (2002) indicate that this brightening depends sensitively on the density and chemical composition of the Venusian atmosphere. Thus, precise measurements of this brightening will provide direct information about the atmospheric structure in the thermosphere and exosphere.

The observed X-ray flux exhibited indications for variability on time scales of minutes. As variability of the solar X-ray flux on these time scales is not uncommon, solar X-rays scattered on Venus are expected to exhibit a similar variability. However, a direct comparison with the solar X-ray flux did not show an obvious correlation. This may be related to the fact that solar X-rays are predominantly emitted from localized regions and that Venus saw a solar hemisphere that was rotated by 46.5–48.0° from the solar hemisphere facing the Earth.

In addition to its proximity to the Sun, the high optical surface brightness of Venus (exceeded only by the Sun) is a particular challenge for X-ray observations, as most X-ray detectors are also sensitive to optical light. Suppression of optical light is usually achieved by optical blocking filters that, however, must not attenuate the X-rays significantly. An alternative and very efficient method for obtaining a clean X-ray signal is to separate the optical flux from the X-ray flux by dispersion. This was one of the techniques applied in the first Chandra observation of Venus (Dennerl et al., 2002). It unambiguously proved that the observed signal is indeed due to X-rays, despite the fact that there is on average only one X-ray photon among $5\times10^{11}$ photons from Venus. The extremely low ratio between the X-ray and optical flux from Venus is mainly caused by the comparatively low X-ray luminosity of the Sun (~$4\times10^{20}$ W; Peres et al., 2000) and the small fluorescence yields of light elements (0.25% for C and 0.85% for O; Krause 1979).

In the first X-ray observation of Venus, no evidence of charge exchange interactions was found. This is in agreement with the sensitivity of the observation, as a charge exchange induced luminosity of ≤ 1.5 MW (Cravens, 2000; Krasnopolsky, 2000; Holmström et al., 2001) would correspond to less than 8 photons. All the observational results, however, are entirely consistent with fluorescent scattering of solar X-rays in the upper Venusian atmosphere.

Thus, X-ray observations of Venus make it possible to study remotely the chemical composition and density structure of the upper atmospheric layers above 100 km. This opens up a novel method of using X-ray observations for monitoring the properties of these regions that are difficult to investigate by other means, and their response to solar activity.

## 7. Mars

The first X-ray observations of Mars were made with the ROSAT satellite on three occasions during April 10–13, 1993, yielding a total exposure of 75 minutes. Mars was not detected in these observations. However, during two observations, Mars was unfavourably placed in the field of view of the ROSAT Position Sensitive Proportional Counter, so that only 35 minutes of exposure were left to search for any X-ray emission.

On 4 July 2001, X-rays from Mars were detected for the first time (Dennerl, 2002). The observation was performed with the ACIS–I detector onboard Chandra. Although the closest approach of Mars to Earth, with a minimum distance of 0.45 AU, had occurred already on 22 June 2001, the Chandra observation was postponed by two weeks. This decision was motivated by the fact that Mars was expected to be an X-ray source mainly due to fluorescent scattering of solar X-rays, and computer simulations of this process by Dennerl had indicated that observing at a non–zero phase angle would result in a diagnostically more valuable image than observing at opposition: while a practically uniform X-ray brightness across the whole planet was expected for a phase angle close to zero, a phase angle of ~15° should result in a characteristic X-ray brightening on the sunward limb. The decision to postpone the Chandra observation was also supported by the favorable fact that Mars was still approaching the perihelion of its orbit, so that its distance from the Earth would increase only slightly to 0.46 AU. Furthermore, the small loss of X-ray photons due to the reduced solid angle would be almost compensated by the fact that Mars would then be closer to the Sun and would intercept more solar radiation.

The Chandra observation (Dennerl, 2002) yielded data of high spatial and temporal resolution, together with spectral information. Mars was clearly detected as an almost fully illuminated disk (Fig. 19), with an indication of the phase effect predicted by the computer simulations. Not only the observed morphology, but also the X-ray luminosity of ~4 MW was fully consistent with fluorescent scattering of solar X-rays in the upper Mars atmosphere. Cravens and Maurellis (2001) had predicted a luminosity of 2.5 MW due to X-ray fluorescence, with an uncertainty factor of about two. The X-ray spectrum was dominated by a single narrow emission line, which was most likely caused by O Kα fluorescence.

No evidence for temporal variability was found. This was also in agreement with fluorescent scattering of solar X-rays, because the solar X-ray flux was quite steady at that time. Dennerl (2002) noticed that the conditions encountered during this observation were favorable for testing the hypothesis of dust–related X-ray emission: scattering of solar X-rays on very small dust particles was one of the early suggestions for explaining the X-ray emission from comets. Wickramasinghe and Hoyle (1996) had noted that X-rays could be efficiently scattered by dust particles, if their size is comparable to the X-ray wavelength. Such attogram dust particles ($\sim 10^{-18}$ g) would be difficult to detect by other means. It might be possible that such particles are present in the upper Mars atmosphere, in particular during episodes of global dust storms.

Incidentally, on June 24 a local dust storm on Mars had originated and expanded quickly, developing into a planet–encircling dust storm by July 11. Such dust storms have been observed on roughly one–third of the perihelion passages during the last decades, but never so early in the Martian year. On July 4, this very vigorous storm had covered roughly one hemisphere – the hemisphere that happened to be visible at the beginning of the Chandra observation. By the end of the observation, which covered one third of a Mars rotation, this hemisphere had mainly rotated away from our view. Thus, a comparison of the Chandra data from both regions would have revealed any influence of the dust storm on the X-ray flux.

There was, however, no change in the mean X-ray flux between the first and second half of the observation, where 150 and 157 photons were detected, respectively (Dennerl, 2002). This implied that, if attodust particles were present in the upper Mars atmosphere, the dust storm did not lead to a local increase in their density high enough to modify the observed X-ray flux significantly. No statement, however, could be made about the situation below ~80 km, as the solar X-rays do not reach these atmospheric layers. While the presence of some attodust in the upper atmosphere could not be ruled out by the Chandra observation, the fact that the ACIS–I spectrum of Mars was dominated by a single emission line showed that any contribution of such particles to the X-ray flux from Mars must be small compared to fluorescence, even in the process of a developing global dust storm.

The computer simulations by Dennerl (2002) showed that scattering of solar X-rays is most efficient between 110 km (along the subsolar direction) and 136 km (along the terminator). This behaviour is similar to Venus, where the volume emissivity was found to peak between 122 km and 135 km (Dennerl et al., 2002). Although Mars was almost fully illuminated, the synthetic images produced for the same phase angle (18.2°) indicated already some brightening on the sunward limb, accompanied by a fading on the opposite limb. While a direct comparison with the observed Mars image (Fig. 19) suffers from low photon statistics, these trends can be seen in the surface brightness profiles (Fig. 20). Thus, the expected limb brightening was actually observed. The close match between the simulated and observed morphology is an argument in favour of X–ray fluorescence as the dominant process responsible for the X-ray radiation. With computer simulations of the X-ray emission due to charge exchange interactions, Holmström et al. (2001) obtained a completely different X-ray morphology.

The most exciting feature in Fig. 20, however, is the gradual decrease of the X-ray surface brightness between 1 and ~3 Mars radii. Although the excess of X-ray photons in this region was only ~ 35 ± 8 relative to the background expected for this area, the spectral distribution of these photons was different from those of Mars. This ruled out the possibility that the halo was caused by, e.g., the optics of the X-ray telescope. Furthermore, the presence of a component with the same spectral distribution as in the halo was indicated in the spectrum of Mars. Within the very limited statistical quality, the spectrum of the halo resembled that of comets. Dennerl et al. (2002) interpreted these findings as evidence for the presence of an X-ray halo around Mars, which is caused by charge exchange interactions between highly charged heavy ions in the solar wind and neutrals in the Martian exosphere.

Triggered by the discovery of cometary X-ray emission (e.g. Lisse et al., 1996; Dennerl et al., 1997; Mumma et al., 1997), the consequences of this process for the X-ray emission of Mars had already been investigated by several authors. Cravens (2000a) predicted an X-ray luminosity of ~0.01 MW. Krasnopolsky (2000) estimated an X-ray emission of ~ $4\times10^{22}$ ph s$^{-1}$. Adopting an average photon energy of 200 eV (e.g. Cravens 1997), this corresponds to an X-ray luminosity of 1.3 MW. Holmström et al. (2001) computed a total X-ray luminosity of Mars due to charge exchange (within 10 Mars radii) of 1.5 MW at solar maximum, and 2.4 MW at solar minimum.

For the X-ray halo observed within 3 Mars radii, excluding Mars itself, the Chandra observation yielded a flux of $(0.9\pm0.4)\times10^{-14}$ erg cm$^{-2}$ s$^{-1}$ in the energy range E=0.5 – 1.2 keV (Dennerl et al., 2002). Assuming isotropic emission, this flux corresponds to a luminosity of 0.5±0.2 MW. This value agrees well with the predictions of Krasnopolsky (2000) and Holmström et al. (2001), in particular when the spectral shape is extrapolated to lower energies. Using the abundances of H, $H_2$, and hot oxygen in the Martian exosphere, Krasnopolsky and Gladstone (2005) compared the observed X-ray luminosity of the Martian X-ray halo with that expected for solar wind charge exchange, and obtained consistent results.

In a detailed computer model, adjusted to the specific circumstances of the Chandra Mars observation, Gunell et al. (2004) confirmed that the contribution from the solar wind charge exchange process to the X-ray emission from the halo is large enough to explain the observed X-ray flux. A direct comparison of the azimuthally averaged radial X-ray brightness profiles, however, showed some discrepancies: the calculated count rates for the halo were higher than the observed ones by a factor between one and three. In a subsequent paper, Gunell et al. (2005) investigated the dependence of the calculated results on the parameters and assumptions used for the simulation. They found that different solar wind models produce X-ray images with significantly different structure. Other uncertainties in the simulation were the density of the neutral Martian exosphere, the composition of the solar wind, the charge exchange cross sections, and the limited spatial resolution of the model. Another uncertainty was the very low statistical data quality, as all the observational evidence about an X-ray halo around Mars was based on only ~35 ± 8 excess photons and was thus near the sensitivity limit of the Chandra observation.

The situation improved considerably with the first observation of Mars with XMM–Newton on November 19–21, 2003. This observation definitively confirmed the presence of the Martian X-ray halo and made a detailed analysis of its spectral, spatial, and temporal properties possible. High resolution spectroscopy of the halo with RGS (Dennerl et al., 2005a) revealed the presence of numerous (~12) emission lines at the positions expected for de-excitation of highly ionized C, N, O, and Ne atoms, strongly resembling a cometary X-ray spectrum. The He-like $O^{6+}$ multiplet was resolved and found to be dominated by the spin–forbidden magnetic dipole transition $2\ ^3S_1 \to 1\ ^1S_0$, confirming charge exchange as the origin of the emission. Thus, this was the first definite detection of charge exchange induced X-ray emission from the exosphere of another planet, providing a direct link to cometary X-ray emission.

In addition to these new results about the Martian X-ray halo emission, the XMM–Newton observation confirmed that the X-ray radiation from Mars itself is mainly caused by fluorescent scattering of solar X-rays: close to Mars, the RGS spectrum was dominated by fluorescence from $CO_2$. Fluorescence from $N_2$ was also observed. XMM–Newton's RGS resolved fine structure in the oxygen fluorescence, which was found to consist of two components of similar flux, resulting from a superposition of several electron transitions in the $CO_2$ molecule (Dennerl et al., 2005a). Further support for the interpretation that the X-rays from Mars itself are caused by fluorescent scattering of solar X-rays comes from the fact that the temporal behaviour of this radiation is well correlated with the solar X-ray flux. Also the Martian X–ray halo exhibited pronounced variability, but, as expected for solar wind interactions, the variability of the halo did not show any correlation with the solar X-ray flux (Dennerl et al., 2005b).

The high spectral dispersion and throughput of XMM–Newton / RGS made it possible to produce X-ray images of the Martian exosphere in individual emission lines, free from fluorescent radiation (Fig. 21). They show extended, anisotropic emission out to ~8 Mars radii, with pronounced morphological differences between individual ions and ionization states. While the emission from ionized oxygen (Fig. 21a-c) appears to be concentrated in two distinct blobs a few thousand kilometers above the Martian poles, with larger heights for $O^{7+}$ than for $O^{6+}$, the emissions from ionized carbon (Fig. 21d-f) exhibit a more band–like structure without a pronounced intensity dip at the position of Mars. For comparison, the morphology of the fluorescent radiation (Fig. 21g-i), is clearly concentrated at the planet. While an interpretation of the fluorescence images is straightforward, an interpretation of the structures in the halo emission is not an easy task, since they depend on many parameters, as listed above (Gunell et al., 2005).

Holmström and Kallio (2004) noted that since crustal magnetizations at Mars are asymmetrically distributed, they will also introduce asymmetries in the solar wind flow around the planet and thus in the X-ray emission. They also remarked that due to the considerable size of the ion gyroradii (~0.3 Martian radii), kinetic effects are very pronounced. Due to its sensitive dependence on so many parameters, the X-ray emission of the Martian halo contains a wealth of valuable information.

## 8. Jupiter

### 8.1. Auroral Emission

Jupiter's ultraviolet auroral emissions were first observed by the International Ultraviolet Explorer (IUE) and soon confirmed by the Voyager 1 Ultraviolet Spectrometer as it flew through the Jupiter system (see Bhardwaj and Gladstone, 2000 for review). X-ray observations of the aurora followed in the same time period (Metzger et al., 1983). The spatial resolution of these early observations both in the X-ray and ultraviolet regions of the electromagnetic spectrum were not adequate to distinguish whether the emissions were linked to source regions in the Io torus of Jupiter's magnetosphere or at larger radial distances from the planet. However, arguments about the large amount of energy required to produce the X-ray emissions by electron bremsstrahlung and the observation by Gehrels and Stone (1983) of a large decrease in the radial phase space distribution of energetic sulfur and oxygen ions 6-8 jovian radii from the planet led Metzger et al. to conclude that the source of the X-ray aurora was energetic heavy ion precipitation from a region just outside the Io plasma torus. Models of the aeronomical effects of energetic electron precipitation (Waite et al., 1983) and energetic oxygen ion precipitation (Horanyi et al., 1988) soon followed in hopes of better constraining the observational attributes that distinguish electron from heavy ion precipitation. The models suggested that most of the aeronomical effects were similar, but that both ultraviolet and X-ray emissions from the precipitating energetic oxygen and sulfur were the most easily recognizable observational marker of ion precipitation. Spectral searches to look for these auroral emission lines were carried out in the ultraviolet (Waite et al., 1988) and at X-ray wavelengths (Waite et al., 1994) using the IUE and ROSAT satellites, respectively. No clear signatures were observed for sulfur near 1256 Å or for oxygen at 1304 Å. Emissions were again seen by ROSAT at X-ray wavelengths, but the spectral resolution was low enough that only a weak preference for line emission as opposed to an electron power law photon distribution was inferred. Furthermore, the spatial resolution of these observations was not significantly better than before. Therefore, the location of the source region and the identity of the responsible precipitating particles remained unconfirmed. A series of observations with ROSAT was carried out over the later half of the 1990's that provided information on the variability of the intensity of the auroral X-rays (Gladstone et al., 1998), better constrained the X-ray charge exchange model of X-ray generation (Cravens et al., 1995), and identified an unexpected correlation with the impact of fragments of the comet Shoemaker-Levy 9 (Waite et al., 1995).

The advent of the Hubble Space Telescope (HST) and the Chandra and XMM-Newton X-ray observatories revolutionized our thinking about Jupiter's aurora. Several investigators began making routine observations of Jupiter's ultraviolet aurora and of the high-latitude emissions that were magnetically-linked to the Galilean moons. The combination was able for the first time to clearly show that Jupiter's aurora originated at radial distances outside the orbits of most of the Galilean moons between 20 and 30 Jovian radii from the planet (cf., Clarke et al., 1998, 2002). That the aurora was magnetically linked to the middle magnetosphere coupled with the high degree of variability of emissions in the dawn sector led Gerard et al. (1994) to speculate that the aurora was driven by plasma shear processes in Jupiter's middle magnetosphere. Theoretical confirmation soon followed, showing that the shears generated in the plasma corotation break down region of the middle magnetosphere (Vasyliunas, 2002) were likely responsible for the auroral generation (Hill, 2001; Bunce and Cowley, 2001), but this implied an electron precipitation source for the ultraviolet aurora that seemed incompatible with the X-ray observations that favored heavy ion precipitation. HST spectral observations at high resolution were invoked to try and settle the issue of electron versus ion precipitation. A search for ultraviolet line emissions from sulfur and oxygen ion precipitation was renewed (Trafton et al., 1994, 1998), but with little more success than before. Where were the heavy ion signatures in the ultraviolet?

Chandra's improved spectral and spatial resolution was to answer the questions. After a false start due to higher sensitivity in the red portion of the visible spectrum than expected, Chandra obtained the first observations of Jupiter's X-ray aurora. Surprisingly the emission was not located in the ultraviolet auroral zone at all, it was located at higher latitudes near Jupiter's pole – regions that mapped to the outer boundary of Jupiter's magnetosphere (Figure 22) and it was pulsing regularly with a forty minute period (see Figure 23) (Gladstone et al., 2002). The periodicity was highly reminiscent of a class of Jupiter radio emissions known as quasi-periodic radio bursts, which had been observed by Ulysses in conjunction with energetic electron acceleration in Jupiter's outer magnetosphere (MacDowall et al., 1993). Furthermore, the high latitude location of the X-ray source region was reminiscent of the position of an ultraviolet auroral storm region where intense auroral emissions were seen to rise to emission levels of over 50 megarayleighs in the $H_2$ Lyman and Werner band system and decline to background levels of a few kilorayleighs on timescales on the order of a minute (Waite et al., 2001). This same polar cap emission region was also identified with an infrared hot spot (Orton et al., 1990) at 7.8 microns and a region of increased high altitude aerosol production (West et al., 2002) probably both associated with the intense auroral particle input that had been seen in this region. Chandra observations continued using both the ASIC-S and the HRC (Elsner et al., 2005a) to get a high signal to noise spectrum of the X-ray emissions and to pin down the auroral variability in time and space, respectively.

The spectral observations from this observational sequence gave a clear indication of heavy ion oxygen precipitation with some weaker indication of sulfur (or carbon) emission (see figure 24) (Elsner et al., 2005a). The line emission from the Jovian X-ray aurora had been confirmed at long last. Although, the X-ray pulsations with a period of forty minutes were not present during this latter observational sequence, variability at similar time scales was seen, but much less organized. Simultaneous radio observations by the Ulysses radio experiment indicated that the radio emission variation demonstrated no strong periodic structure as on past occasions. Nonetheless, it was disappointing that no statistical correlation could be found (Elsner et al., 2005a). Simultaneous observations at ultraviolet wavelengths by HST and with Chandra at X-ray wavelengths were much more successful. Time-tagged observations at both wavelengths allowed the identification of a Jovian polar cap auroral flare. However, the spatial correlation was not as expected. The X-rays did increase in time in a manner consistent with the ultraviolet flare, but rather than peak at the ultraviolet flare location they were peaked in a morphologically associated region, the "kink", which most likely magnetically maps to the dusk flank of Jupiter's magnetosphere (see Figure 25) (Elsner et al., 2005a).

A clear spectral signature of oxygen line emission and a high-latitude source region that mapped to near the magnetopause boundary presented other difficulties - as long as the X-ray aurora was at near auroral latitudes there had been sufficient phase space density of energetic sulfur and oxygen ions in the middle magnetosphere to account for the X-ray emissions, but X-ray emissions at high latitudes linked to the outer magnetosphere, which did not have sufficient phase space density of energetic heavy ions to provide the measured X-ray intensity. Acceleration of energetic ions was invoked to increase the phase space distribution, but now the question was whether the acceleration involved magnetospheric heavy ions or solar wind heavy ions (in the intervening years Cravens (1997) had shown that unexpected X-ray emissions at comets could be explained by solar wind charge exchange of heavy ions, a suggestion first made for Jupiter (Horanyi et al., 1988)). Cravens et al. (2003) explored this acceleration process and demonstrated that both sources would work, but required megavolt accelerations. However, this acceleration level was exactly what was needed to explain the energetic electrons seen by Ulysses (McKibben et al., 1993) and apparently associated with the quasi-periodic radio emissions. Cravens et al. (2003) also indicated a marker that might be used to distinguish a magnetospheric source from a solar wind source - the solar wind source would produce copious quantities of Lyman alpha from the acceleration of solar wind protons, although they might be energetic enough to bury the Lyman alpha emission deep enough within Jupiter's magnetosphere to not be seen.

Subsequent spectral observations at medium and high resolution by XMM-Newton (Branduardi-Raymont et al., 2004, 2006a,b) indicated the presence of very strong He-like oxygen X-ray emission lines from Jupiter's aurorae; an early suggestion that carbon may be a better candidate than sulfur for explaining other spectral features at lower energies was not confirmed by further observations. However, for the first time XMM-Newton identified a higher energy component in the auroral spectra (see figure 26), and found it to be variable on timescales of days: its spectral shape is consistent with that predicted from bremsstrahlung of energetic electrons precipitating from the magnetosphere (Figure 27). The variability suggests a link to changes in the energy distribution of the electrons, and may be related to the intense solar activity reported at the time (November 2003).

So is the solar wind or the magnetosphere responsible for the heavy ions? The answer may be that both contribute. Bunce et al. (2004) have put forth a theory that suggest that magnetic reconnection near the cusp of Jupiter's magnetosphere may be responsible for the acceleration of the heavy ions and from analogy with the Earth, both solar wind and magnetospheric plasma are accelerated in this process. Many questions remain concerning the details of temporal variation and spatial mapping, as well as the existence and fate of accelerated heavy particles, ions and electrons within Jupiter's aurora. However, at last a plausible working hypothesis exists that links the X-ray observations to a generation mechanism. Perhaps one day in-situ Jupiter observations will finally complete the story.

### 8.2. Non-Auroral (Disk) Emission

The existence of low-latitude disk X-ray emission from Jupiter was first recognised in ROSAT observations (Waite et al. 1997). The X-rays were initially thought to be the result of precipitation of energetic S and O ions out of Jupiter's inner radiation belts into the planet's atmosphere. This view was supported by a suggested correlation between low magnetic field regions and enhanced emission. However, a further correlation between visible and X-ray limb brightness was taken to indicate that a solar-driven process may be at work (Gladstone et al. 1998). Two alternative mechanisms were then proposed, which could explain the low-latitude ROSAT measurements well (Maurellis et al. 2000): they are elastic scattering of solar X-rays by atmospheric neutrals and fluorescent scattering of C K-shell X-rays off methane molecules located below the Jovian homopause.

The clearest view yet of Jupiter's X-ray emissions was obtained in the year 2000 with a Chandra HRC-I observation (Fig. 28, Gladstone et al. 2002). In addition to resolving two auroral 'hot spots' in the polar regions, the HRC-I image clearly displays a relatively uniform distribution of X-ray emission covering the whole of Jupiter's disk at low latitude. However, no information on the X-ray spectrum was obtainable at this stage. Observations by XMM-Newton in April and November 2003 provided clear evidence of differences in spectral shape between the low-latitude disk emission and the aurorae. This is vividly illustrated in Fig. 29 (Branduardi-Raymont et al. 2006a, b, this issue) where combined EPIC spectra of the aurorae and of the low-latitude disk are displayed: the disk spectrum peaks at higher energies (0.7 – 0.8 keV) than the aurora (0.5 – 0.6 keV) and lacks the high energy component (above ~ 3 keV) present in the latter. Branduardi-Raymont et al. (2006b, this issue, Fig. 2) show EPIC images from the Nov. 2003 XMM-Newton observation in narrow spectral bands centered on the $O^{6+}$, $O^{7+}$, $Fe^{16+}$ and $Mg^{10+}$ emission lines: the $O^{6+}$ emission peaks clearly at the North and (more weakly) South auroral spots, $O^{7+}$ extends to lower latitudes, with an enhancement at the North spot, while MgXI and especially $Fe^{16+}$ display a more uniform distribution over the planet's disk.

The disk X-ray spectrum measured by XMM-Newton is well fitted by a coronal plasma model with temperature in the range 0.4 – 0.5 keV and solar abundances, after including additional $Mg^{10+}$ and $Si^{12+}$ emission (at 1.35 and 1.86 keV respectively): the presence of these lines in the spectrum is a likely consequence of solar activity, which we know was enhanced over the Oct.–Nov. 2003 period. The XMM-Newton spectral data, then, appear to give support to the view that Jupiter's low-latitude disk X-rays are scattered solar radiation. Similar results showing the disk emission to be harder than the auroral emission, were reached from the February 2003 Chandra ACIS-S observation of the planet (Bhardwaj et al., 2004; Elsner et al., 2005a). These Chandra observations appear to confirm a correlation between regions of low magnetic field strength and somewhat higher soft X-ray count rate, suggesting the possibility of a secondary component in addition to the scattered solar X-ray flux. Unlike the auroral X-ray emission, X-ray emission from Jupiter's disk does not show any variability on timescales from 10-100 m.

The conclusions derived from spectral studies of Jupiter are strengthened by the observation, again in November 2003 by XMM-Newton, of similar day-to-day variability in the solar and Jovian equatorial X-ray fluxes. A large solar X-ray flare occurring on the Jupiter-facing side of the Sun is found to have a corresponding feature in Jupiter's disk X-ray lightcurve (see Fig. 30, from Bhardwaj et al., 2005a). This finding lends support to the view that indeed Jupiter's low latitude X-rays are largely scattered radiation of solar origin. Recent calculations of the effects of albedo (Cravens et al., 2006) are consistent with the observed fluxes, under the hypothesis that scattering, elastic and fluorescent, is at the origin of the observed emission. These results, however, do not rule out the presence of other source(s) of low-latitude X-ray emission, e.g., precipitation of energetic S and O ions from Jupiter's radiation belts, especially into regions of low magnetic field.

## 9. Saturn

By analogy with Jupiter, X-ray emission from Saturn might be expected since it possesses a magnetosphere with energetic electrons and ions within it. The detection of X-rays from Jupiter by the Einstein observatory only prompted the search for Saturnian X-rays (cf. Bhardwaj and Gladstone, 2000). Early attempts to detect X-ray emission from Saturn with Einstein in 1979 December (Gilman et al., 1986) and with ROSAT in 1992 April (Ness and Schmitt, 2000) were negative and marginal, respectively.

Saturnian X-rays were unambiguously observed by XMM-Newton in October 2002 (Ness et al. 2004a) and by the Chandra X-Ray Observatory in April 2003 (Ness et al. 2004b). The detections by the XMM-Newton EPIC-pn and the Chandra ACIS-S were based on 162 photons in each observation with only 56 and 50, respectively, expected from background (Ness et al. 2004a,b). However, the observed X-ray flux from Saturn was much higher during Oct. 2002 ($1.6 \times 10^{-14}$ erg cm$^{-2}$ s$^{-1}$) than in April 2003 ($0.68 \times 10^{-14}$ erg cm$^{-2}$ s$^{-1}$). The X-rays seen by the Chandra ACIS-S appeared concentrated in the low-latitudes with no apparent brightening near the polar regions (Ness et al., 2004b). Ness et al (2004a,b) found statistically acceptable fits for simple spectral models, but were unable to attribute a specific physical mechanism for Saturn's X-ray emission.

In January 2004, Saturn was again observed by the Chandra ACIS-S in two exposures, one on 20 January and the other on 26-27 January, each observation lasted for about one full Saturn rotation (Bhardwaj et al. 2005b). These observations showed (Fig. 31) that X-rays from Saturn are highly variable – a factor of 2 to 4 variability in brightness over one week. These observations also revealed X-rays from Saturn's south polar cap on January 20 (see Fig. 31, left panel), which are not evident in the January 26 observation (see Fig. 31, right panel) and in earlier Chandra observations (Ness et al. 2004b). X-rays from the south polar cap region were present only in the 0.7-1.4 keV energy band, in contrast with Jupiter's X-ray aurora for which the emission is mostly in the bands 0.3-0.4 keV and 0.6-0.7 keV (Branduardi-Raymont et al., 2004, Elsner et al., 2005a). Any emission from the north polar cap region was blocked by Saturn's rings. It seems unlikely that the X-ray emission from the south polar cap is auroral in nature, and more likely that they are an extension of the disk X-ray emission.

An X-ray flare was detected from the disk of Saturn during the Chandra observation on 20 January 2004 (Bhardwaj et al., 2005b). Taking light travel time into account, this X-ray flare from Saturn coincided with an M6-class flare emanating from a sunspot that was clearly visible from both Saturn and Earth.. Moreover, the lightcurve for the flare from Saturn was very similar to that of the solar X-ray flare (Fig. 32). This was the first direct evidence suggesting that Saturn's disk X-ray emission is principally controlled by processes happening on the Sun (Bhardwaj et al. 2005b). A good correlation has been observed between Saturn X-rays and F10.7 solar activity index, suggesting a solar connection. As is the case for Jupiter's disk, X-ray emission from Saturn seems likely to be due to the scattering of the incident solar X-ray flux (Bhardwaj et al. 2005b).

The spectrum of X-rays from Saturn's disk is very similar to that from Jupiter's disk (Bhardwaj et al. 2006, in preparation). Saturn's disk spectrum measured on Jan. 20, 2004 is quite similar to that measured on April 14-15, 2003 in the 0.3-0.6 keV range. However, at energies 0.6–1.2 keV the former is stronger by a factor of 2 to 4. This is probably due to the M6-class solar X-ray flare on January 20, with a corresponding hardening of the solar X-ray flux (and hence of Saturn's X-ray emission). This is supported by spectral fitting results (Bhardwaj et al., 2006, in preparation) showing that the best-fitting MEKAL model requires a higher temperature for the Jan. 20, 2004, data (0.69 keV) than for the April 14-15, 2003, data (0.39 keV).

Thus, observations of Saturn's X-ray emissions suggest that they are variable on timescales of hours to weeks to months and are mostly produced by scattering of solar X-rays by atmospheric species in its upper atmosphere (Bhardwaj, 2006; Cravens et al. 2006). The X-ray power emitted from Saturn's disk is roughly one-fourth of that from Jupiter's disk, which is consistent with Saturn being twice as far as Jupiter from Sun and Earth.

## 10. Comets

The discovery of high-energy X-ray emission in 1996 from C/1996 B2 (Hyakutake) has revealed the existence of a new class of X-ray emitting objects (Lisse et al., 1996). Observations since 1996 have shown that the very soft (E < 1 keV) emission is due to an interaction between the solar wind and the comet's atmosphere, and that X-ray emission is a fundamental property of comets. Theoretical and observational work has demonstrated that charge exchange collisions of highly charged solar wind ions with cometary neutral species is the best explanation for the emission. Now a rapidly changing and expanding field, the study of cometary X-ray emission appears to be able to lead us to a better understanding of a number of physical phenomena: the nature of the cometary coma, other sources of X-ray emission in the solar system, the structure of the solar wind in the heliosphere, and the source of the local soft X-ray background.

The observed characteristics of the emission can be organized into the following 4 categories: (1) spatial morphology, (2) total X-ray luminosity, (3) temporal variation, and (4) energy spectrum. Any physical mechanism that purports to explain cometary X-ray emission must account for all of these characteristics.

X-ray and EUV images of C/1996 B2 (Hyakutake) made by the ROSAT and EUVE satellites look very similar (Lisse et al., 1996; Mumma et al., 1997) (Figure 33). Except for images of C/1990 N1 (Dennerl et al. 1997) and C/Hale-Bopp 1995 O1 (Krasnopolsky et al., 1997), all EUV and X-ray images of comets have exhibited similar spatial morphologies. The emission is largely confined to the cometary coma between the nucleus and the Sun; almost no emission is found in the extended dust or plasma tails. The peak X-ray brightness gradually decreases with increasing cometocentric distance r with a dependence of about $r^{-1}$ (Krasnopolsky, 1997; Dennerl, 1997). The brightness merges with the soft X-ray background emission (McCammon and Sanders 1990; McCammon et al., 2002) at distances that exceed $10^4$ km for weakly active comets, and can exceed $10^6$ km for the most luminous (Dennerl et al., 1997; Figure 34). The region of peak emission is crescent-shaped with a brightness peak displaced towards the Sun from the nucleus (Lisse et al., 1996, 1999). The distance of this peak from the nucleus appears to increase with increasing values of the gas production rate, Q, and for Hyakutake was located at $r_{peak} \approx 2 \times 10^4$ km (Figure 33b).

The observed X-ray luminosity, $L_x$, of C/1996 B2 (Hyakutake) was $4 \times 10^{15}$ ergs s$^{-1}$ (Lisse et al. 1996) for an aperture radius at the comet of $1.2 \times 10^5$ km. [Note that the photometric luminosity depends on the energy bandpass and on the observational aperture at the comet. The quoted value assumes a ROSAT photon emission rate of $P_X \approx 10^{25}$ s$^{-1}$ (0.1 – 0.6 keV), in comparison to Krasnopolsky et al.'s (2000) EUVE estimate of $P_{EUV} \approx 7.5 \times 10^{25}$ s$^{-1}$ (0.07 - 0.18 keV and 120,000 km aperture).] A positive correlation between optical and X-ray luminosities was demonstrated using observations of several comets having similar gas ($Q_{H2O}$) to dust (Afρ, following A'Hearn et al. 1984) emission rate ratios (Figure 35) (Lisse et al. 1997b, 1999, 2001, Dennerl et al. 1997, Mumma et al. 1997; Krasnopolsky et al. 2000). $L_x$ correlates more strongly with the gas production rate $Q_{gas}$ than it does with $L_{opt} \sim Q_{dust} \sim$ Afρ (Figures 34, 35). Particularly dusty comets, like Hale-Bopp, appear to have less X-ray emission than would be expected from their overall optical luminosity $L_{opt}$. The peak X-ray surface brightness decreases with increasing heliocentric distance r, independent of Q (Dennerl et al. 1997), although the total luminosity appears roughly independent of r. The maximum soft X-ray luminosity observed for a comet to date is $\sim 2 \times 10^{16}$ erg s$^{-1}$ for C/Levy at 0.2 – 0.5 keV (Dennerl et al. 1997; Lisse et al. 1999, 2001; Figure 35).

Photometric lightcurves of the X-ray and EUV emission typically show a long-term baseline level with superimposed impulsive spikes of a few hours' duration, and maximum amplitude 3 to 4 times that of the baseline emission level (Lisse et al. 1996, 1999, 2001). Figure 36 demonstrates the strong correlation found between the time histories of the solar wind proton flux (a proxy for the solar wind minor ion flux), the solar wind magnetic field intensity, and a comet's X-ray emission, for the case of comet 2P/Encke 19997 (Lisse et al. 1999). Neugebauer et al. (2000) compared the ROSAT and EUVE luminosity of C/1996 B2 (Hyakutake) with time histories of the solar wind proton flux, oxygen ion flux, and solar X-ray flux, as measured by spacecraft residing in the solar wind. They found the strongest correlation between the cometary emission and the solar wind oxygen ion flux, a good correlation between the comet's emission and the solar wind proton flux, but no correlation between the cometary emission and the solar X-ray flux.

Until 2001, all published cometary X-ray spectra had very low spectral energy resolution ($\Delta E/E \sim 1$ at 300–600 eV) and the best spectra were those obtained by ROSAT for C/1990 K1 (Levy) (Dennerl et al. 1997) and by BeppoSAX for comet C/ 1995 O1 (Hale-Bopp) (Owens et al. 1998). These observations were capable of showing that the spectrum was very soft (characteristic thermal bremsstrahlung temperature kT ~0.23 ± 0.04 keV) with intensity increasing towards lower energy in the 0.01 to 0.60 keV energy range, and established upper limits to the contribution of the flux from K-shell resonance fluorescence of carbon at 0.28 keV and oxygen at 0.53 keV. However, even in these "best" spectra, continuum emission (such as that produced by the thermal bremsstrahlung mechanism) could not be distinguished from a multi-line spectrum, such as would result from the SWCX mechanism. Non-detections of comets C/Hyakutake, C/Tabur, C/Hale-Bopp, and 55P/Temple-Tuttle using the XTE PCA (2-30 keV) and ASCA SIS (0.6-4 keV) imaging spectrometers were consistent with an extremely soft spectrum (Lisse et al. 1996, 1997).

Higher resolution spectra of cometary X-ray emission have now appeared in the literature. The Chandra X-ray Observatory (CXO) measured soft X-ray spectra from comet C/1999 S4 (LINEAR) (Lisse et al. 2001) over an energy range of 0.2 - 0.8 keV, and with a full width half maximum energy resolution of ΔE = 0.11 keV (Figure 37). The spectrum is dominated by line emission from $C^{+4}$, $C^{+5}$, $O^{+6}$, and $O^{+7}$ excited ions, not by continuum. A spectrum of comet C/1999 T1 (McNaught-Hartley) (Krasnopolsky et al. 2002) showed similar line emission features, with a somewhat higher ratio of $O^{+6}$ to $O^{+7}$ emission, and emission due to $Ne^{+9}$. A new spectrum of comet 2P/Encke (Lisse et al. 2005) shows a very different ratio of line emission in the $C^{+4}$, $C^{+5}$, $O^{+6}$, and $O^{+7}$ lines, due to the collisionally thin nature of the low activity coma, and the unusual post-shock charge state of the solar wind at the time of observation. Line emission is also found in XMM-Newton spectra of comet C/1999 T1 (McNaught-Hartley) and, more recently, in CXO spectra of C/2000 WM1 (LINEAR) and C/2002 C1 (Ikeya-Zhang) (Dennerl et al. 2003, private communication). An XMM-Newton spectrum of C/2000 WM1 (LINEAR) shows characteristic SWCX X-ray signatures in unprecedented detail (Dennerl et al. 2003).

From other work, there are suggestions of charge exchange line emission from other species than $C^{+4}/C^{+5}$, $O^{+6}/O^{+7}$, and $Ne^{+9}$. A re-analysis of archival EUVE Deep Survey spectrometer spectra (Krasnopolsky and Mumma 2001) suggests EUV line emission features from comet C/1996 B2 (Hyakutake) due to $O^+$, $O^{+5}$, $O^{+4}$, $C^{+4}$, $O^{+6}$, $C^{+5}$, $He^+$, $Ne^{+7}$ (Figure 38a, b). Krasnopolsky et al. have suggested emission lines due to Mg and Si in C/McNaught-Hartley (Krasnopolsky et al. 2002), and $He^{+2}$ in C/Hale-Bopp (Krasnopolsky et al. 1997), although these remain unconfirmed and controversial due to the sensitivity of the results on the details of the instrumental background subtraction. Lisse et al. (2005) find possible emission due to $N^{+6}$ at 425 eV contributing to a reduced 380/450 eV ratio.

A large number of explanations for cometary X-rays were suggested following the discovery paper in 1996. These included thermal bremsstrahlung emission due to solar wind electron collisions with neutral gas or dust in the coma (Bingham et al. 1997; Dawson et al. 1997; Northrop 1997; Northrop et al. 1997; Uchida et al. 1998; Shapiro et al. 1999), microdust collisions (Ibadov, 1990; Ip and Chow, 1997), K-shell ionization of neutrals by electron impact (Krasnopolsky, 1997), scattering or fluorescence of solar X-rays by cometary gas or by small dust grains (Lisse et al. 1996; Wickramasinghe and Hoyle 1996, Owens et al. 1998), and by charge exchange between highly ionized solar wind ions and neutral species in the cometary coma (SWCX) (Cravens 1997a; Häberli et al. 1997; Wegmann et al. 1998; Kharchenko and Dalgarno 2000; Kharchenko et al. 2003; Schwadron and Cravens 2000). Early evaluation of these various mechanisms (Dennerl et al. 1997; Krasnopolsky 1997; Lisse et al. 1999) favored only three of them: the SWCX mechanism, thermal bremsstrahlung, and scattering of solar radiation from very small (i.e., attogram—$10^{-19}$ g) dust grains, with SWCX mechanism suggested as the dominant process.

A significant problem with mechanisms involving solar wind electrons (i.e., bremsstrahlung or K-shell ionization) is that the predicted luminosities are too small by factors of 100 – 1000 because the fluxes of high-energy electrons near comets are too low (Krasnopolsky 1997, Cravens 2000a, 2002). Furthermore, X-ray emission has been observed out to great distances from the nucleus, beyond the bow shock (Figure 34), and the thermal energy of unshocked solar wind electrons at these distances is about 10 eV. No emission has ever been found to be associated with the plasma tail of a comet. Finally, the new, high resolution spectra demonstrating multiple atomic lines are inconsistent with a continuum-type mechanism or a mechanism producing only a couple of K-shell lines as the primary source of cometary X-rays.

Mechanisms based on dust grains also have a number of problems. It was known since the discovery of cometary X-ray radiation in 1996 that Rayleigh scattering of solar X-ray radiation from ordinary cometary dust grains (i.e., about 1 µm in size) cannot produce the observed luminosities – the cross section for this process is too small. A potential solution to this problem is to invoke a population of very small, attogram ($10^{-19}$ g) grains with radii on the order of the wavelengths of the observed X-ray radiation, 10 to 100 Å, which can resonantly scatter the incident X-ray radiation. The abundance of such attogram dust grains is not well understood in comets, as they are undetectable by remote optical observations; however, there were reports from the VEGA Halley flyby of a detection of an attogram dust component using the PUMA dust monitor (Vaisberg et al. 1987, Sagdeev et al. 1990). However, the statistical studies of the properties of several comets (Figures 34, 35) demonstrate that X-ray emission varies with a comet's gas production rate and not the dust production rate (Dennerl et al. 1997, Lisse et al. 1999, 2001). Further, the cometary X-ray lightcurves (Lisse et al. 1996, 1999, 2001; Neugebauer et al. 2000) correlate with the solar wind ion flux and not with solar X-ray intensity. Finally, dust scattering mechanisms cannot account for the pronounced lines seen in the new high resolution spectra – emission resulting from dust scattering of solar X-rays should mimic the Sun's X-ray spectral continuum, with some blueing added, similar to what is observed in the terrestrial atmosphere for Rayleigh scattering of sunlight (Krasnopolsky, 1997).

The SWCX mechanism requires that the observed X-ray emission is driven by the solar wind flux and that the bulk of the observed X-ray emission be in lines. Localization of the emission to the sunward half of the coma, solar wind flux-like time dependence, and line emission dominated spectral signature of the observed emission all strongly point to the solar wind charge exchange mechanism as being responsible for cometary X-rays.

Numerical simulations of the solar wind interaction with Hyakutake including SWCX have been used to generate X-ray images. A global magnetohydrodynamic (MHD) model (Haberli et al. 1997) and a hydrodynamic model (Wegmann et al. 1998) were used to predict solar wind speeds and densities and the X-ray emission around a comet. The simulated X-ray images are similar to the observed images. Recent work by Wegmann et al. (2004) has shown that by determining the location of the emission maximum in the collisionally thick case, the neutral gas production rate can be determined in 5 comets observed by ROSAT and XMM. Using comet C/2000 WM1, Wegmann and Dennerl (2005) used the location of rapid changes in the first derivative of the projected surface brightness with distance from the nucleus to determine the position of the cometary bow shock.

On the other hand, it is not clear that the emission pattern always follows the plasma structures. Newer work by Bodewits (Bodewits et al., 2004; Lisse et al., 2005) suggests that the crescent-shaped, sunward offset morphology is found only for comets with a coma dense enough to be in the collisionally thick regime - for low activity comets, the emission will be maximum wherever the coma has its maximum density, typically at the nucleus (Fig 33b). This may explain the unusual emission morphologies seen in comets like d'Arrest (Mumma et al. 1997), and 2P/Encke 2003 (Lisse et al. 2005).

Up until now, the temporal variation of the solar wind dominated the observed behavior on all but the longest timescales of weeks to months. A "new" form of temporal variation has recently been demonstrated in the Chandra observations of comet 2P/Encke 2003 (Lisse et al. 2005), wherein the observed X-ray emission is modulated at the 11.1 hr period of the nucleus rotation (Fig 36b). Rotational modulation of the signal should only be possible in collisionally thin (to SWCX) comae with weak cometary activity, where a change in the coma neutral gas density can directly affect the power density of cometary X-ray.

Model SWCX spectra are in good agreement with the low resolution X-ray spectra of cometary X-ray emission, and the line centers of the high resolution spectra have been successfully predicted using SWCX theory (Figures 37 and 38; Lisse et al. 1999, 2001; Krasnopolsky and Mumma 2001; Weaver et al. 2002). The application of the SWCX model to comets has entailed a number of approaches to date. Some work has included only a few solar wind species but used a careful cascading scheme (Haberli et al., 1997). Other approaches have used a simple cascading scheme and simple collision cross sections, but included a large number of solar wind ions and charge states (Wegmann et al. 1998, Schwadron and Cravens 2000). Kharchenko and Dalgarno (2000) and Kharchenko et al. (2003) have treated the atomic cascading process more carefully than other modelers, although the spatial structure of the solar wind-cometary neutral interaction in their model was highly simplified. They predicted the existence of a large number of atomic lines, including the following: $O^{5+}$ ($1s^2 5d \rightarrow 1s^2 2p$) at 106.5 eV, $C^{+4}$ ($1s2s \rightarrow 1s^2$) at 298.9 eV, $C^{+5}$ ($2p \rightarrow 1s$) at 367.3 eV, $C^{+5}$ ($4p \rightarrow 1s$) at 459.2 eV, and $O^{+6}$ (1s2p ->1s$^2$) at 568.4 eV. At least some of these lines appear in the best cometary X-ray spectra to date, the CXO ACIS-S spectra of C/1999 S4 (LINEAR) and C/1999 T1 (McNaught-Hartley): the peak measured near 0.56 keV is certainly the combination of 3 closely spaced helium-like O$^{+6}$ (1s2p and 1s2s ->1s$^2$) transitions (which comes from SWCX of solar wind O$^{+7}$) and the line located at 0.32 keV is due to helium-like C$^{+4}$ (1s2p and 1s2s ->1s$^2$). Similar identifications have been made for the EUVE spectrum of C/1996 B2 (Hyakutake) (Krasnopolsky and Mumma 2001).

Mainly driven by the solar wind, cometary X-rays provide an observable link between the solar corona, where the solar wind originates, and the solar wind where the comet resides. Once we have understood the SWCX mechanism's behavior in cometary comae in sufficient detail, we will be able to use comets as probes to measure the solar wind throughout the heliosphere. This will be especially useful in monitoring the solar wind in places hard to reach with spacecraft – such as over the solar poles, at large distances above and below the ecliptic plane, and at heliocentric distances greater than a few AU (Lisse et al. 1996, 2001, Krasnopolsky et al. 2000). For example, ~1/3 of the observed soft X-ray emission is found in the 530-700 eV oxygen O$^{+7}$ and O$^{+6}$ lines; observing photons of this energy this will allow studies of the oxygen ion charge ratio of the solar wind, which is predicted to vary significantly between the slow and fast solar winds (Neugebauer et al. 2000; Schwadron and Cravens 2000; Kharchenko and Dalgarno 2001).

**11. Io Plasma Torus**

CXO observations on 25-26 November 1999 and 18 December 2000 detected a faint diffuse source of soft X-rays from the region of the Io plasma torus (IPT) (Elsner et al. 2002). The 2000 CXO image, obtained with the HRC-I camera (Figure 39), exhibited a dawn-dusk asymmetry similar to that seen in the EUV (Hall et al. 1994; Gladstone and Hall 1998). Figure 40 shows the background-subtracted CXO/ACIS-S IPT spectrum for 25-26 November 1999. This spectrum shows evidence for line emission centered on 574 eV (very near a strong O$^{+6}$ line), together with a very steep continuum spectrum at the softest X-ray energies. There could be contributions from other charge states, since plasma torus models (Bagenal et al. 1992, Bagenal 1994) consist mostly of ions with low charge states, consistent with photoionization and ion-neutral charge exchange in a low-density plasma and neutral gas environment. Keeping in mind that the calibration of the ACIS energy response is most uncertain at these very low energies, the corresponding power emitted by the IPT was ~0.12 GW. Elsner et al. (2002) were unable to provide a convincing physical mechanism for the observed IPT X-ray emissions, although they noted that bremsstrahlung from nonthermal electrons might account for a significant fraction of the continuum X-rays. The 2003 CXO observations also observed X-ray emission from the IPT, although at a fainter level than in 1999 or 2000. The morphology exhibited the familiar dawn-dusk asymmetry. These IPT data are not yet fully analyzed and will be reported on in a future paper.

## 12. Galilean Satellites

The 1999 and 2000 observations also discovered X-ray emission from the Galilean satellites (Elsner et al. 2002, see Figure 41). These satellites are very faint when observed from Earth orbit and the detections of Io and Europa, while statistically very significant, were based on ~10 photons each, a tribute to the high spatial resolution and low background of the CXO. The energies of the detected X-ray events ranged between 300 and 1890 eV and appeared to show a clustering between 500 and 700 eV, suggestive of oxygen K-shell fluorescent emission. The estimated power of the X-ray emission was 2 MW for Io and 3 MW for Europa. There were also indications of X-ray emission from Ganymede. X-ray emission from Callisto seems likely at levels not too far below the CXO sensitivity limit, since the magnetospheric heavy ion fluxes (Cooper et al. 2001) are an order of magnitude lower than at Ganymede and Europa, respectively. No Galilean satellite was detected during the 24-26 February 2003 CXO observations. However, both Jupiter's disk and the IPT were about 1/2 as bright as in the earlier observations; reduction in the emission from the moons by a similar factor is likely the cause of the failure to detect them in the new observations.

The emissions from Europa are best explained by energetic H, S, and O ion bombardment of the icy surface and subsequent fluorescent emission from the deposition of energetic particle energy in the top 10 microns of the surface, which is optically thin to outgoing X-ray emission. For a stopping range of 10 microns in unit density $H_2O$ the corresponding ion energies for H, S, and O are 540 keV, 18 MeV, and 10 MeV respectively from the SRIM (Stopping and Range of Ions in Matter) model (Ziegler et al., 1985) now updated for 2003 (see http://www.srim.org/index.htm). Higher energy ions passing through this layer also deposit energy in decreasing amounts with increasing energy. For electrons the stopping energy in this layer is 20 keV, and most of the electron energy flux is at much higher energy (Cooper et al. 2001), so the minimum-ionizing magnetospheric electrons deposit relatively little energy for X-ray production in this layer but do contribute to the scattered continuum from deposition at greater depths. However, much of the low-energy electron flux may be deflected away from the surface (Saur et al. 1998) by electric and magnetic fields associated with ionospheric currents.

The incident flux on the moons is made up mostly of energetic H, O, and S ions (Paranicas et al. 1999; Paranicas et al. 2002a; Cooper et al. 2001) with a lesser amount of He (Mauk et al. 2004). Although Europa has an ionosphere that partly deflects low-energy (< 10 keV) magnetospheric ions (Saur et al. 1998; Paranicas et al. 2002b), more-energetic (> 10 keV) ions preferentially reach its surface and produce a large flux of X-rays (Johansson et al., 1995). Since higher charge states lose their energy more quickly, the more penetrating MeV ions, which rapidly lose their remaining atomic electrons on initial entry into the ice before reaching stopping range depths of microns to centimeters (increasing with energy) become particularly important for X-ray production. The relatively larger gyroradii for these ions means that the effective cross-section of Europa for impact of ions producing X-ray fluorescent lines can be larger than its physical size (Pospieszalska and Johnson, 1989), although the magnetosphere-moon interaction is complex and some ions within the gyroradius distance do pass downstream without impact. Magnetohydrodynamic (MHD) simulations have been done on the plasma-moon interaction but the needed hybrid modeling for energetic ion interactions is only now in progress. Initial calculations of Elsner et al. (2002) showed that estimates of the expected X-ray flux due to K shell fluorescence of the oxygen in the ice on Europa, using measurements of the expected H, O, and S flux at Europa and taking into account absorption in the surface layers, were within a factor of three of the measured X-ray flux. They noted that O Kα X-rays at 525 eV have an optical depth in ice of about 10μ; according to irradiation models (Paranicas et al., 2001; Cooper et al., 2001), ion energy losses dominate those from electrons in this layer.

An intriguing possibility is placement of an imaging X-ray spectrometer on board a mission to the Jupiter system (Elsner et al. 2005b,c). In orbit around a Galilean satellite, for example Europa, and although immersed in a fierce radiation environment, such an instrument would be able to map the elemental abundances of the surface for elements from C through Fe. During lengthy maneuvers to place the spacecraft in orbit about Europa, the X-ray instrument would be ideally placed to make unprecedented studies of Jupiter's X-ray auroral and low-latitude X-ray emission and of that from the IPT as well. It would also be able to measure elemental surface abundances of the other Galilean satellites averaged over their surfaces or possibly a handful of large-scale regions. An imaging X-ray spectrometer would also be very useful on missions to other solar system bodies such as the Moon, Mercury, the moons of Mars, asteroids, and comets.

**13. The Rings of Saturn**

The rings of Saturn are one of the most fascinating objects in our solar system. The main ring system, from inside out, consists of the D (distance from Saturn, 1.11–1.235 $R_S$; Saturn radius $R_S$ = 60,330 km), C (1.235–1.525 $R_S$), B (1.525–1.95 $R_S$), and A (2.025–2.27 $R_S$) rings. These are followed by the fainter F, G, and E rings, which span 2.324–8.0 $R_S$. The rings are known to be made of mostly water ($H_2O$) ice.

Using the Chandra ACIS-S observations in January 2004, Bhardwaj et al. (2005c) recently reported the discovery of X-ray emission from the rings of Saturn, confined to the narrow (~130 eV wide) energy band 0.49-0.62 keV (Fig. 42). This result was highly significant statistically, with 28 and 14 photons detected from the rings in this energy band on Jan. 20 and Jan 26-27, 2004, with only 3 and 1.5, respectively, expected from background. This band is centered on the oxygen Kα fluorescence line at 0.53 keV, suggesting that fluorescent scattering of solar X-rays from oxygen atoms in the $H_2O$ icy ring material is the likely source mechanism (Bhardwaj et al. 2005c). The X-ray power emitted by the rings in the 0.49–0.62 keV band on January 20 was 84 MW, which is about one-third of that emitted from Saturn's disk in the 0.24–2.0 keV band (Bhardwaj et al. 2005a). Figure 43 shows the X-ray image of Saturn and its rings in January, 2004, in the 0.49-0.63 keV energy band. Similar to the emission from Saturn's disk, the ring X-ray emission is highly variable, being a factor of 2-3 brighter on Jan. 20 than on Jan. 26-27 (Bhardwaj et al. 2005c).

Ness et al. (2004b) also reported possible X-ray emission on one side of the rings in their 14-15 April, 2003, observations. Bhardwaj et al. (2005c) reanalyzed the Chandra ACIS-S3 Saturn observation of April 2003 in the same manner as their 2004 January observations and found a clear detection of X-rays from rings with features similar to those in January 2004 observations. During 2003 April 14-15 the X-ray power emitted by the rings in the 0.49–0.62 keV band was about 70 MW.

There is an apparent asymmetry in the X-ray emission from the east (morning) and west (evening) ansae of the rings. However, the significance of this apparent asymmetry decreases when the data sets of Jan 2004 and April 2003 are conbined. By analogy with the spokes phenomenon, Bhardwaj et al. (2005c) suggested a possible mechanism producing asymmetry via meteoritic bombardment.

## 14. Asteroids

Asteroids, also known as "minor bodies" or "minor planets", are rocky and metallic objects orbiting the Sun but too small to be considered planets (the largest asteroid Ceres is 450 km in radius). They are believed to left over from the material that formed the planets. While asteroids are often considered primitive bodies, in fact the amount of evolution they have undergone varies widely (Chapman et al., 2004).

X-rays from asteroids have been studied by experiments on two rendezvous missions. The first was the X-ray/Gamma-Ray Spectrometer (XGRS) on the Near Earth Rendezvous (NEAR)-Shoemaker Mission to asteroid 433 Eros (Trombka et al., 1997; Starr et al., 2000). The second was the X-ray Spectrometer (XRS) on the Hayabusa mission to asteroid 25143 Itokawa (Okada et al., 2005). An attempt to detect X-rays from asteroid 1998 WT24 using 10 ks of Chandra observation on Dec. 11, 2001, was unsuccessful.

NEAR-Shoemaker entered Eros orbit on Feb. 14, 2000, then completing its one-year long mission around it. Eros is approximately 33x13x13 kilometers in size and looks like a "fat banana". It is the second largest near-Earth asteroid and its "day" is 5.27 hours long. Eros has a heavily cratered surface with one side dominated by a huge, scallop-rimmed gouge, and the opposite side by a conspicuous raised sharp-rimmed crater. The X-ray spectrometer for the XGRS functioned in ~1-10 keV energy range in order to determine the elemental composition of the Eros' surface. The XRS observed the asteroid in low orbit (<50 km) during 2 May – 12 August, 2000, and again during 12 Dec. 2000 - 12 Feb., 2001. Results from the earlier of these two time periods were first reported by Trombka et al. (2000). Later analysis from both time periods was reported by Nittler et al. (2001). These studies suggest that elemental ratios for Mg/Si, Al/Si, Ca/Si, and Fe/Si are most consistent with a primitive chondrite and give no evidence of global differentiation. The S/Si ratio is considerably lower than that for a chondrite and is most likely due to surface volatilization. More recently, results from a new calibration of X-ray spectrometer have been reported by Lim et al. (2005). The major result is that the conclusion that the surface S/Si relative abundance of Eros is subchondritic has proven robust under the new calibration. The new Fe/Si relative abundance is slightly higher than the 2001 value, although it is still within 1-$\sigma$ of many ordinary chondrites. The overall conclusion is that Eros is broadly "primitive" in its chemical composition and has not experienced global differentiation into a core, mantle and crust, and that the observed departures from chondritic S/Si and Fe/Si are caused by surface effects.

Hayabusa reached the asteroid 25143 Itokawa on 12 Sept. 2005. Results from the XRS experiment (Okada et al. 2005) on this mission will come out soon in the near future.

**15. The Heliosphere**

Neutral atoms from the local interstellar cloud surrounding the heliosphere can freely enter our solar system (Frisch, 1998). The main neutral species are atomic hydrogen and atomic helium. Other species such as Ne, O, N, C and Ar (Gloeckler, 1996) have also been detected, but since their densities are low compared to those of hydrogen and helium, they are not considered in the SWCX models reviewed here.

The density of interstellar hydrogen is ~0.15 cm$^{-3}$ and the inflowing hydrogen interacts weakly with the plasma near the nose of the heliopause (Frisch, 1998), creating the "hydrogen wall" with somewhat enhanced H densities. Because of this interaction the hydrogen slows down and heats up. Closer to the Sun in the inner heliosphere, the interstellar hydrogen is subjected to solar radiation pressure and the gravitational pull of the Sun. These processes create a cavity in the H distribution around the Sun where interstellar hydrogen is non-existent. Because interstellar hydrogen is ionized as it passes the Sun, its density is greater upwind than downwind.

The density of interstellar helium is about 10% that of interstellar hydrogen. The helium does not interact with the plasma as it enters the heliosphere and the Sun's radiation pressure has minimal effect on it. Consequently interstellar helium can be found closer to the Sun than interstellar hydrogen. Neutral interstellar helium also gets ionized as it passes the Sun; however, due to the gravitational pull, there is gravitational focusing, giving rise to the high density helium cone in the down wind direction.

Cravens et al. (2001) created a simple SWCX model to study the time-dependent behavior of X-ray emissions due to SWCX with interstellar neutral helium and hydrogen. X-ray emission due to SWCX is dependent on the solar wind flux and the density of the neutrals the solar wind charge exchanges with. For a more detailed explanation of the model see the section on the Geocorona. Cravens et al. found that due to the large attenuation factor of interstellar hydrogen (i.e., the cavity mentioned earlier), any time variations in the solar wind flux gets averaged out in the X-ray emission due to SWCX with interstellar hydrogen. Consequently, even if there is considerable time variation in the solar wind flux, there is little time variation in the associated X-ray emissions from the H. However, the attenuation effect for helium is smaller, and for the same solar wind flux variations, greater time variations can be seen in helium-related X-ray emissions.

Robertson and Cravens (2003) improved on the earlier Cravens et al. (2001) model by using the Fahr (1971) hot model for the distribution of the interstellar species in the heliosphere. They found that for the same solar wind flux and from the same observation point (Earth), the look direction into the heliosphere makes a significant difference in the calculated SWCX X-ray intensities. Figure 44 shows X-ray intensities from SWCX with interstellar and geocoronal neutrals for a look direction from Earth that is in the downwind direction (intersecting the helium cone), where the hydrogen density is at a minimum and the helium densities are at a maximum. X-ray emissions for the same solar wind flux but for an upwind look direction are shown in Fig. 45. In both cases the time variations in X-ray emission due to SWCX with interstellar helium are particularly noticeable. Note the high magnitudes of the total and "helium" intensities in the down wind direction. The X-ray emission due to SWCX with interstellar hydrogen shows little variation for either of the look directions, but its magnitude is about twice as high in the upwind direction than in the downwind direction.

Pepino et al. (2004) calculated the soft X-ray spectra of heliospheric X-ray emission from the SWCX mechanism including both interstellar helium and hydrogen during solar maximum conditions. Luminosity maps of SWCX with interstellar helium and hydrogen were created for both fast and slow solar wind conditions. The helium focusing cone was very distinct in the maps and Pepino et al. found that about 80% of the down wind X-ray intensity comes from the cone. In the slow solar wind the X-ray emission power was about 3 times larger than that of the fast solar wind, and they concluded that regions where the slow solar wind is prevalent will have more intense radiation than regions where the fast solar wind is prevalent. Significant differences between the spectra of SWCX with either species were also found (see figure 46).

The origin of the Long Term Enhancements (LTE), observed during the ROSAT sky-survey and mentioned in the geocorona section, were not understood when they were first observed. Cravens (2000b) noticed that the variations in LTE showed a good correlation with the measured time variations of the solar proton flux, which suggested that the SWCX mechanism, previously applied to cometary X-ray emission, might indeed be responsible for the LTE. Cravens et al. (2001) created a simple model in which both charge exchange with interstellar hydrogen and helium as well as charge exchange with neutral geocoronal hydrogen was explored. He fed a time-varying solar wind flux into his charge exchange model and looked at the resulting X-ray intensities due to SWCX. There was great correlation between the X-ray intensities and the LTE. The LTE is the time varying component of the SWCX inside the solar system. However, Robertson and Cravens (2003) reasoned that a steady state component of the SWCX-related X-ray emission could not be removed from the measured soft X-ray background and thus remained in the published soft X-ray background maps. They modeled the steady state component of the LTE, by using the same look directions and solar wind flux for the period of observation of ROSAT and created a map of what the steady state observation would look like and compared it to the ROSAT soft X-ray map. They concluded that about 25%-50% of the ROSAT soft X-ray background intensities are of heliospheric origin rather than "cosmic" or from the local hot interstellar bubble (Lallement, 2004).

Lallement (2004) also modeled the non-variable component of the SWCX-heliospheric X-ray emission in order to assess its relative contribution to the cosmic X-ray background. If this component can be removed from the ROSAT all sky-survey, a map of the local hot interstellar bubble can be obtained. Lallement used a scaling factor to determine the maximum heliospheric contribution to the sky map. She concluded that there is considerably less hot gas emission in the local bubble than previously thought, which reduces the pressure discrepancy between the local bubble and our heliosphere by a factor of 4.

**15.1. Astrospheric Charge Exchange**

Just as charge exchange X-rays are emitted throughout the heliosphere, similar emission must occur within the astrospheres of other stars with highly ionized stellar winds that are located within interstellar gas clouds that are at least partially neutral. Although very weak, this emission in principle offers the opportunity to measure mass-loss rates and directly image the winds and astrospheres of other main sequence late-type stars. Imaging would provide information on the geometry of the stellar wind, such as whether outflows are primarily polar, azimuthal, or isotropic, and whether or not other stars have analogs of the slow (more ionized) and fast (less ionized) solar wind streams. Enhanced emission in the relatively dense neutral gas outside the stellar-wind termination shock would indicate the size of the astrosphere and the orientation of the relative star-cloud motion. The neutral gas density could be inferred from the falloff of CX emission at large distances from the star, which is proportional to $1/\tau_{CX} = 1/\sigma_{CX} n_H$, where $\sigma_{CX}$ is the charge exchange cross section of the relevant ion and $n_H$ is the neutral hydrogen (or helium) density. Wind velocity can also be determined from the hardness ratio of the H-like Lyman emission, as discussed in Section 17.

The method of mass-loss detection using CX emission is described by Wargelin and Drake (2001) and was first applied to Proxima Centauri (M5.5 dwarf flare star, distance 1.30 parsecs) using the Chandra X-ray Observatory (Wargelin and Drake, 2002). The null detection of a CX halo implied an upper limit to the mass–loss rate of 14 times the solar rate of $2 \times 10^{-14}$ solar masses per year. Wood et al. (2002) derived an upper limit for Prox Cen of 0.2 times the solar rate using H Ly $\alpha$ absorption profiles; that indirect but more sensitive technique yields a typical mass-loss rate of order 1 to 10 times the solar rate for most late-type dwarf stars. In order to directly observe the CX halos around other stars, an increase in collecting area of roughly a factor of 100 over the area for Chandra will be needed. Such X-ray observatories are planned (Constellation-X, XEUS) but will not launch until late in the next decade.

**16. Diffuse Soft X-ray Background and Implications for Planetary X-ray Studies**

Observation of X-ray emission from the solar system must be considered in the light of the cosmic diffuse X-ray background, a source that while constant in time is strongly variable across the sky both in intensity and in spectral shape. Lacking useful velocity information, it is not possible to distinguish the origin of a specific photon in any given observation, a difficulty which is exacerbated by the fact that the cosmic background is itself made up of a number of distinct components. To complicate the issue further, most of these components also vary independently across the sky. Since we observe in the real world, there are also instrumental backgrounds which can be comparable to the cosmic fluxes depending on which instrument is being used and what energy band is being observed.

Much of the problematic nature of observations of solar-system objects in X-rays is relaxed if the object's emission is limited in angular extent, i.e., effectively a point source. Planetary X-ray emission falls into this regime as the field of view of modern X-ray observatories is typically ~10 arc minutes or greater. Cometary emission, on the other hand, can fill the entire field of view making it very difficult to determine the "zero level" of an observation (the surface brightness of all unwanted components, both cosmic and detector in origin). Solar wind charge exchange emission from interactions with either interplanetary neutrals within the solar system or with exospheric material in Earth's magnetosheath, except in cases of significant temporal variation during the period of the observation, is indistinguishable from the cosmic background. Even during periods of temporal variation, the spectral analysis of the variable part (high state) of the spectrum has caveats. For example, the time variable portion of the quiescent SWCX emission is unlikely to have zero flux and may be spectrally variable in shape as well. Thus, differencing the spectra from high and low states may not actually give the spectrum of the excess in the high state.

The bulk of the cosmic X-ray emission noted in the following is thermal in origin with spectra dominated by lines, many of the same lines which are also observed in solar-system emission. The clearest case of this is the SWCX emission in an XMM-Newton observation reported by Snowden et al. (2004). The SWCX emission spectrum is dominated by $O^{6+}$ and $O^{7+}$ lines, with the $O^{7+}$ line well in excess of the cosmic flux. This raises the issue of significant cross contamination at the detailed spectral level as the oxygen lines in particular are used as a standard diagnostic of temperature, density, and ionization equilibrium in astrophysical plasmas.

Figures 47 and 48 provide views of the cosmic background in the ¼ keV and ¾ keV bands from ROSAT All-Sky Survey (RASS), the energy range where much of the solar-system X-ray emission can be found. The difference in structure is real and an indication of its complexity. Part of the variation in surface brightness of the two bands is due to absorption as the Galactic column density is in the interesting range, i.e., the Galaxy is neither completely optically thin or thick for either energy band but their optical depths differ by an order of magnitude. However, much of the variation is due to the emission regions in the Galaxy being thermal in nature and having a variety of temperatures, and therefore different emission band ratios. These maps show the irreducible backgrounds upon which solar-system objects must be observed.

But what is the cosmic X-ray background composed of? There are many distinct components contributing significantly to the cosmic X-ray flux which originate beyond the solar system. The closest is referred to as the Local Hot Bubble (LHB, Snowden et al. 1998 and references therein) and is an irregularly shaped region of ~$10^6$ K plasma surrounding the Sun and extending from a few tens of parsecs in the Galactic plane to a couple hundred parsecs out of the plane. The cavity in the neutral hydrogen of the Galactic disk which contains this plasma is well mapped by interstellar absorption line studies (Sfier et al., 1999, and references therein) but was first identified as a local deficit in the Galactic ISM column density (Knapp, 1975). Emission from the LHB contributes effectively all of the observed ¼ keV flux in the Galactic plane to roughly half at higher latitudes, but only a small fraction of the flux observed in the ¾ keV band. The origin of the LHB is likely one or more supernovae in the past 5-10 million years (e.g., Maíz-Apellániz, 2001; Knie et al., 1999) and we are observing the fossil radiation.

One optical depth in the ¼ keV band is ~$1 \times 10^{20}$ H I $cm^{-3}$ which is roughly twice the minimum Galactic column density at high latitudes allowing the observation of emission from above the disk but only from limited distances within it. Therefore, the majority of the rest of the emission observed at ¼ keV, which is not attributable to distinct objects, is at high latitudes and likely originates in the lower halo of the Galaxy where the structure on the sky is due both to intrinsic emission variation as well as absorption. Origins for the hot plasma responsible for this emission, also at ~$10^6$ K, include *in situ* supernovae (Shelton, 1998) and "chimneys" from within the disk.

From the cosmic background point of view, SWCX emission has contributed significant contamination in the ¼ keV band. The so-called Long-Term Enhancements (LTEs, Snowden et al. 1995) observed by the RASS and later identified as SWCX by Cravens (1997) and Cox (1998) were at times comparable to the cosmic flux. Figure 47 (lower panel) shows the RASS ¼ keV map without the LTEs removed. While the RASS data were processed to remove as much of the SWCX emission as possible, the amount of residual flux "contaminating" the maps is not well known but may be ~25% of the minimum flux in the plane, or even higher (Lallement, 2004).

The ¾ keV band map (Figure 48, upper panel) exhibits an entirely different structure than that at ¼ keV. At these and higher energies the sky consists of a relatively flat background dominated by distinct objects, some of relatively large solid angle. Emission in the Galactic center direction is from the superposition of the Galactic bulge (e.g., Snowden et al. 1997, see below) and Loop I. Loop I is a nearby superbubble at a distance of ~150 pc with a radius of ~100 pc subtending roughly 90º on the sky. The North Polar Spur, the bright plume of emission rising out of the Galactic plane at a longitude of 25º is the limb-brightened edge of this superbubble. Loop I is powered by the stellar winds and supernovae of the Sco-Oph OB associations and has a temperature of a ~$3 \times 10^6$ K.

Other distinct relatively large solid angle extended objects in the ¾ keV map include the Eridanus enhancement at l,b~200º,-30º (powered by the Orion OB associations), the Cygnus superbubble at l,b~80º,0º (which may be a superbubble or the serendipitous association of a number of smaller supernova remnants, SNRs), and the Vela and Puppis SNRs at l,b~265º,-5º. While optical depths in this band are an order of magnitude smaller than at ¼ keV, much of the Galactic disk is still hidden from us as can be seen by the absorption through along the plane, particularly at l,b~30º,0º.

The Galactic X-ray bulge provides the other half of the confused emission in the Galactic center direction. Its emission is most clearly seen in the southern hemisphere in Figure 48. It has a scale height of 1.9 kpc and a radial extent over the Galactic plane of 5.4 kpc (Snowden et al., 1997). At higher energies there is a Galactic ridge which is confined to the Galactic disk (scale height of ~100 pc but covers the longitude range +/- 45°, Kaneda et al. 1997). The Galactic bulge may tie into an extended halo, or not, as the evidence is weak for such a structure. The relatively isotropic background most clearly seen at high latitudes in Figure 2 has at least two components. About half is comprised of unresolved emission from distinct objects at cosmological distances, i.e., the large population of AGN. Most of the rest is known to originate in the local universe as it has a significant flux of oxygen emission at zero redshift (McCammon et al. 2002). The likely origin is therefore either a Galactic halo or the local group. Perhaps 10% of the flux is also due to unresolved point sources, this time stars in our own Galaxy (Kuntz and Snowden 2001).

Figure 49 provides an example of how the cosmic spectrum varies over the sky. Plotted are data from a relatively bright region within Loop I and from a typical region at high latitude. At energies less that 1.5 keV the spectra are dominated by lines and complexes of lines which are smeared out by the intrinsic energy resolution of the detectors. The spectra agree well with each other at energies above 2 keV where they are dominated by the extragalactic background from the superposition of unresolved AGN. The fact that the spectra are also relatively consistent below 0.5 keV is an artifact of the detector response that is dropping rapidly at lower energies. At ¼ keV the high-latitude spectrum would be almost three times brighter than the "bright" region. Also plotted in Figure 49 is a spectrum from the same high-latitude region taken during a period of significant SWCX emission. The SWCX spectrum is also strongly dominated by line emission in this energy range with no distinguishing characteristics compared to the spectra of the cosmic background.

**17. Laboratory Simulation of X-Rays from Planets and Comets**

The discoveries of X-ray emission from planets and comets have resulted in intensive laboratory research efforts. The realization that much of the emission is due to charge exchange processes has especially spurred novel research in X-ray spectroscopy. This research has expanded existing laboratory X-ray astrophysics experiments, which have traditionally focused on X-ray production by electron-ion collisions. A review of laboratory X-ray astrophysics was given recently by Beiersdorfer (2003). In the following we focus on laboratory measurements targeted to produce data for explaining X-ray observations of cometary and planetary atmospheres and the solar and stellar heliospheres produced by charge exchange.

**17.1. K-Shell X-ray Spectra**

Some of the first charge-exchange-induced emission spectra relevant to cometary X-ray emission were obtained by Greenwood et al. (2000, 2001) using low-resolution solid-state detectors. They, for example, studied bare $Ne^{10+}$ ions from an electron cyclotron resonance (ECR) source at the Jet Propulsion Laboratory interacting with $H_2O$, He, $H_2$ and $CO_2$. The data were obtained at ion-neutral collision energies (70 keV) relevant to the high-energy solar wind. Similarly, Beiersdorfer et al. (2001) measured the X-ray emission from the interaction of bare $Ne^{10+}$ ions and neutral neon using an electron beam ion trap at the Lawrence Livermore National Laboratory. These data were obtained at much lower ion-neutral collision energies (~ 220 eV). Combining their data with those from Greenwood et al., Beiersdorfer et al. (2001) showed that the shape of the X-ray emission spectrum is dependent on the ion velocity and thus represents a diagnostic for the ion velocity.

Even such rather low-resolution laboratory measurements have demonstrated that X-ray spectra formed by charge exchange are different from those formed by electron-impact excitation. In the case of the spectra from H-like ions, this is mainly seen in the relative magnitude of the various components of the Lyman series. Because charge exchange captures electrons into levels with high principal quantum number $n$, emission from this high-$n$ level can be very prominent. We illustrate this in Fig. 50, where we compare the emission from $Fe^{25+}$ produced by charge exchange and by direct electron-impact excitation. These spectra were recently measured on an electron beam ion trap at the Lawrence Livermore National Laboratory by Wargelin et al. (2005). The emission from high-$n$ levels are clearly enhanced relative to the $n = 2 \rightarrow n = 1$ Ly-$\alpha$ emission in the spectrum formed by charge exchange; by contrast, the emission from the high-$n$ levels is considerably less prominent in the spectrum formed by electron-impact excitation. This spectral signature may aid in identifying charge exchange as a dominant or contributing mechanism in line formation in a variety of situations.

High-resolution laboratory measurements have now revealed some of the details of the emission near the limit of the Lyman series (Beiersdorfer et al. 2003, 2005b). Replacing the rather low-resolution solid-state detectors used in the above-mentioned measurements with a microcalorimeter has enabled Beiersdorfer et al. to resolve the Lyman lines, as illustrated in Fig. 51. This figure shows the Lyman lines of $O^{7+}$ formed either in the interaction of $O^{8+}$ ions with $CH_4$ or with $N_2$. The emission near the series limit is quite different. In fact, the spectrum produced in the interaction with $CH_4$ shows an additional line, the $n = 6 \rightarrow 1$ Lyman-$\varepsilon$ line, which is absent in the spectrum produced in the interaction with $N_2$. The reason is that the ionization potential of $CH_4$ is lower than that of $N_2$, allowing capture into a higher $n$ level, as discussed by Beiersdorfer et al. (2003). The measurements indicate that high-resolution spectra of the series limit may be used as a diagnostic of the interaction gas. Reliable modeling of the laboratory spectra is, however, not yet possible, as discussed by Wargelin et al. (2005) and Otranto et al. (2006).

Unlike hydrogen-like ions, helium-like ions do not exhibit strong emission near the series limit (Beiersdorfer et al. 2001; Tawara et al. 2001; Beiersdorfer et al. 2003; Wargelin et al. 2005; Beiersdorfer 2005). The reason is that most electron capture proceeds into triplet levels, which are prohibited from decaying directly to the $1s^2$ $^1S_0$ ground state of the helium-like ion. Nevertheless, the spectra of He-like ions formed by charge exchange differ considerably from those formed by electron-impact excitation. Laboratory measurements showed that the tell-tale difference is in the $n = 2 \to n = 1$ emission. In spectra excited by electron collisions, the strongest line emanates from the singlet level: $1s2p\ ^1P_1 \to 1s^2\ ^1S_0$. In spectra excited by charge exchange, the strongest lines are those emanating from the triplet levels: $1s2p\ ^3P_1 \to 1s^2\ ^1S_0$, $1s2p\ ^3P_2 \to 1s^2\ ^1S_0$, and $1s2s\ ^3S_1 \to 1s^2\ ^1S_0$. This is illustrated in Fig. 52, which shows spectra of He-like $Ar^{16+}$ taken by Beiersdorfer et al. (2005a) at the National Spherical Tokamak Experiment (NSTX).

The spectrum of helium-like $Ar^{16+}$ formed by charge exchange could be reproduced by theory with high accuracy (Beiersdorfer et al. 2005a). This is, however, not the case for most other spectra of helium-like ions measured in the laboratory so far (cf. Beiersdorfer et al. 2003). For most helium-like spectra the situation is similar to that already discussed for hydrogen-like ions. The difficulty in modeling the spectra may arise from the fact that the interaction is not with atomic hydrogen but is with gases that allow for the capture of more than one electron during the charge exchange process. In principle, any neutral gas other than atomic hydrogen can contribute more than one electron to charge exchange, requiring that theory produce models that are way more complex than currently available.

The role of double and multiple electron capture in forming X-ray emission spectra has recently been investigated by Ali et al. (2005). They studied the interaction of bare $Ne^{10+}$ ions produced at the ECR source at the University of Nevada Reno with neutral neon. Their measurements employed a so-called reaction microscope which allowed triple coincidence measurements between the emitted x rays, the target ions, and the projectile ions. This way they could determine whether a given x ray corresponded to the capture of one or more electrons by the $Ne^{10+}$ ions. They found that a single electron is captured in less than 60% of all collisions. More than one electron is captured in almost 40% of all collisions. The additional electron(s) is (are) typically given off again by autoionization, so that the net charge of the $Ne^{10+}$ ions drops by only one, the same as in single electron capture. However, the X-ray emission spectrum looks different depending on whether one or more electrons were initially captured. The reason is that in order to expel the additional electron(s) by autoionization, the remaining electron has to drop to a lower level. Therefore, the remaining electron has less energy for conversion into X-rays. The difference in the X-ray emission following single-electron capture and multi-electron capture is evident in the spectra shown in Fig. 53. These measurements show that unless the reaction gas is neutral hydrogen, the spectral emission formed by charge exchange is a superposition of spectra formed by single-electron and multi-electron capture. These measurements also showed that true double capture, in which two electrons are captured but neither is expelled by autoionization, is rare. Using the crossed-beam apparatus at Groningen, Lubinski et al. (2001), however, showed that double capture dominates over single capture in very low-energy collisions of $He^{2+}$ ions with neutral gases. As a result, the ratio of the Lyman-$\alpha$ line in $He^+$ formed by single electron capture to that of the $1s2p\ ^1P_1 \to 1s^2\ ^1S_0$ transition in neutral helium is dependent on the collision energy. Bodewits et al. (2004) used this fact to estimate the velocity of the solar wind interacting with comet Hale-Bopp from observations of these two transitions taken with the Extreme Ultraviolet Explorer satellite.

Although a lot more work will need to be done to understand all the mechanisms involved in charge exchange, the laboratory measurements of the K-shell emission of the solar wind heavy ions $C^{5+}$, $C^{6+}$, $N^{6+}$, $N^{7+}$, $O^{7+}$, $O^{8+}$, and $Ne^{9+}$ undergoing charge exchange have been very successfully used to simulate the X-ray emission of comet C/1999 S4 (LINEAR) (Beiersdorfer et al., 2003) observed with Chandra. Excellent agreement between the simulation and the Chandra observations in the energy range above 300 eV was achieved.

**17. 2. L-Shell X-ray Spectra**

So far our discussion has concentrated on K-shell emission. L-shell emission lines or lines from even higher shells may be just as or even more prominent in spectra from planetary and cometary atmospheres. For example, the same model calculation mentioned above that produced good agreement with the K-shell emission of $Ar^{16+}$ (Beiersdorfer et al. 2005a) predicts numerous transitions terminating in higher shells, as illustrated in Fig. 54. Few of these transitions in the higher-$n$ shells have been observed in the laboratory. Those that have been observed in the laboratory typically fall into the extreme ultraviolet or visible range. For example, a variety of such lines have been employed as diagnostics in tokomaks (Stratton et al. 1990; von Hellermann et al. 1995; Whyte et al. 1998), and the crossed-beam technique has been used to study the emission from lithium-like or beryllium-like ions in the extreme ultraviolet regime (Bliek et al. 1998; Lubinski et al. 2000; Ehrenreich et al. 2005). These systems are close analogs to the hydrogen-like and helium-like ions. However, additional lines appear in these spectra because of the fact that the $n = 2$ ground configuration may assume both an $\lambda = 0$ and an $\lambda = 1$ angular momentum state.

Measurements of the X-ray emission from more complicated L-shell ions, for example, from $Fe^{21+}$ which may play a role in extrasolar systems with stellar winds hotter than those of our Sun, have been attempted at the Livermore electron beam ion trap facility (Beiersdorfer et al. 2000). However, like the measurement of the L-shell emission of krypton ions performed at the electron beam ion source at the Macdonald Laboratory at the Kansas State University (Tawara et al. 2002), these measurements still suffer from a lack of high resolution. Experiments are under way, which utilize high-resolution microcalorimeters to overcome these limitations (Canton et al. 2005).

**18. SUMMARY**

Except for the Jovian system (and possibly Earth's aurora), all solar system X-ray emission (except the Sun) is powered by the solar wind and/or solar X-ray emission. Charge-exchange is often an important process. What happens in individual cases then depends on distance from the Sun and on the local environment. Table 2 summaries our current knowledge of the X-ray emissions from the solar system bodies that have been observed to produce soft X-rays. Several other solar system bodies, which include Titan,

Uranus, Neptune, and inner-icy satellites of Saturn, are also expected to be X-ray sources, but are probably too faint to be detected even with modern X-ray observatories.

Table 2. Summary of the characteristics of soft X-ray emission from solar system bodies

| Object | Emitting Region | Power Emitted[a] | Special Characteristics | Possible Production Mechanism |
|---|---|---|---|---|
| Earth | Auroral atmosphere | 10-30 MW | Correlated with magnetic storm and substorm activity | Bremsstrahlung from precipitating electrons, and characteristic line X-rays |
| Earth | Non-auroral atmosphere | 40 MW | Correlated with solar X-ray flux | Scattering of solar X-rays by atmosphere |
| Jupiter | Auroral atmosphere | 0.4-1 GW | Pulsating (~20-60 min) X-ray hot spot in north polar region | Energetic ion precipitation from magnetosphere and/or solar wind + electron bremsstrahlung |
| Jupiter | Non-auroral atmosphere | 0.5-2 GW | relatively uniform over disk | Resonant scattering of solar X-rays + possible ion precipitation from radiation belts |
| Moon | Dayside / Geocornonal (Nightside) | 0.07 MW | Correlated with solar X-rays / Nightside emissions are ~1% of the dayside | Scattering and fluorescence due to solar X-rays by the surface elements on dayside. / SWCX with geocorona |
| Comets | Sunward-side coma | 0.2-1 GW | Intensity peaks in sunward direction ~$10^5$-$10^6$ km ahead of cometary nucleus | SWCX with cometary neutrals |
| Venus | Sunlit atmosphere | 50 MW | Emissions from ~120-140 km above the surface | Fluorescent scattering of solar X-rays by C and O atoms in the atmosphere |
| Mars | Sunlit atmosphere / exosphere | 1-4 MW / 12 MW | Emissions from upper atmosphere at heights of 110–130 km / emissions extend out to ~8 Mars radii | Fluorescent scattering of solar X-rays by C and O atoms in the upper atmosphere / SWCX with Martian corona |
| Io | Surface | 2 MW | Emissions from upper few microns of the surface | Energetic Jovian magnetospheric ions impact on the surface |
| Europa | Surface | 1.5 MW | Emissions from upper few microns of the surface | Energetic Jovian magnetospheric ions impact on the surface |
| Io Plasma Torus | Plasma torus | 0.1 GW | Dawn-dusk asymmetry observed | Electron bremsstrahlung + ? |
| Saturn | Sunlit disk | 0.1-0.4 GW | Varies with solar X-rays | scattering of solar X-rays + Electron bremsstrahlung ? |
| Rings of Saturn | Surface | 80 MW | Emissions confined to a narrow energy band around at 0.53 keV. | fluorescent scattering of solar X-rays by atomic oxygen in $H_2O$ ice + ? |

| Heliosphere | Entire heliosphere | $10^{16}$ W | Emissions vary with solar wind variation | SWCX with heliospheric neutrals |

[a]The values quoted are values at the time of observation. X-rays from all bodies are expected to vary with time. For comparison the total X-ray luminosity from the Sun is $10^{20}$ W.
SWCX ≡ Solar wind charge exchange = charge exchange of heavy highly ionized solar wind ions with neutrals.

**Acknowledgements**

Most of the results summarized in this review are due to observations with the Chandra and XMM-Newton X-Ray Observatories, and several of the authors are grateful for grants from those programs. The work of Peter Beiersdorfer was performed under the auspices of the Department of Energy by Lawrence Livermore National Laboratory under contract W-7405-ENG-48 and supported by NASA's Planetary Atmospheres Program.

Saur, J., Strobel, D. F., and Neubauer, F. M., 1998. Interaction of the Jovian magnetosphere with Europa: Constraints on the neutral atmosphere. J. Geophys. Res. 103(E9), 19947-19962.

Schmitt, J. H. M. M., Snowden, S. L., Aschenbach, B., Hasinger, G., Pfeffermann, E., Predehl, P., Trümper, J., 1991. A soft X-ray image of the Moon. Nature 349, 583-587.

Schunk, R. W. and A. F. Nagy, 2000. Ionospheres: Physics, Plasma Physics, and Chemistry, Cambridge Univ. Press, Cambridge.

Schwadron, N. A., Cravens, T. E., 2000. Implications of solar wind composition for cometary X-rays, Astrophys. J., 544, 558-566.

Singhal, R.P., Chakravarty, S.C., Bhardwaj, A., Prasad, B., 1992. Energetic electron precipitation in Jupiter's upper atmosphere. J. Geophys. Res., 97, 18245-18256.

Sfeir, D. M.; Lallement, R.; Crifo, F.; Welsh, B. Y. 1999, Mapping the contours of the Local bubble: preliminary results, Astron. Astrophys., 346, 785-797.

Shapiro, V. D., R. Bingham, J. M. Dawson, Z. Dobe, B. J. Kellett, and D. A. Mendis 1999. Energetic Electrons Produced by Lower Hybrid Waves in the Cometary Environment and Soft X Ray Emission: Bremsstrahlung and K Shell Radiation, J. Geophys. Res. 104, 2537-2554.

Sharber J.R., et al., 1993. Observations of the UARS particle environment monitor and computation of ionization rates in the middle and upper atmosphere during a geomagnetic storm, Geophys. Res. Lett., 20, 1319.

Shelton, R. L. 1998, Simulations of Supernova Remnants in Diffuse Media and Their Application to the Lower Halo of the Milky Way. I. The High-Stage Ions, Astrophys. J., 504, 785-804.

Sletten, A., Stadsnes, J., Trefall, H., 1971. Auroral-Zone X-ray Events and their Relation to Polar Magnetic Substorms, J. Atmos. Solar Terr. Phys., 33, 589-604.

Smith, D.M., L.I. Lopez, R. Lin and C. P. Barrington-Leigh, 2005. Science, 307,

Snowden, S. L., Collier, M. R., and Kuntz, K. D., 2004. XMM-Newton Observation of Solar Wind Charge Exchange Emission. Astrophys. J., 610, 1182.

Snowden, S. L., D. McCammon, D. N. Burrows, and J. A. Mendenhall, 1994. Analysis procedures for ROSAT XRT/PSPC observations of extended objects and the diffuse background. Astrophys. J., 424, 714.
66

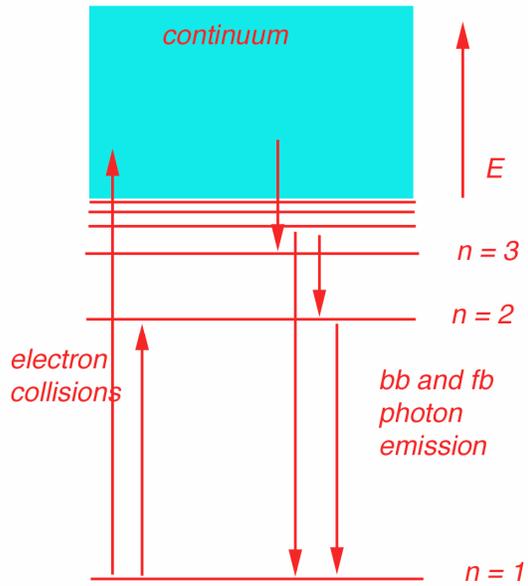

**Figure 1.** Schematic of energy levels for a hydrogen-like atom. Free-free (ff), bound-free (fb), and bound-bound (bb) transitions are indicated.

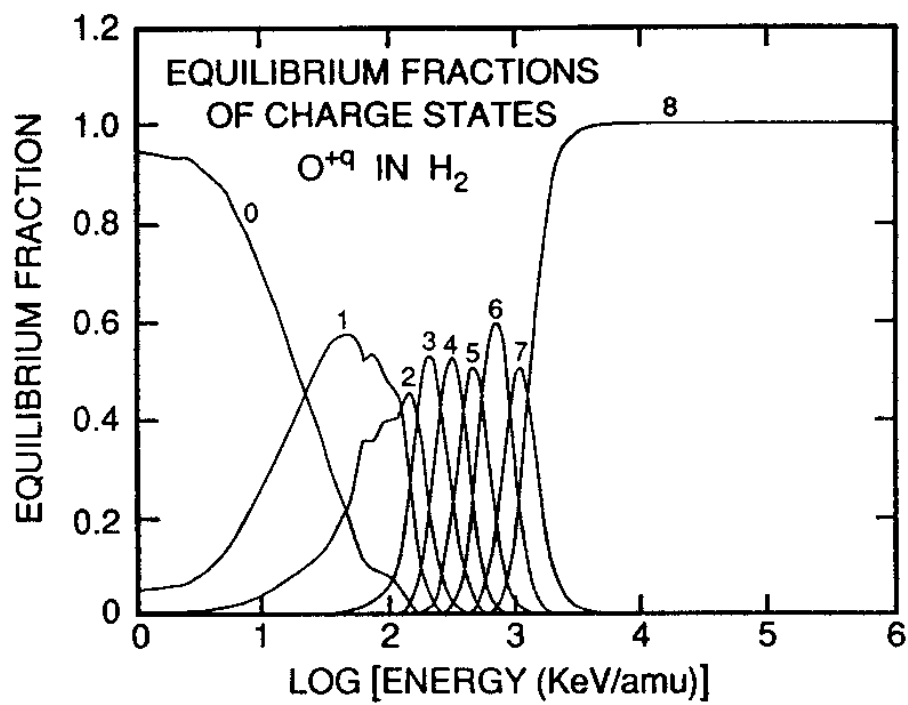

**Figure 2.** Equilibrium fractions for highly-charged oxgyen ions (labeled with the charge q) interacting with molecular hydrogen gas a function of the log base-10 of the oxygen beam energy (units of keV/nucleon). From *Cravens et al.* (1995).

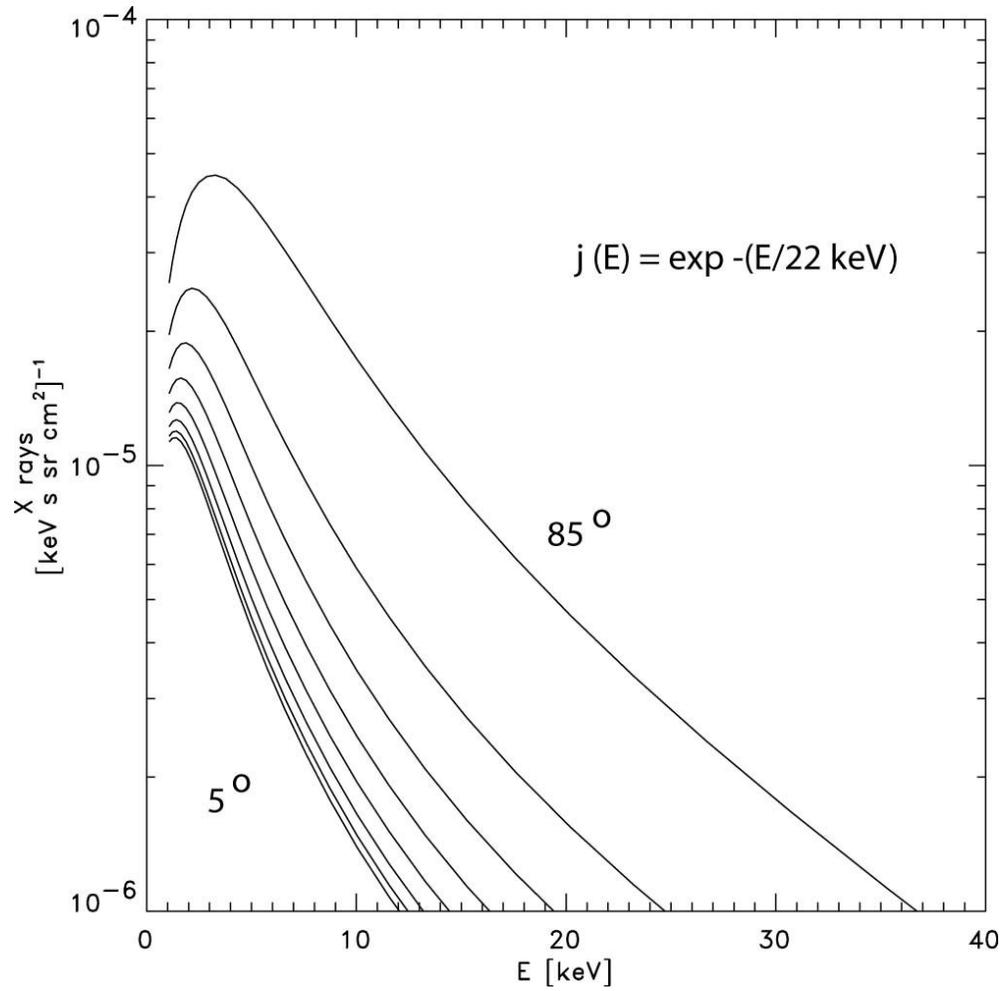

**Figure 3.** X-ray spectra in eight 10° zenith angle intervals (center angles from 5° to 80°) produced by an exponential distribution of electrons with e-folding energy of 22 keV, isotropic in the downward hemisphere, based on a model by Lorence (1992).

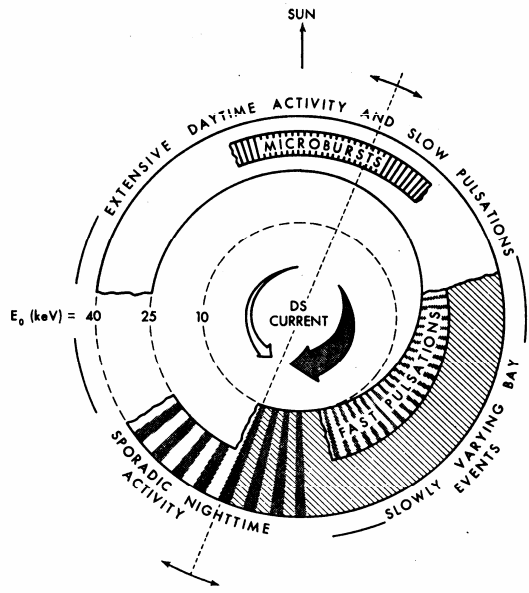

**Figure 4.** Diurnal pattern of spectral character of energetic (50-250 keV) electron precipitation suggested by bremsstrahlung X-ray observations in the auroral zone (from Barcus and Rosenberg, 1966).

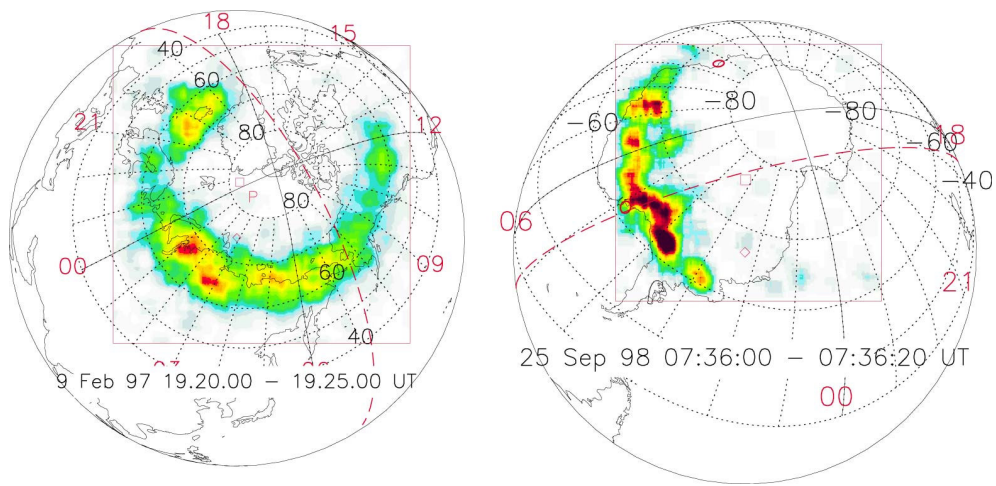

**Figure 5.** Auroral X-ray images of the Earth from the Polar PIXIE instrument. To the left: 5 minutes accumulation of X-rays in the energy range 8.1-19.7 keV from the northern hemisphere on February 9, 1997. To the right: 20 seconds accumulation of X-rays in the energy range 2.7-9.6 keV on September 25, 1998 from the southern hemisphere. 00, 06, 12 and 18 denotes the magnetic local times.

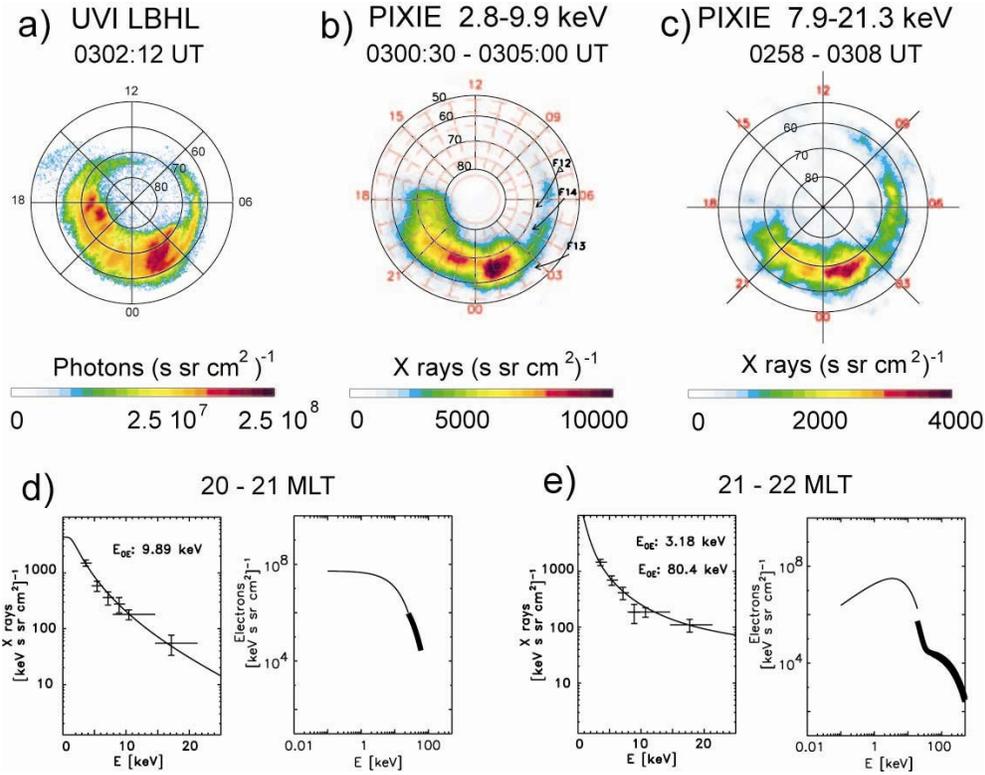

**Figure 6.** (a) UVI and (b, c) PIXIE images in two different energy bands, from July 31, 1997. (d): Left: The measured X-ray energy spectrum where an estimated X-ray spectrum produced by a single exponential electron spectrum with e-folding energy 9.89 keV is shown to be the best fit to the measurements. Right: The electron spectrum derived from UVI and PIXIE, where thin line is UVI contribution, thick line is PIXIE contribution. Both plots are averages within a box within 20-21 magnetic local time and 64°-70° magnetic latitude. (e): Same as (d) but within 21-22 MLT, where X rays produced by a double exponential electron spectrum is shown to be the best fit to the X-ray measurements (from Østgaard et al, 2001).

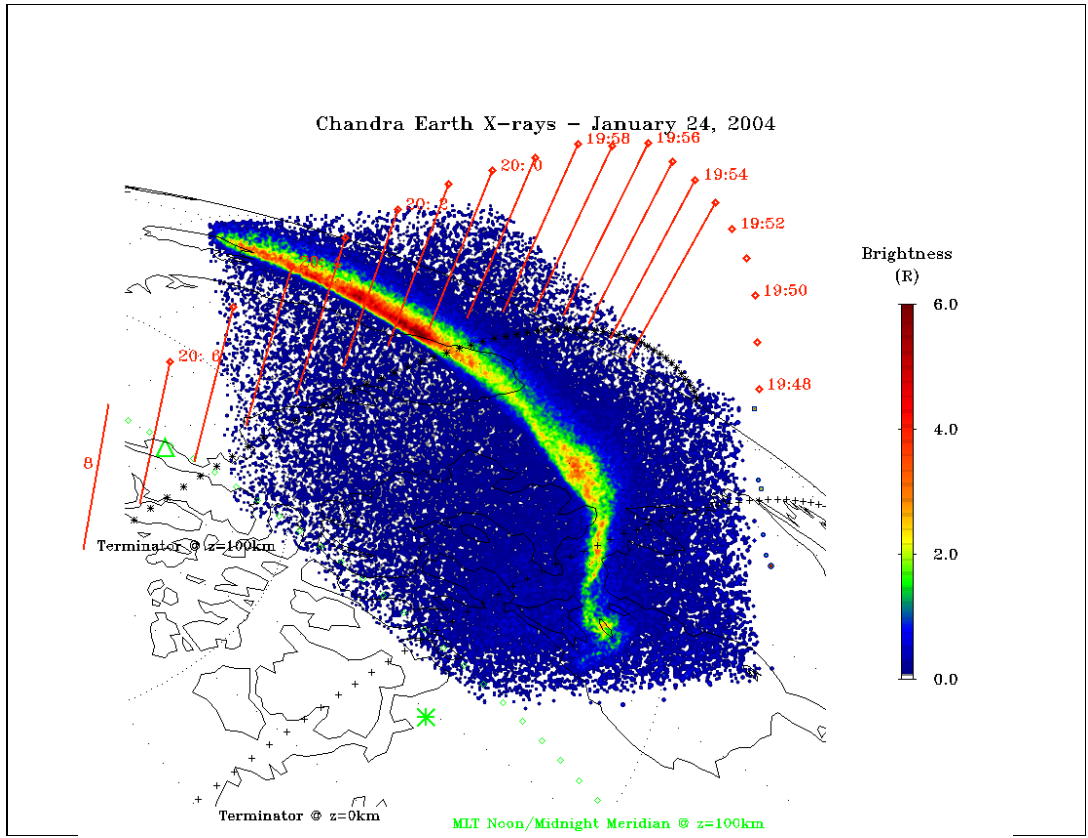

**Figure 7.** Chandra HRC-I X-ray image of auroral region on January 24, 2004 showing a bright arc. The orbital location of satellite DMSP F13 is shown by red diamonds, with 2-minute time ticks and vertical lines extending down to an altitude of 100 km (from Bhardwaj et al., 2006).

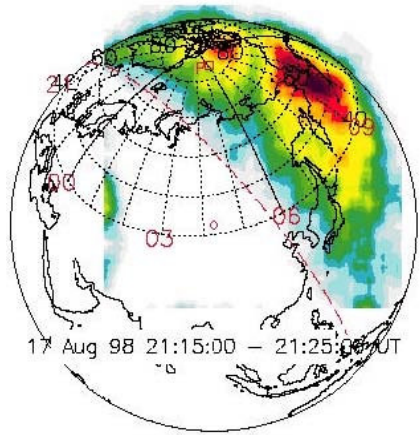

**Figure 8.** X-ray image of Earth from the Polar PIXIE instrument for energy range 2.9-10.1 keV obtained on August 17, 1998, showing the dayside X rays during a solar X-ray flare. The grid in the picture is in corrected geomagnetic coordinates, and the numbers shown in red are magnetic local time. The terminator at the surface of the Earth is shown as a red dashed line.

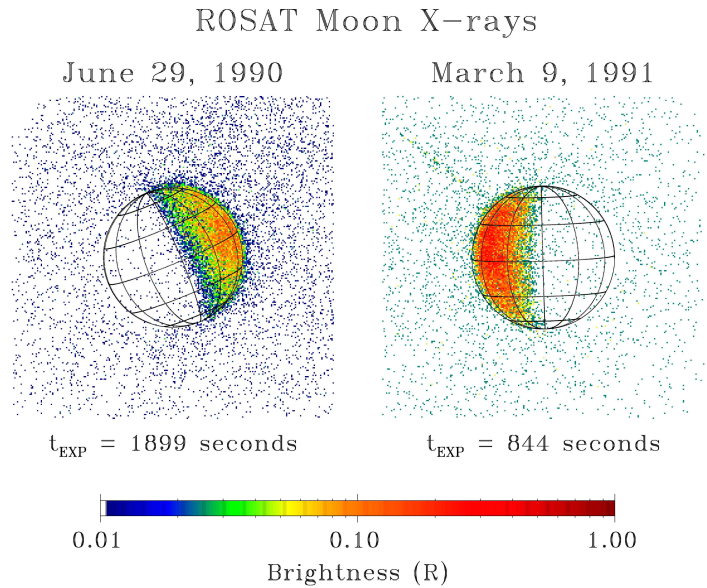

**Figure 9.** ROSAT soft X-ray (0.1–2 keV) images of the Moon at first (left side) and last (right side) quarter. The dayside lunar emissions are thought to be primarily reflected and fluoresced sunlight, while the faint night side emissions are foreground due to charge exchange of solar wind heavy ions with H atoms in Earth's exosphere. The brightness scale in R assumes an average effective area of 100 $cm^2$ for the ROSAT PSPC over the lunar spectrum.

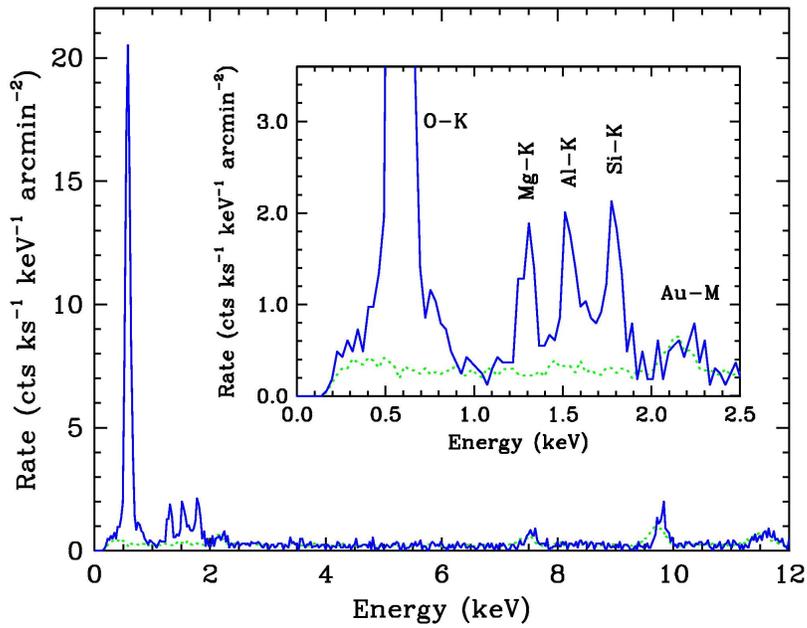

**Figure 10.** Chandra spectrum of the bright side of the Moon. The green dotted curve is the detector background. K-shell fluorescence lines from O, Mg, Al, and Si are shifted up by 50 eV from their true values because of residual optical leak effects. Features at 2.2, 7.5, and 9.7 keV are intrinsic to the detector. From Wargelin et al. 2004.

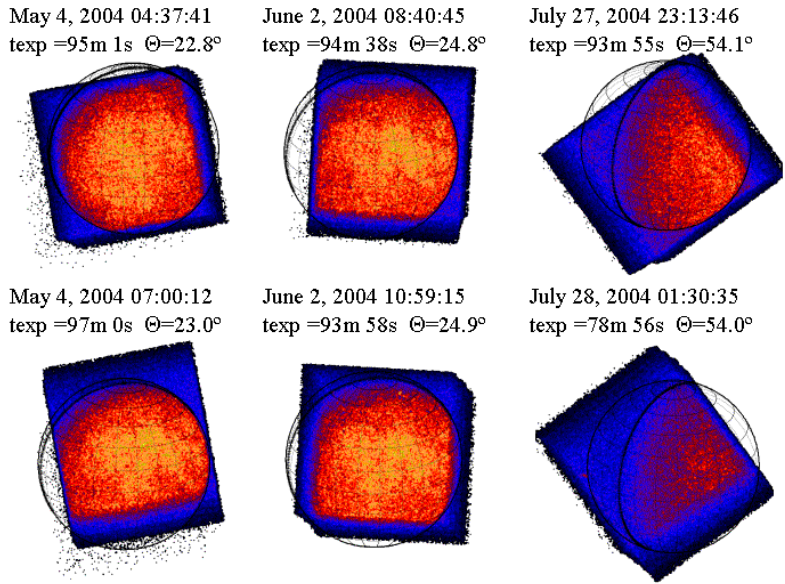

**Figure 11.** Moon observations by the Chandra HRC-I camera, which images 0.1–10 keV photons with a point-spread function of 0.5" (about 10× better than ROSAT). Two images each at phase angles of 23º, 25º, and 54º were obtained, for a total of 33.2 ks. The mean energy of the detected photons is of order 500 eV. The Moon just overfills the 30'×30' HRC-I field of view. These preliminary X-ray images show a clear, though low-contrast, albedo reversal with respect to images in visible light.

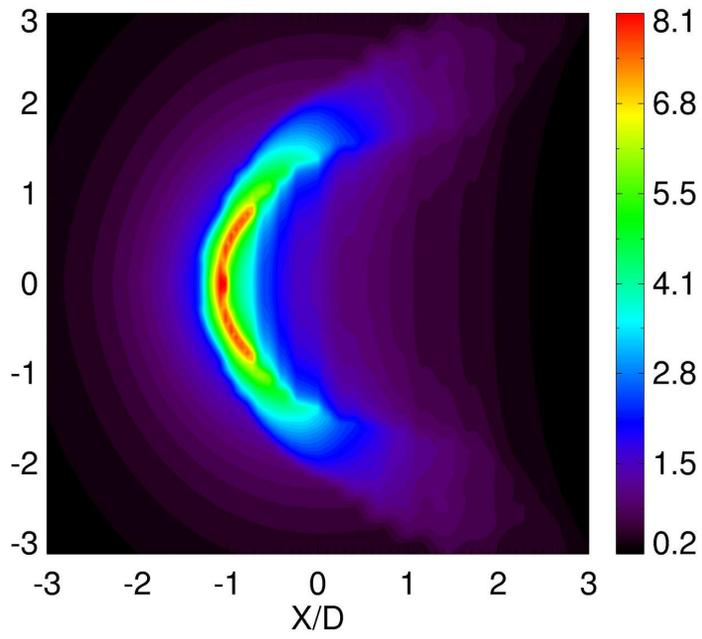

**Figure 12.** Image of the X-Ray intensity as observed from the Earth's flanks in the equatorial plane. Units are keV cm$^{-2}$ s$^{-1}$ sr$^{-1}$. R and X are coordinates in the image plane. From Robertson and Cravens (2003).

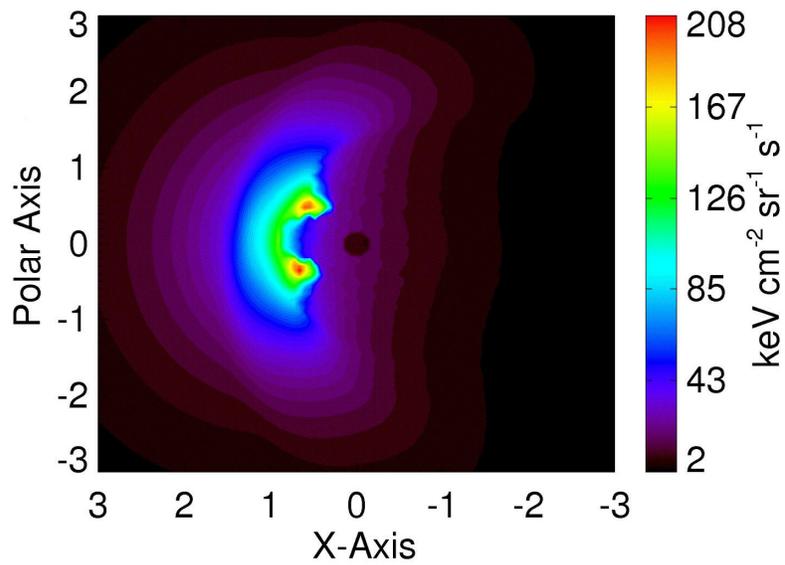

**Figure 13:** Similar image of X-Ray intensities as observed from the Earth's flanks in the equatorial plane during the March 31$^{st}$ 2001 CME. Units are in D (distance to the magnetopause). The cusps are included. From Robertson et al. (2005).

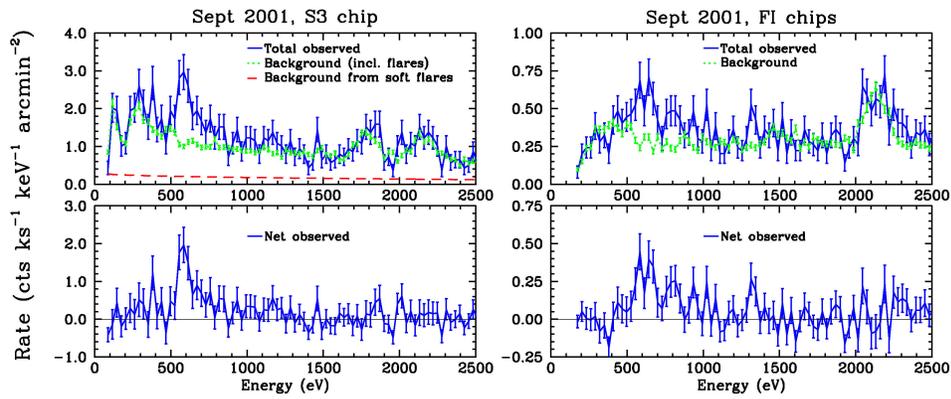

**Figure 14:** Observed and background-subtracted spectra from the September 2001 Chandra observation of the dark side of the Moon, with 29-eV binning. Left panel is from the higher-QE but lower-resolution ACIS S3 CCD; right panel shows the higher-resolution ACIS front-illuminated (FI) CCDs. Oxygen emission from charge exchange is clearly seen in both spectra, and energy resolution in the FI chips is sufficient that O Lyman α is largely resolved from O Kα. High-n H-like O Lyman lines are also apparent in the FI spectrum, along with what is likely Mg Kα around 1340 eV. From Wargelin et al. (2004).

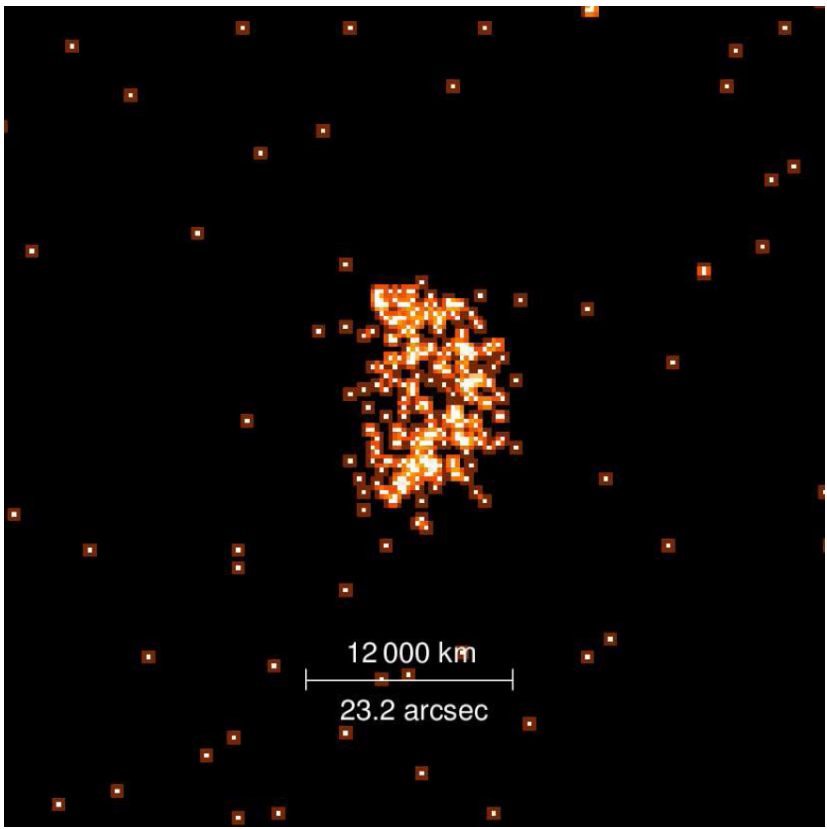

**Figure 15:** First X–ray image of Venus, obtained with Chandra ACIS–I on 13 January 2001. The X–rays result mainly from fluorescent scattering of solar X–rays on C and O in the upper Venus atmosphere, at heights of 120 – 140 km. In contrast to the Moon, the X–ray image of Venus shows evidence for brightening on the sunward limb. This is caused by the fact that the scattering takes place on an atmosphere and not on a solid surface (from Dennerl *et al.*, 2002).

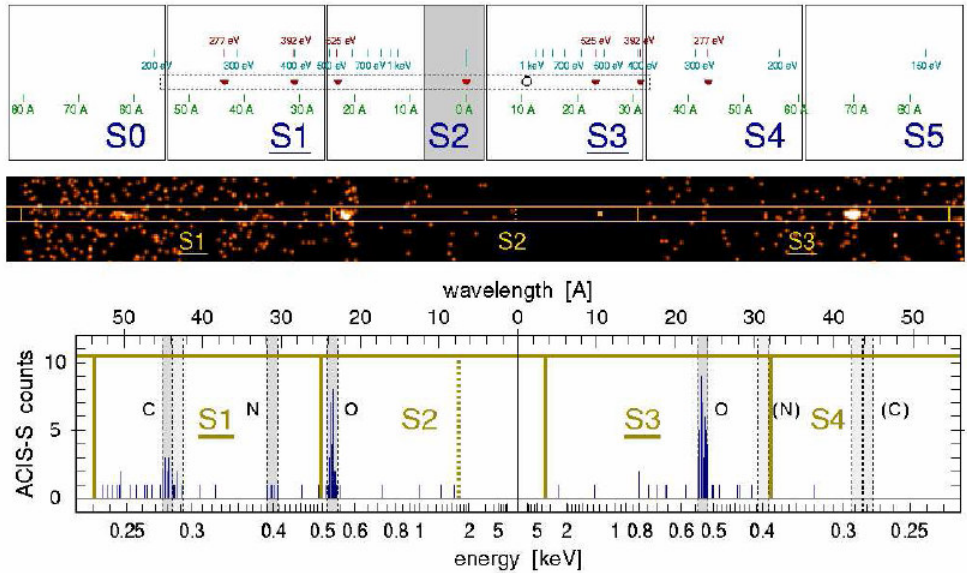

**Figure 16: a)** Expected LETG spectrum of Venus on the ACIS–S array. Energy and wavelength scales are given along the dispersion direction. Images of Venus are drawn at the position of the C, N, and O fluorescence lines, with the correct size and orientation. The dashed rectangle indicates the section of the observed spectrum shown below. **b)** Observed spectrum of Venus, smoothed with a Gaussian function with $\sigma = 20″$. The two bright crescents symmetric to the center are images in the line of the O-K$\alpha$ fluorescent emission, while the elongated enhancement at left is at the position of the C-K$\alpha$ fluorescent emission line. The Sun is at bottom. **c)** Spectral scan along the region outlined above. Scales are given in keV and Å. The observed C, N, and O fluorescent emission lines are enclosed by dashed lines; the width of these intervals matches the size of the Venus crescent (22.8″) (from Dennerl *et al.*, 2002).

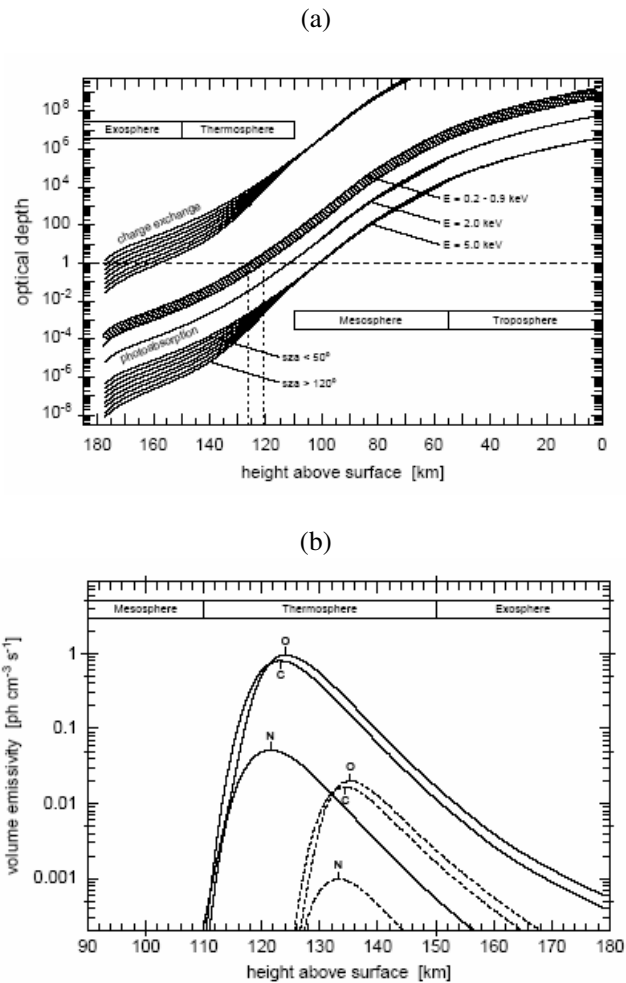

**Figure 17: a)** Optical depth of the Venus model atmosphere with respect to charge exchange (above) and photoabsorption (below), as seen from outside. The upper/lower boundaries of the hatched area refer to energies just above/below the C and O absorption edges. For better clarity the dependence on the solar zenith angle (sza) is only shown for $E = 5.0$ keV; the curves for the other energies refer to sza $< 50°$. The dashed line, at $\tau = 1$, marks the transition between the transparent ($\tau < 1$) and opaque ($\tau > 1$) range. For a specific energy, the optical depth increases by at least 12 orders of magnitude between 180 km and the surface. For charge exchange interactions a constant cross section of $3 \times 10^{-15}$ cm$^2$ was assumed. **b)** Volume emissivities of C, N, and O K$\alpha$ fluorescent photons at zenith angles of zero (subsolar, solid lines) and 90° (terminator, dashed lines). The height of maximum emissivity rises with increasing solar zenith angles because of increased path length and absorption along oblique solar incidence angles. In all cases maximum emissivity occurs in the thermosphere, where the optical depth depends also on the solar zenith angle (a) (from Dennerl *et al.*, 2002).

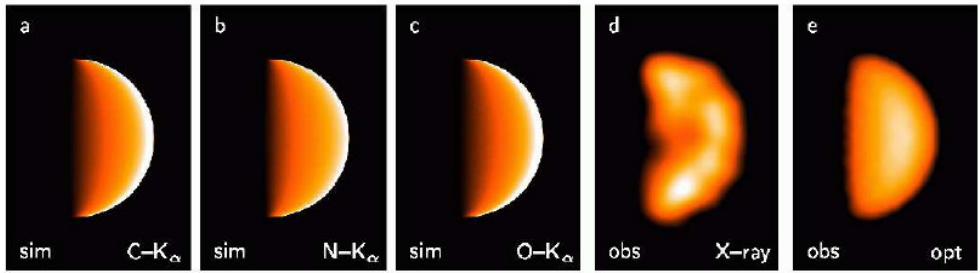

**Figure 18**: **a – c)** Simulated X–ray images of Venus at C–Kα, N–Kα, and O–Kα, for a phase angle φ = 86.5°. The X–ray flux is coded in a linear scale, extending from zero (black) to $1.2 \times 10^6$ ph cm$^{-2}$ s$^{-1}$ (a), $5.2 \times 10^4$ ph cm$^{-2}$ s$^{-1}$ (b), and $1.6 \times 10^6$ ph cm$^{-2}$ s$^{-1}$ (c), (white). All images show considerable limb brightening, especially at C–Kα and O–Kα. **d)** Observed X–ray image: same as Fig.15, but smoothed with a Gaussian filter with σ = 1.8″ and displayed in the same scale as the simulated images. This image is dominated by O–Kα fluorescence photons. **e)** Optical image of Venus, taken 20 hours before the ACIS–I observation. (from Dennerl *et al.*, 2002).

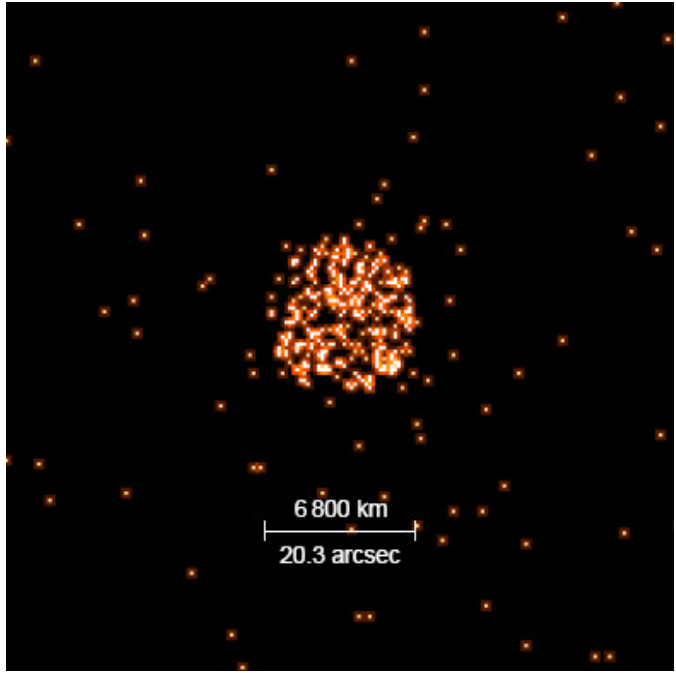

**Figure 19:** First X–ray image of Mars, obtained with Chandra ACIS–I. The X–rays result mainly from fluorescent scattering of solar X–rays on C and O in the upper Mars atmosphere, at heights of 110 – 130 km, similar to Venus. The X–ray glow of the Martian exosphere is too faint to be directly visible in this image. (from Dennerl 2002).

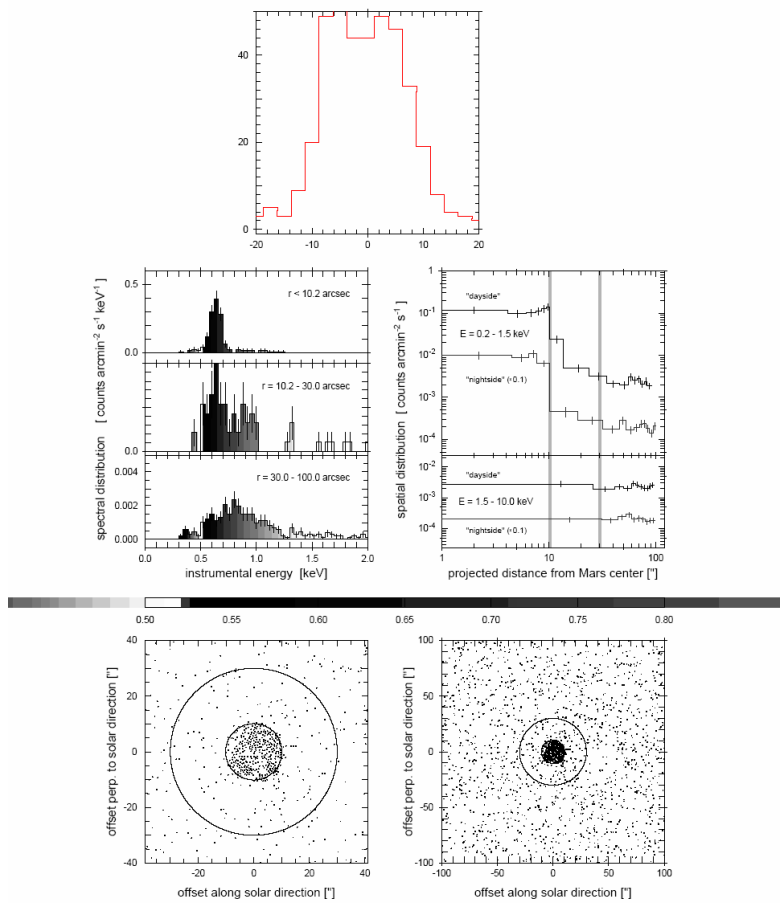

**Figure 20:** Spatial distribution of the photons around Mars in the soft ($E = 0.2 - 1.5$ keV) and hard ($E = 1.5 - 10.0$ keV) energy range, in terms of surface brightness along radial rings around Mars, separately for the "dayside" (offset along projected solar direction >0) and the "nightside" (offset <0); note, however, that the phase angle was only 18.2°. For better clarity the nightside histograms were shifted by one decade downward. The bin size was adaptively determined so that each bin contains at least 28 counts. The thick vertical lines enclose the region between one and three Mars radii. (from Dennerl 2002).

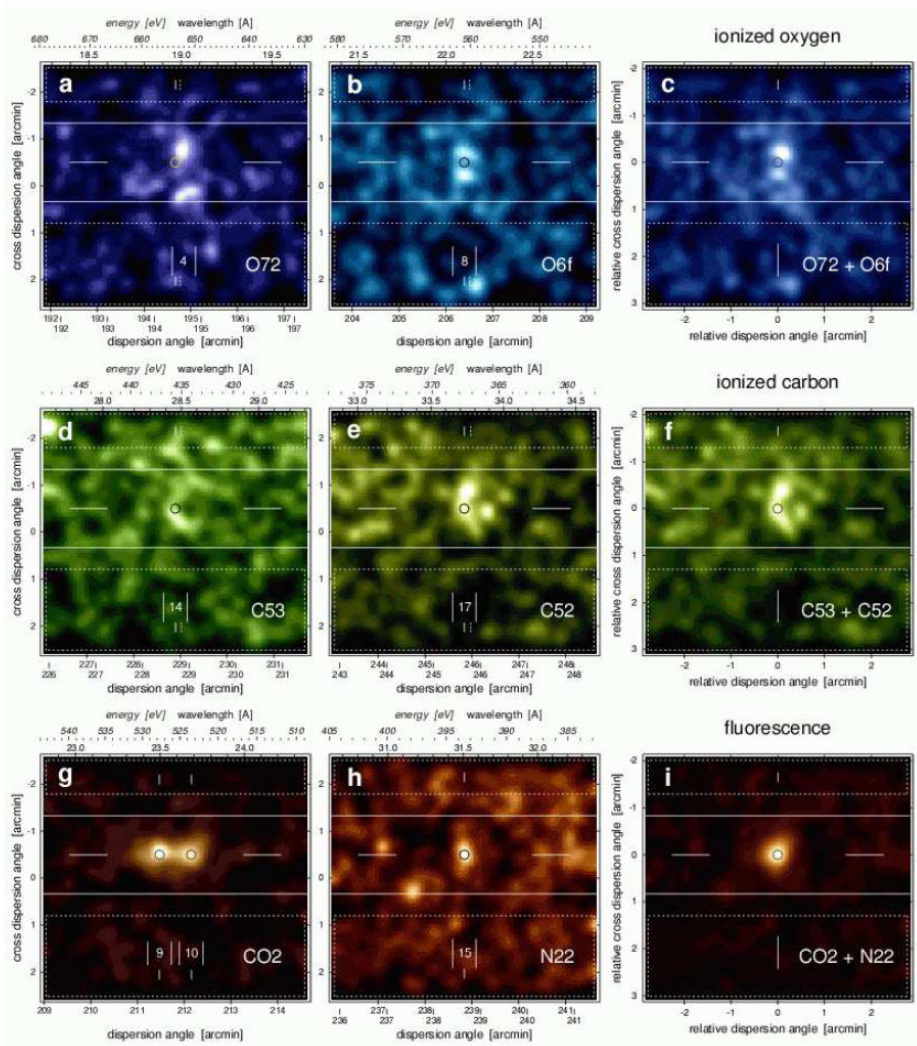

**Figure 21:** RGS images of Mars and its halo in the individual emission lines of ionized oxygen (top row), ionized carbon (middle row), and fluorescence of $CO_2$ and $N_2$ molecules (bottom row). The images were corrected for exposure variations, were binned into 2"×2" pixels and smoothed with a Gaussian function with $\sigma$ = 8"×8". All are displayed at the same angular scale; the dynamic scale, however, was individually adjusted. (from Dennerl *et al.*, 2005a).

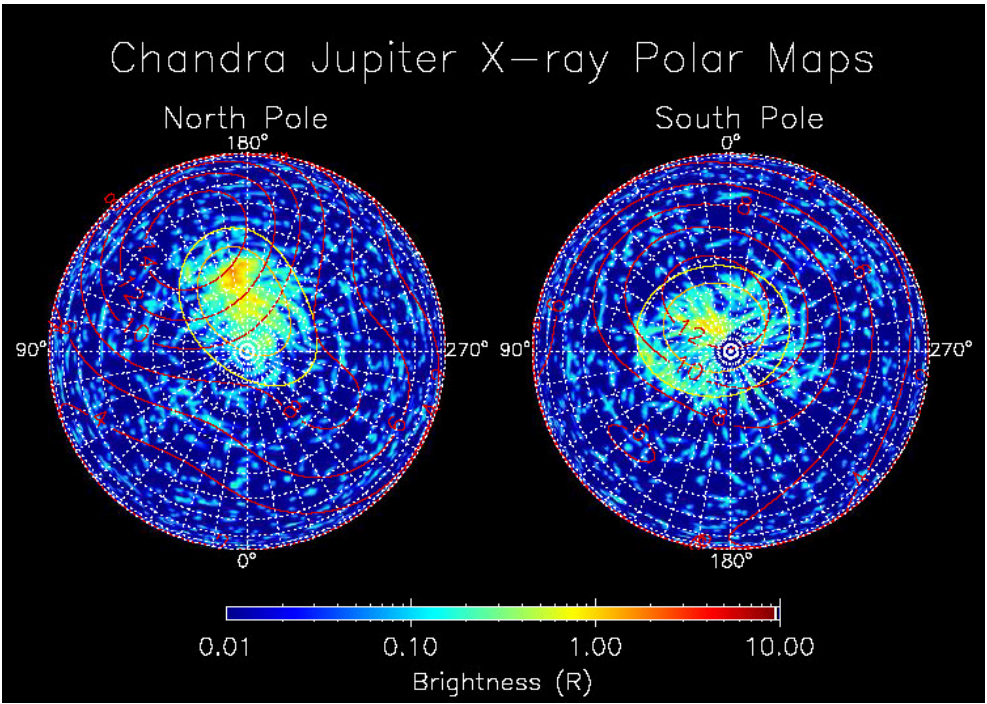

**Figure 22**: Jovian X-ray morphology first obtained with Chandra HRC-I on 18 Dec. 2000, showing bright X-ray emission from the polar 'auroral' spots, indicating the high-latitude position of the emissions.

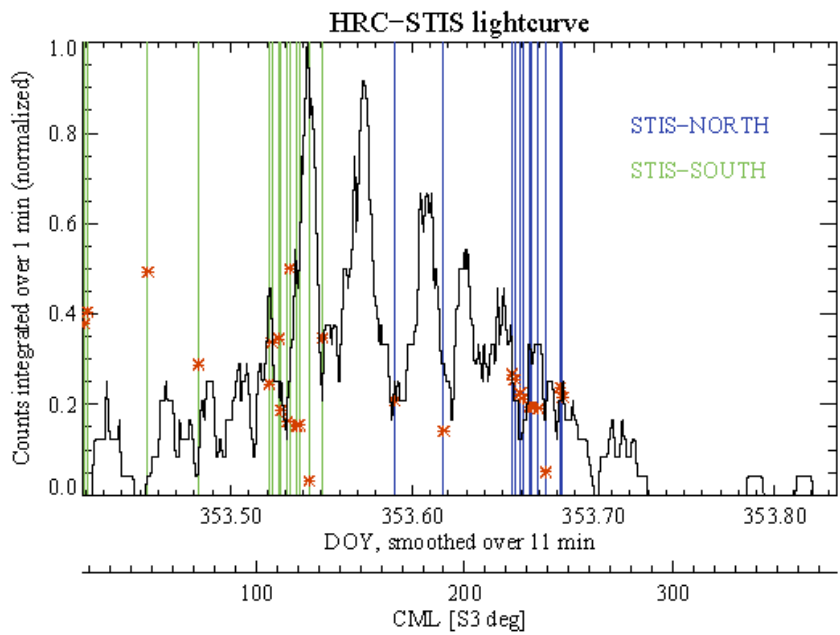

**Figure 23:** X-ray intensity variation with time indicating the 40 m periodicity at high northern latitudes of Jupiter.

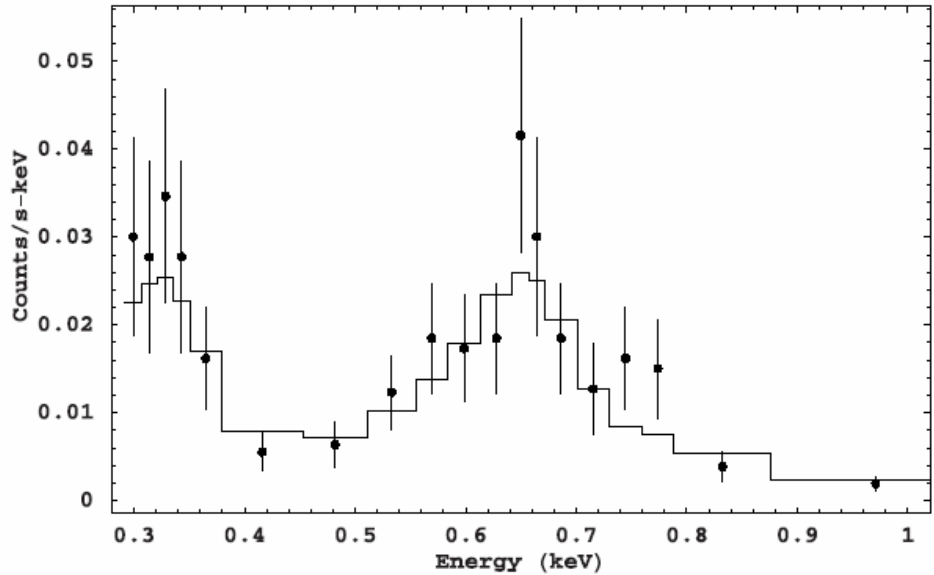

**Figure 24**: Fit with 2 added VAPEC models of the north auroral zone emission between 300 eV and 1 keV. The fitting parameters are the plasma temperature of oxygen and sulfur, the ratio of sulfur over oxygen and a normalization factor. Vapec model consider a species in collisional equilibrium. Chi squared is 11.51, the reduced chi-squared is 0.767. S/O is 16.6 times the solar value, the oxygen temperature is 355 eV and the sulfur temperature is 172 eV.

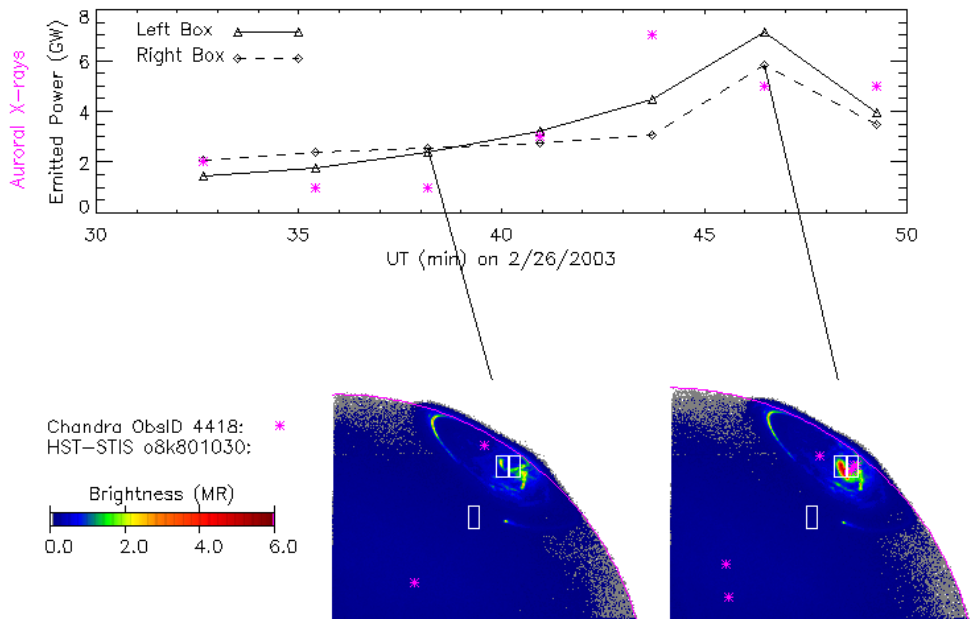

**Figure 25**: The relationship between ultraviolet emission and X-ray emission in a bright Jovian polar flare.

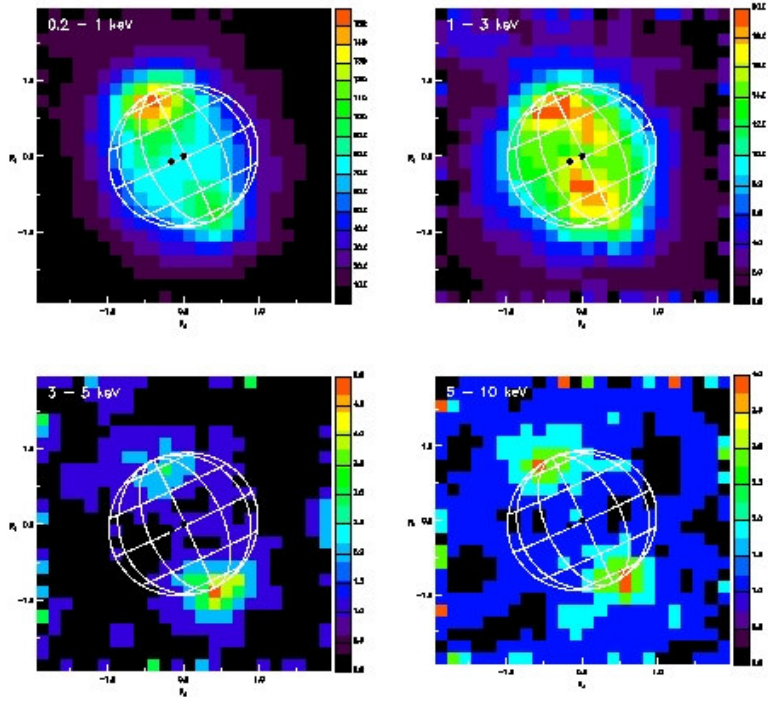

**Figure 26**: Smoothed *XMM-Newton* EPIC CCD images of Jupiter in different energy bands, highlighting the presence of a high energy component in the aurorae. From top left, clockwise: 0.2 – 1.0 keV, 1 – 3 keV, 5 – 10 keV and 3 – 5 keV. The color scale bar is in units of EPIC counts.

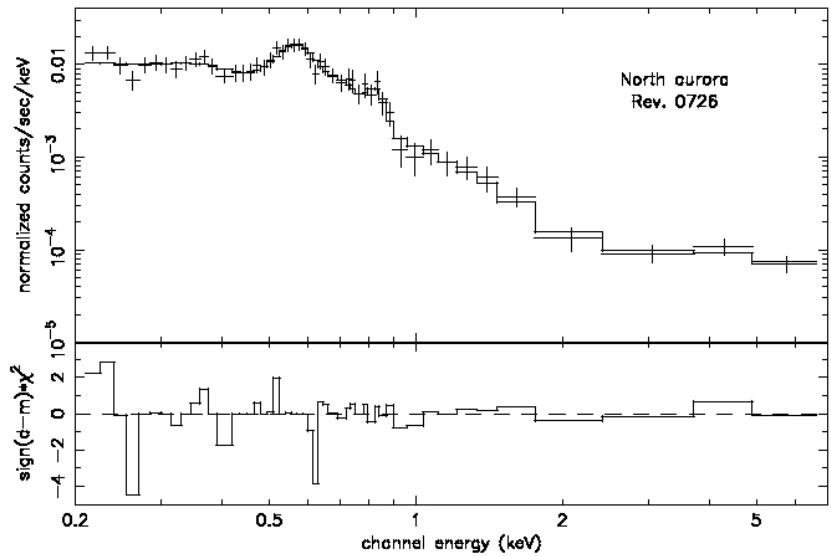

**Figure 27**: The *XMM-Newton* EPIC spectrum of Jupiter's North aurora in November 2003. The best fit comprises a thermal bremsstrahlung component (kT = 0.4 keV) and four emission lines at low energies (including strong OVII at 0.57 keV); a power law, predominant at high energies, accounts for what is believed to be an electron bremsstrahlung component.

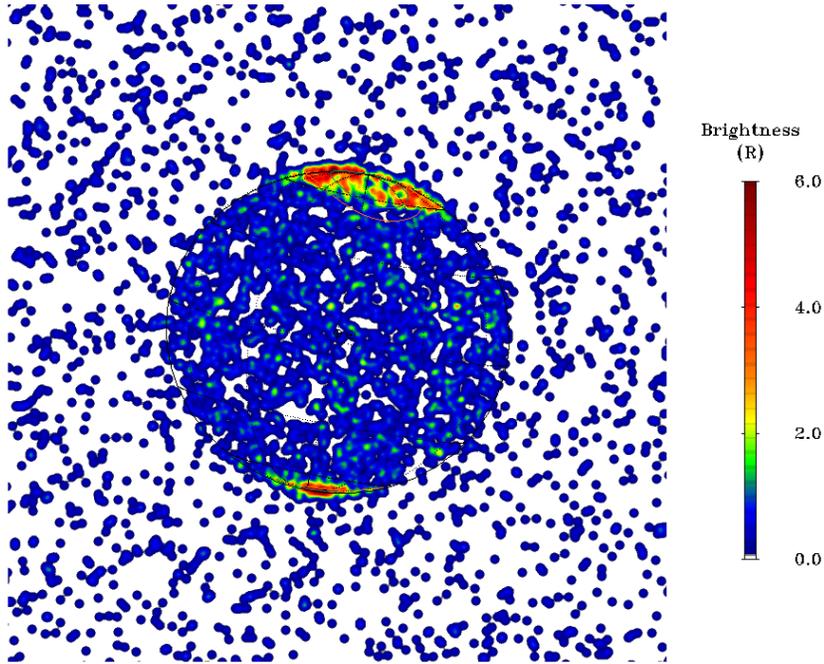

**Figure 28**. Jovian X-ray morphology first obtained with Chandra HRC-I on 18 Dec. 2000, showing bright X-ray emission from the polar 'auroral' spots, indicating the high-latitude position of the emissions, and a uniform distribution form the low-latitude 'disk' regions. (from Gladstone et al. 2002).

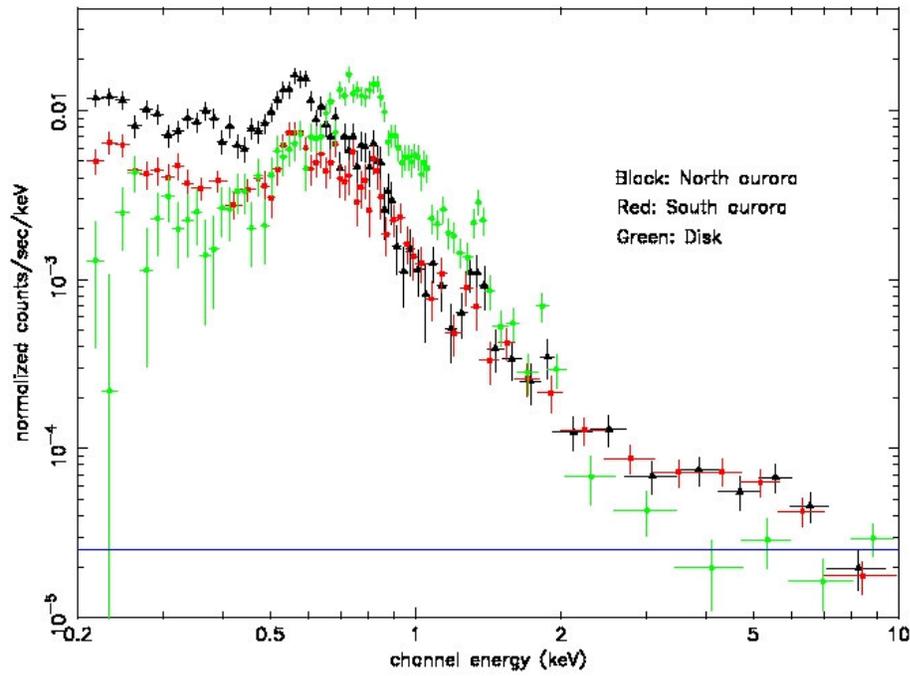

**Figure 29**. Combined XMM-Newton EPIC spectra from the Nov. 2003 observation of Jupiter. Data points for the North and South aurorae are in black and red respectively. In green is the spectrum of the low-latitude disk emission. Differences in spectral shape between the aurorae and the disk emission are very clear (see text for details).

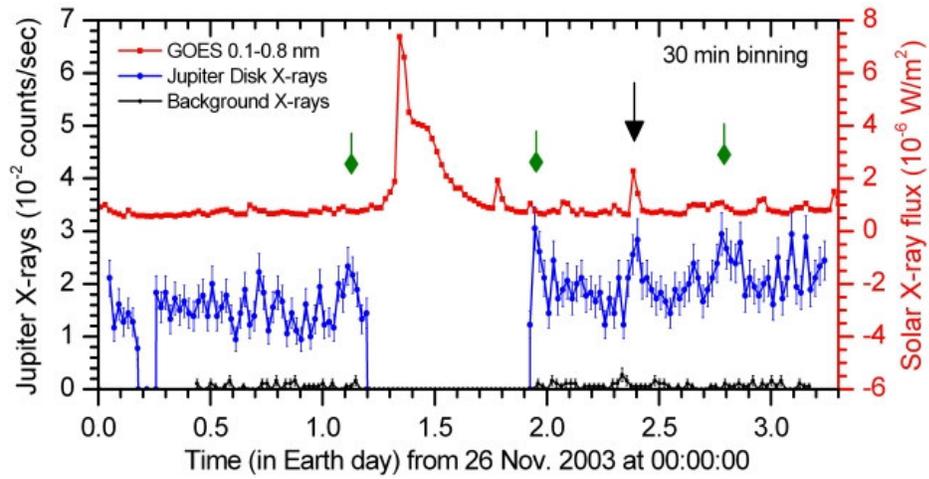

**Figure 30.** Comparison of Jupiter's disk X-ray lightcurve in Nov. 2003 (blue) with GOES 10 0.1-0.8 nm solar X-ray data (red), after shifting to account for light travel time delay. The black arrow (at 2.4 days) refers to the time of the largest solar flare visible from both, Earth and Jupiter. [from Bhardwaj et al. 2005a].

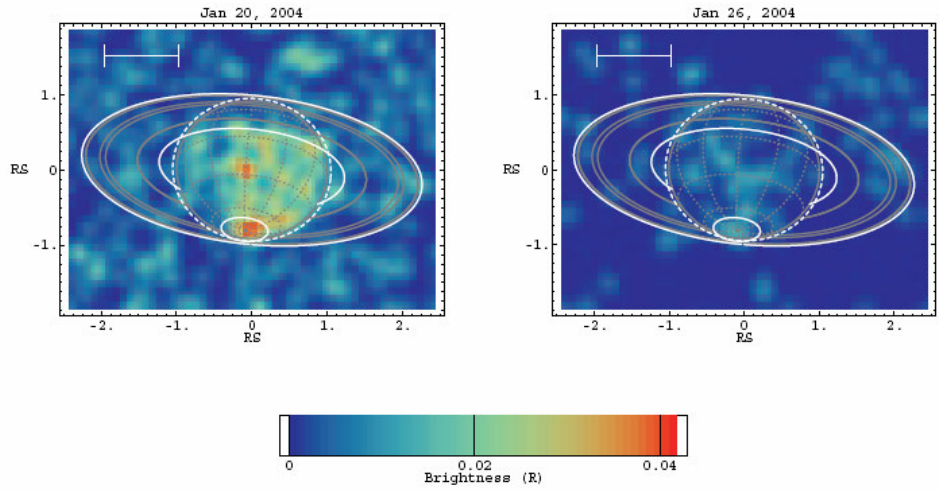

**Figure 31.** Chandra ACIS X-ray 0.24–2.0 keV images of Saturn on January 20 and 26, 2004. Each continuous observation lasted for about one full Saturn rotation. The horizontal and vertical axes are in units of Saturn's equatorial radius. The white scale bar in the upper left of each panel represents 10″. The two images, taken a week apart and shown on the same color scale, indicate substantial variability in Saturn's X-ray emission. [from Bhardwaj et al. 2005b].

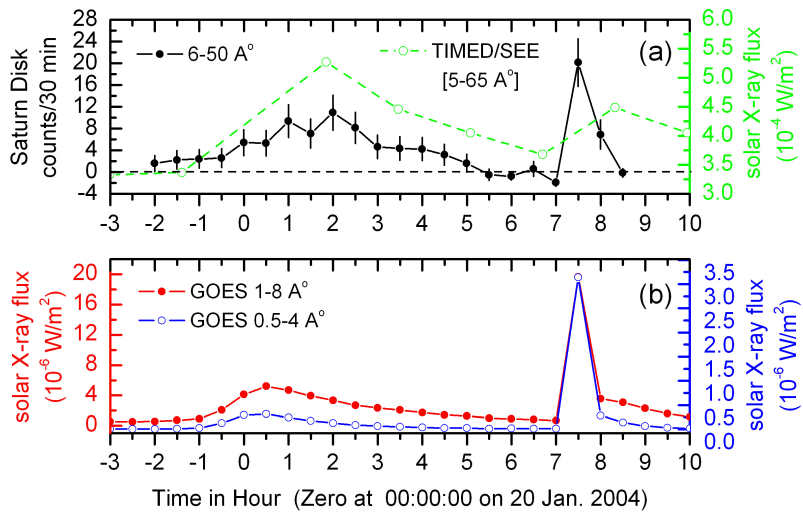

**Figure 32.** Light curve of X-rays from Saturn and the Sun on 2004 January 20. All data are binned in 30-minute increments, except for the TIMED/SEE data, which are 3 minute observation-averaged fluxes obtained every orbit (~12 measurements per day). (*a*) Background-subtracted low-latitude (nonauroral) Saturn disk X-rays (0.24–2.0 keV) observed by Chandra ACIS, plotted in black (after shifting by -2.236 hr to account for the light-travel time difference between Sun-Saturn-Earth and Sun-Earth). The solar 0.2–2.5 keV fluxes measured by TIMED/SEE are denoted by open green circles and are joined by the green dashed line for visualization purpose. (*b*) Solar X-ray flux in the 1.6–12.4 and 3.1–24.8 keV bands measured by the Earth-orbiting GOES-12 satellite. A sharp peak in the light curve of Saturn's disk X-ray flux—an X-ray flare—is observed at about 7.5 hr, which corresponds in time and magnitude with an X-ray solar flare. In addition, the temporal variation in Saturn's disk X-ray flux during the time period prior to the flare is similar to that seen in the solar X-ray flux. [from Bhardwaj et al. 2005b].

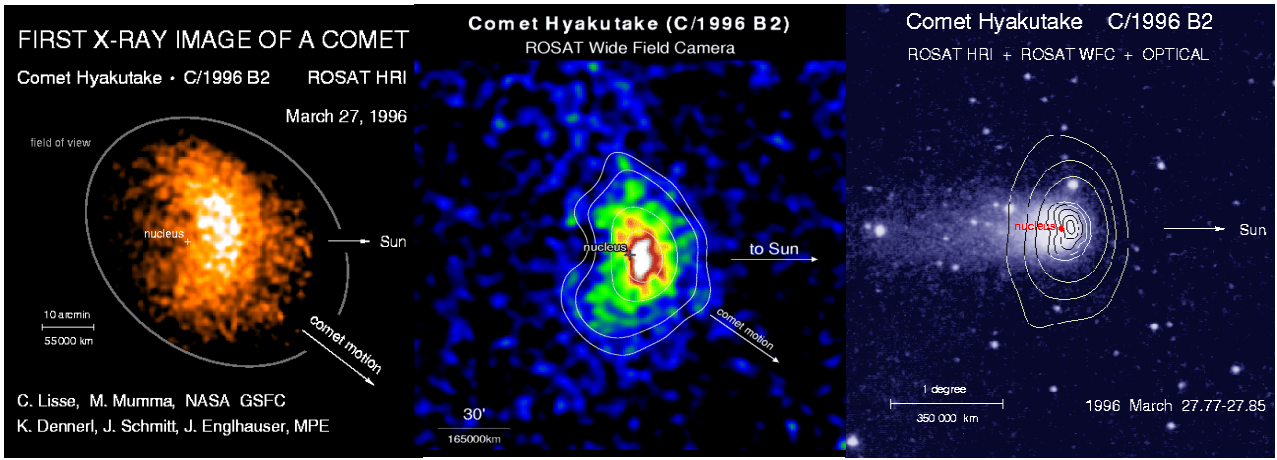

**Fig. 33a**

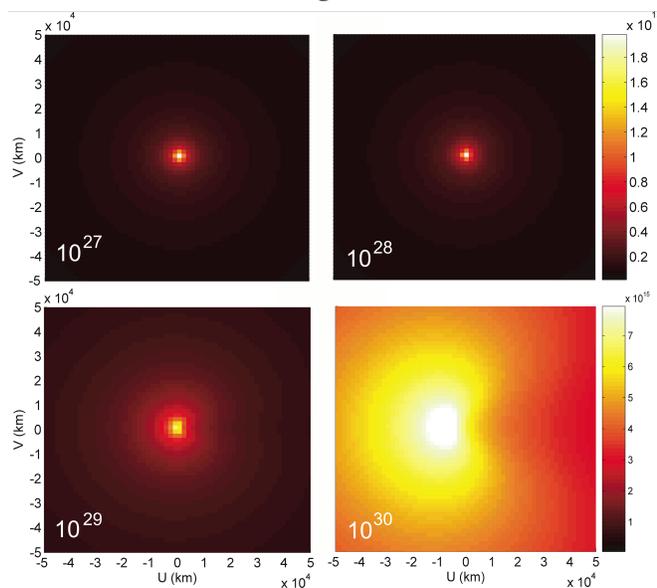

**Fig. 33b**

**Figure 33.** Cometary X-ray Emission Morphology. **a)** Images of C/Hyakutake 1996B2 on 26 - 28 March 1996 UT: **Left)** ROSAT HRI 0.1 - 2.0 keV X-ray, **middle)** ROSAT WFC .09 - 0.2 keV extreme ultraviolet, and **Right)** visible light, showing a coma and tail, with the X-ray emission contours superimposed. The Sun is towards the right, the plus signs mark the position of the nucleus, and the orbital motion of the comet is towards the lower right in each image. From Lisse et al. (1996). **b)** Morphology as a function of comet gas production rate (given in terms of molecules sec-1 in the lower right of each panel), following Bodewits et al. (2005). Note the decreasing concentration of model source function and the increasing importance of diffuse halo emission in the extended coma as the gas production rate increases. from Lisse et al. (2005).

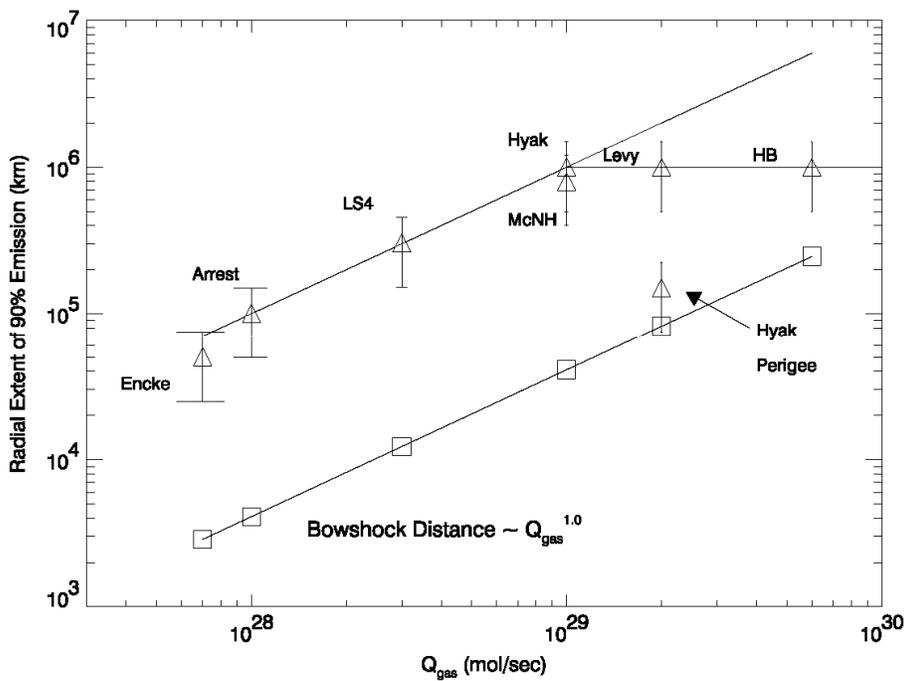

**Figure 34.** Spatial extent of the emission vs. the comet's outgassing rate. Plot of the estimated gas production rate $Q_{gas}$ vs. radial distance from the comet nucleus required to encircle 95% of the total observed cometary X-ray flux. Upper curve - broken power law with radial extent $\sim Q_{gas}^{1.00}$ up to $Q_{gas} \sim 10^{29}$ mol sec$^{-1}$ and $\sim 10^6$ km for higher values fits the imaging data well. Lower curve - estimated radius of the bowshock for each observation, allowing for variable cometary outgassing activity and heliocentric distance (boxes). X-ray emission has been found outside the bowshock for all comets except for C/1996 B2 (Hyakutake) in March 1996, which is clearly an outlier from the best-fit trend.

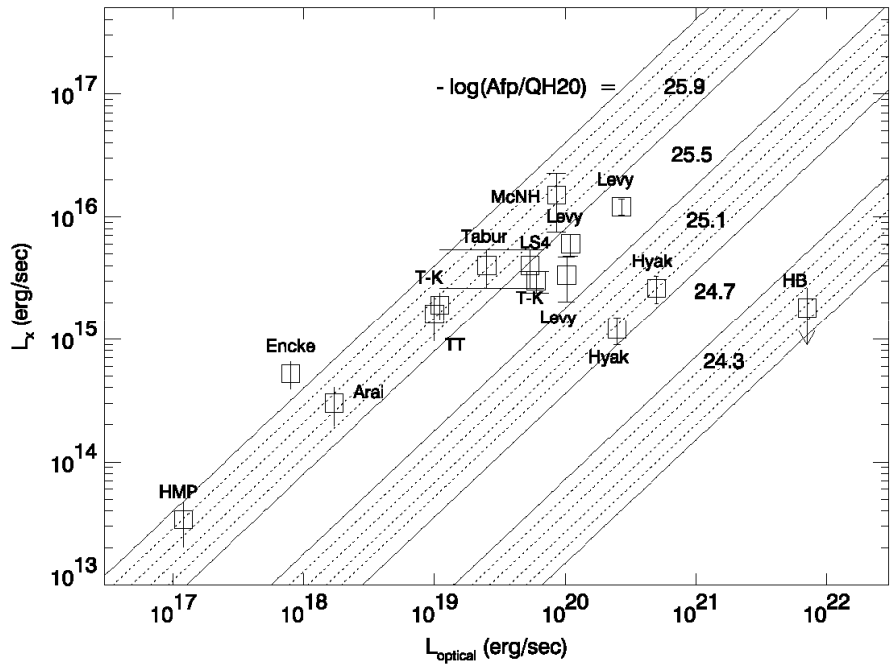

**Figure 35.** X-ray vs. optical luminosity plot (following Dennerl et al. 1997) for the 8 detected ROSAT comets and the Chandra comets C/1999 S4 (LINEAR) and C/1999 T1 (McNaught-Hartley) observed at 1- 3 AU. Groups of equal emitted dust mass to emitted gas mass ratio (D/G), as measured in the optical by the ratio $Af\rho/Q_{H2O}$, are also shown. For Encke and other "gassy", optically faint comets, the resulting slope $L_x/L_{opt}$ is roughly constant. Above $L_{opt} \sim$ few x $10^{19}$ erg s$^{-1}$, however, $L_x$ appears to reach an asymptote of $\sim$ 5 x $10^{16}$ erg s$^{-1}$ - the extremely active, bright, and "dusty" comets C/Levy 1990, C/1996 B2 (Hyakutake), and C/Hale-Bopp 1995 O1 are little brighter in the X-ray than C/Tabur 1996, P/Tempel-Tuttle 1998, or C/1999 S4 (LINEAR). A possible explanation is that this is the total amount of solar wind ions available within the neutral coma radius of $\sim 10^6$ km at 1 AU (Lisse et al. 2001). It is also possible that the relatively large amounts of dust in these comets, as noted from their increasing D/G ratio, may be somehow inhibiting the CXE process.

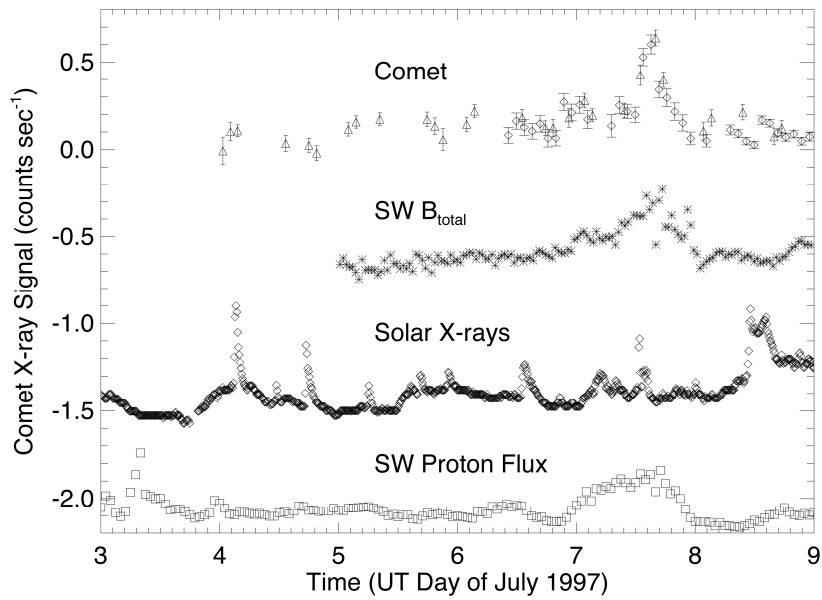

**Fig. 36a.**

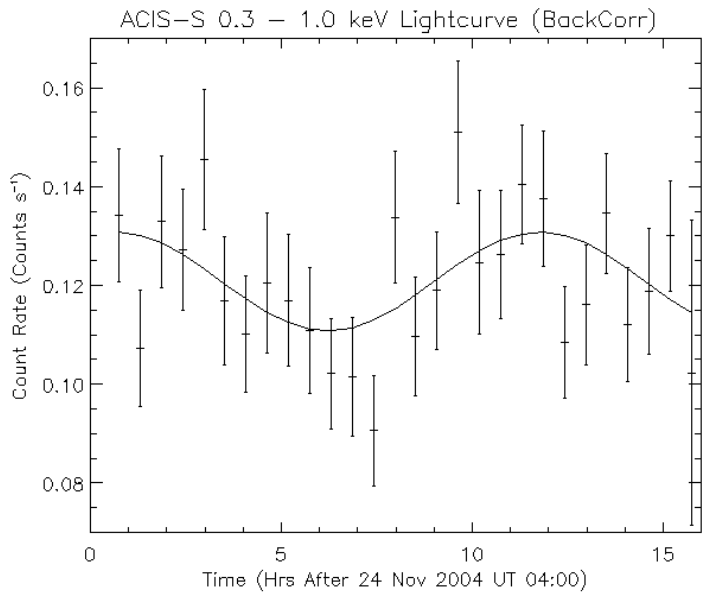

**Fig. 36(b)**

**Figure 36.** Temporal trends of the cometary X-ray emission. **a)** Lightcurve, solar wind magnetic field strength, solar wind proton flux, and solar X-ray emission for 2P/Encke 1997 on 4-9 July 1997 UT. All error bars are ± 1σ. Δ - HRI light curve, 4-8 July 1997. ◊ - EUVE scanner Lexan B light curve 6 - 8 July 1997 UT, taken contemporaneously with the HRI observations, and scaled by a factor of 1.2. Also plotted are the WIND total magnetic field $B_{total}$ (*), the SOHO CELIAS/SEM 1.0 - 500 Å solar X-ray flux (◊), and the SOHO CELIAS solar wind proton flux (boxes). There is a strong correlation between the solar wind magnetic field/density and the comet's emission. There is no direct correlation between outbursts of solar X-rays and the comet's outbursts. After Lisse *et al.* 1997a. **b)** High time resolution lightcurve, total Chandra ACIS-S 0.3 – 1.0 keV count rate versus time, grouped by the 30 pointings of the spacecraft on the comet for 2P/Encke on 24 Nov 2003. Crosses – as observed count rate, with solar wind flux trend removed. Plotted error bars are 1s. Solid curve – best fit lightcurve using the 11.1 hour period of Fernandez *et al.* 2005 and Belton et al. 2005. The best-fit X-ray sinusoid has a 20% (peak to peak) amplitude versus the mean of 0.121 cps, and a $\chi^2_n$ of 1.04, versus a $\chi^2_n$ of 1.4 for no variation with time, which can be rejected at the 98% C.L. Modulation due to nucleus rotation is clearly seen.

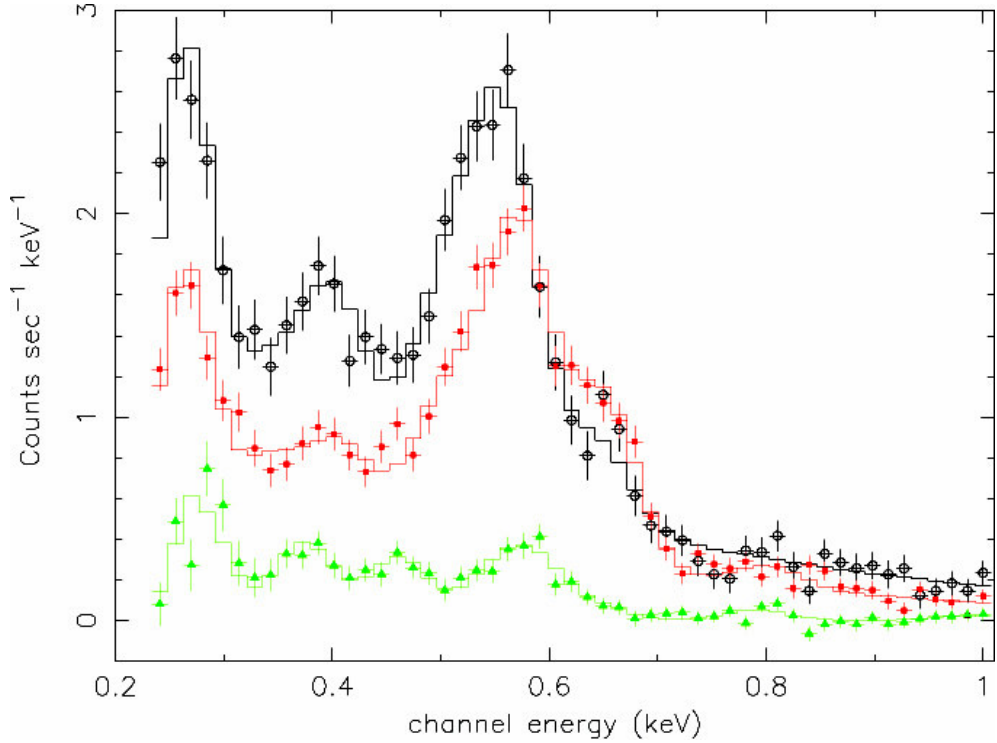

**Figure 37.** Chandra ACIS medium resolution CCD X-ray spectra of the X-ray emission from three comets. All curves show ACIS-S3 measurements of the 0.2 – 1.5 keV pulse height spectrum, as measured in direct detection mode. with ±1σ error bars and the best-fit emission line + thermal bremsstrahlung model convolved with the ACIS-S instrument response as a histogram. The positions of several possible atomic lines are noted. Pronounced emission due to OVIII and OVII is evident at 560 and 660 eV, and for CVI, CV, and NVI emission lines at 200 - 500 eV . Best-fit model lines at 284, 380, 466, 552, 590, 648, 796, and 985 are close to those predicted for charge exchange between solar wind $C^{+5}$, $C^{+6}$, $C^{+6}/N^{+6}$, $O^{+7}$, $O^{+7}$, $O^{+8}$, $O^{+8}$, and $Ne^{+9}$ ions and neutral gases in the comet's coma. (Red) ACIS spectra of C/LINEAR 1999 S4 (circles), from Lisse *et al.* (2001). (Black) Comet McNaught-Hartley spectra (squares), after Krasnopolsky *et al.* 2003. (Green) 2P/Encke spectrum taken on 24 Nov 2003, multiplied by a factor of 2. The C/1999 S4 (LINEAR) and C/McNaught-Hartley 2001 observations had an average count rate on the order 20 times as large, even though Encke was closer to Chandra and the Earth when the observations were being made. Note the 560 complex to 400 eV complex ratio of 2 to 3 in the two bright, highly active comets, and the ratio of approximately 1 for the faint, low activity comet Encke. from Lisse et al. (2005).

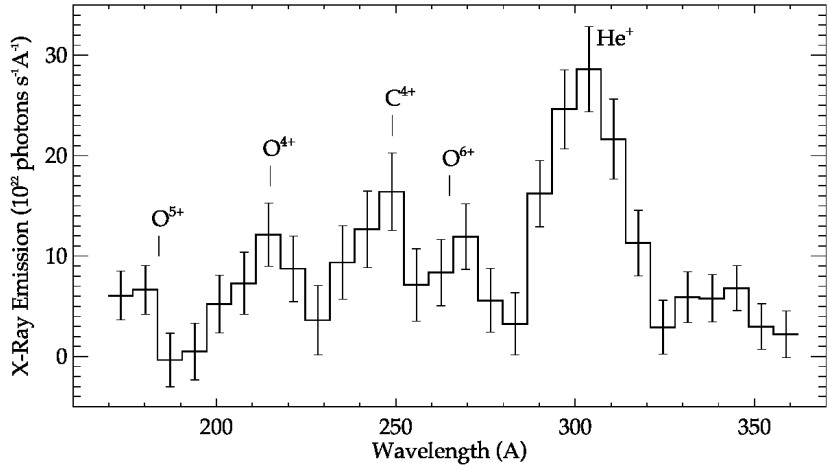

**Fig. 38(a).**

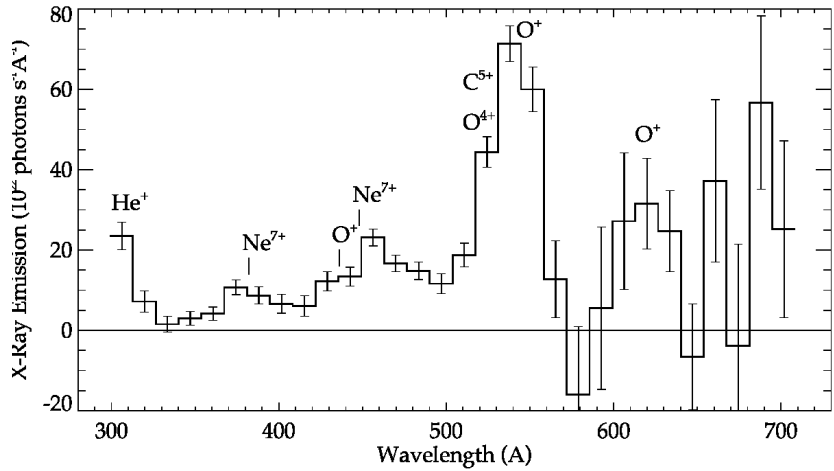

**Fig. 38(b).**

**Figure 38**. EUVE observations of line emission from C/1996 B2 (Hyakutake), following Krasnopolsky and Mumma 2001. **a)** MW (middle wavelength) 0.034 - 0.073 keV spectrum on March 23, 1996. **b)** LW (long-wavelength) 0.018 - 0.04 keV. The extreme ultraviolet spectra are clearly dominated by line emission. The best agreement with CXE model predictions are for the $O^{+4}$, $O^{+6}$, $C^{+4}$, and $Ne^{+7}$ lines.

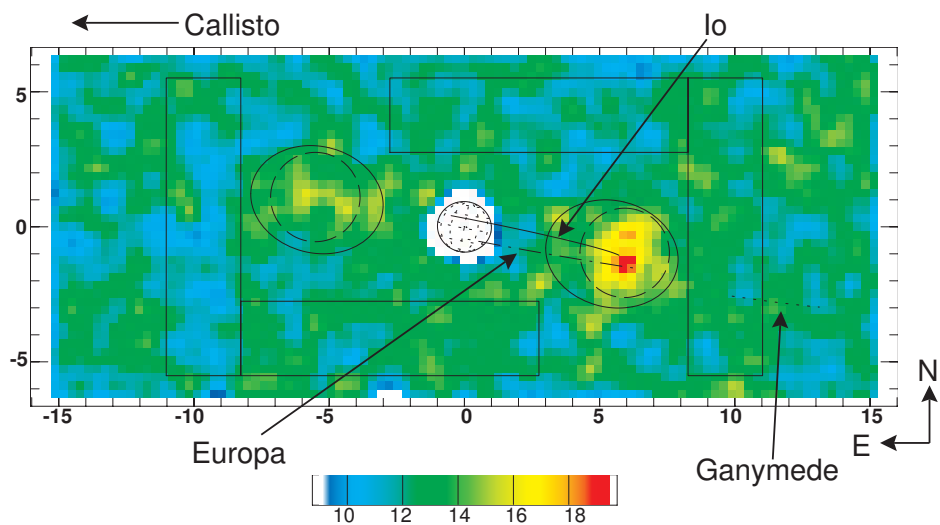

**Figure 39.** CXO/HRC-I image of the IPT (2000 December 18). The image has been smoothed by a two-dimensional Gaussian with σ = 7.38'' (56 HRC-I pixels). The axes are labeled in units of Jupiter's radius, $R_J$, and the scale bar is in units of smoothed counts per image pixel. The paths traces by Io, Europa, and Ganymede are marked on the image. Callisto is off the image to the dawn side, although the satellite did fall within the full microchannel plate field of view. The regions bounded by rectangles were used to determine background. The regions bounded by dashed circles or solid ellipses were defined as source regions.

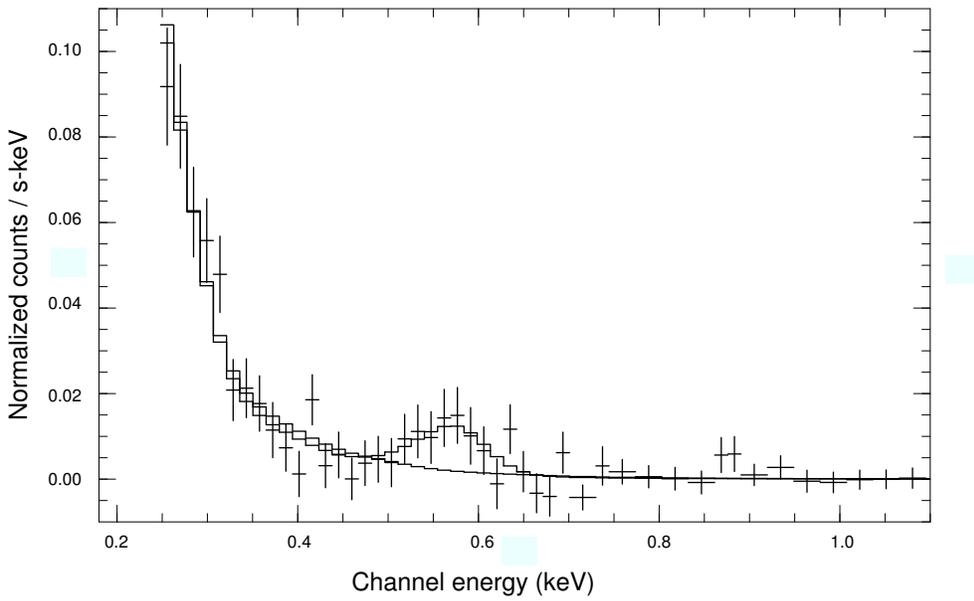

**Figure 40.** CXO spectrum for the Io Plasma Torus from November 1999. The solid line presents a model fit for the sum of a power-law spectrum and a Gaussian line, while the dashed line represents just a pure power law spectrum. The line is consistent with K-shell flurorescent emission from oxygen ions.

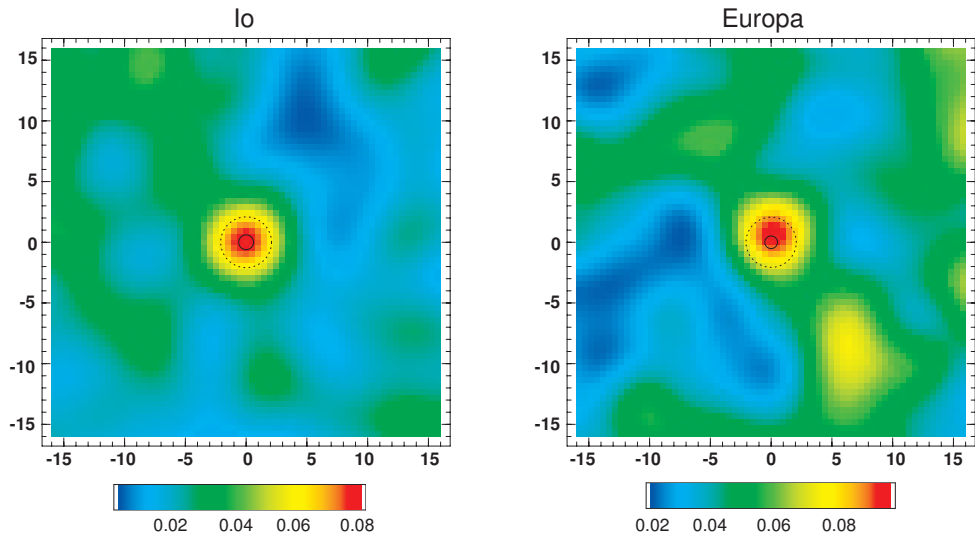

**Figure 41.** CXO images of Io and Europa (0.25 keV < E < 2.0 keV) from November, 1999 (Elsner et al. 2002). The images have been smoothed by a two-dimensional gaussian with $\sigma = 2.46$ arcsec (5 detector pixels). The axes are labeled in arcsec (1 arcsec $\simeq$ 2995 km) and the scale bar is in units of smoothed counts per image pixel (0.492 by 0.492 arcsec). The solid circle shows the size of the satellite (the radii of Io and Europa are 1821 km and 1560 km, respectively), and the dotted circle the size of the detect cell.

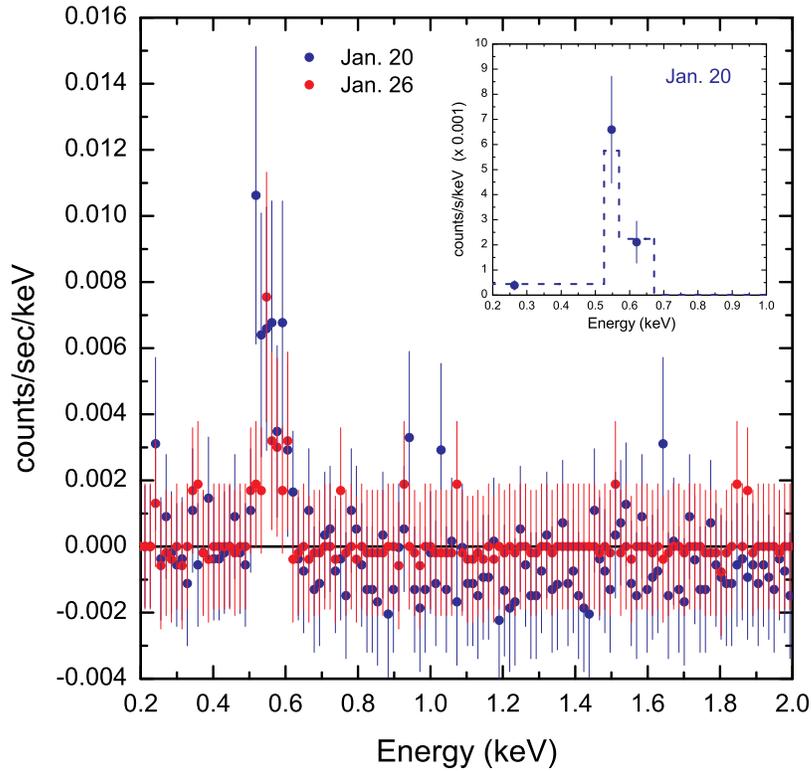

**Figure 42.** Background-subtracted Chandra ACIS-S3–observed X-ray energy spectrum for Saturn's rings in the 0.2–2.0 keV range on 2004 January 20 and 26–27. The cluster of X-ray photons in the ~0.49–0.62 keV band suggests the presence of the oxygen Kα line emission at 0.53 keV in the X-ray emission from the rings. The inset shows a Gaussian fit (peak energy = 0.55 keV, σ = 140 eV), shown by the dashed line, to the ACIS-observed rings' spectrum on January 20. Each spectral point (filled circle with error bar) represents ≥10 measured events. The spectral fitting suggests that X-ray emissions from the rings are predominantly oxygen Kα photons.

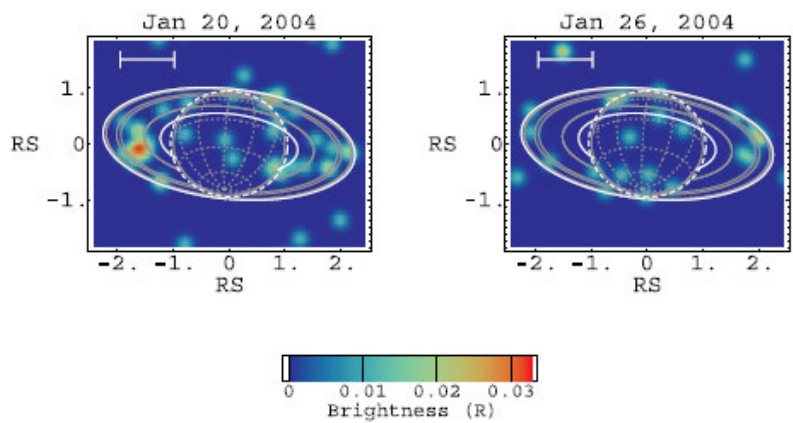

**Figure 43.** Chandra ACIS X-ray images of the Saturnian system in the 0.49–0.62 keV band on 2004 January 20 and 26–27. The X-ray emission from the rings is clearly present in these restricted energy band images (see Fig. rings1); the emission from the planet is relatively weak in this band (see Fig. S1 for an X-ray image of the Saturnian system in the 0.24–2.0 keV band).

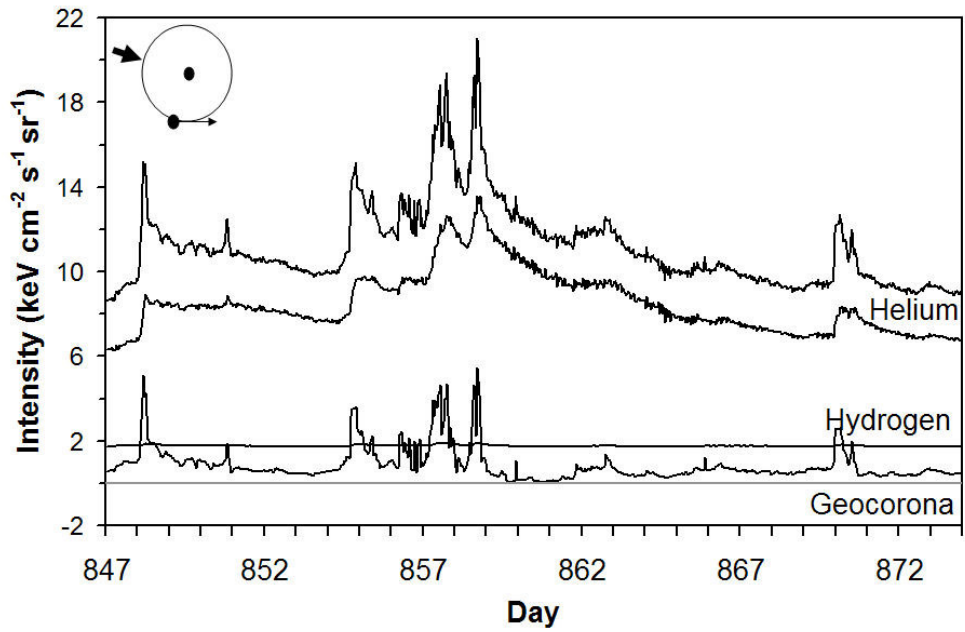

**Figure 44**: SWCX X-ray intensities for a downwind direction. The direction of the interstellar wind is indicated with the thick arrow. The center of the circle is the Sun. The location of the Earth and the look direction are also indicated. The top curve is the total intensity and the labeled curves indicate the individual contributions to the total.

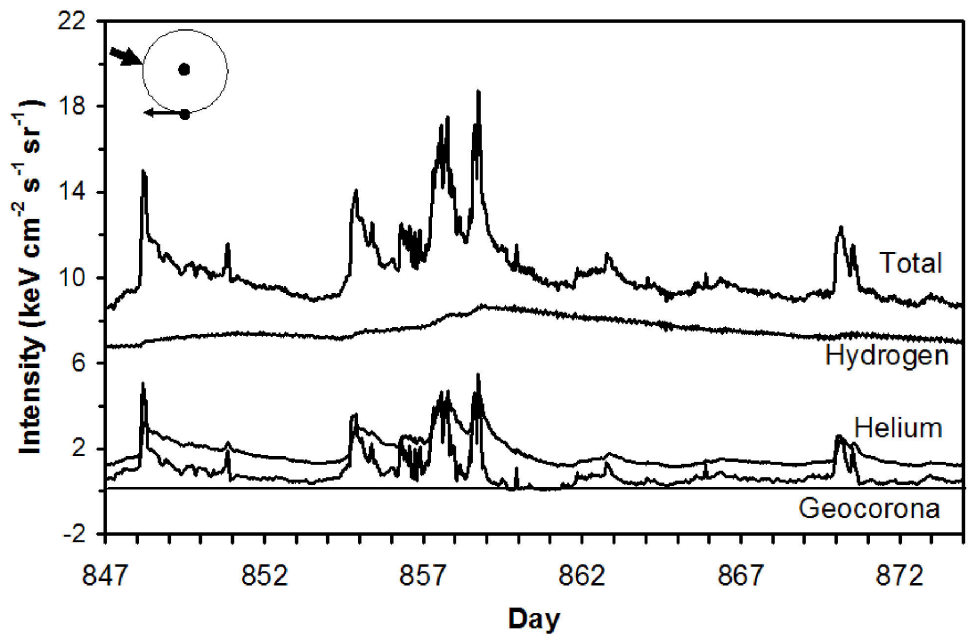

**Figure 45**: Similar to figure 44, but for an upwind look direction.

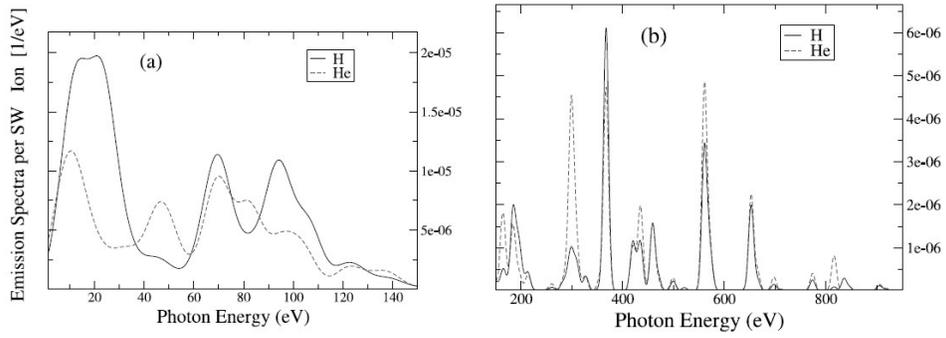

**Figure 46**: Predicted spectra of EUV and soft X-ray emission due to charge exchange between the slow solar wind and neutral hydrogen and helium for low photon energy (a) and higher photon energy (b). Number of photons is normalized to a single solar wind ion. [Pepino et al., 2004].

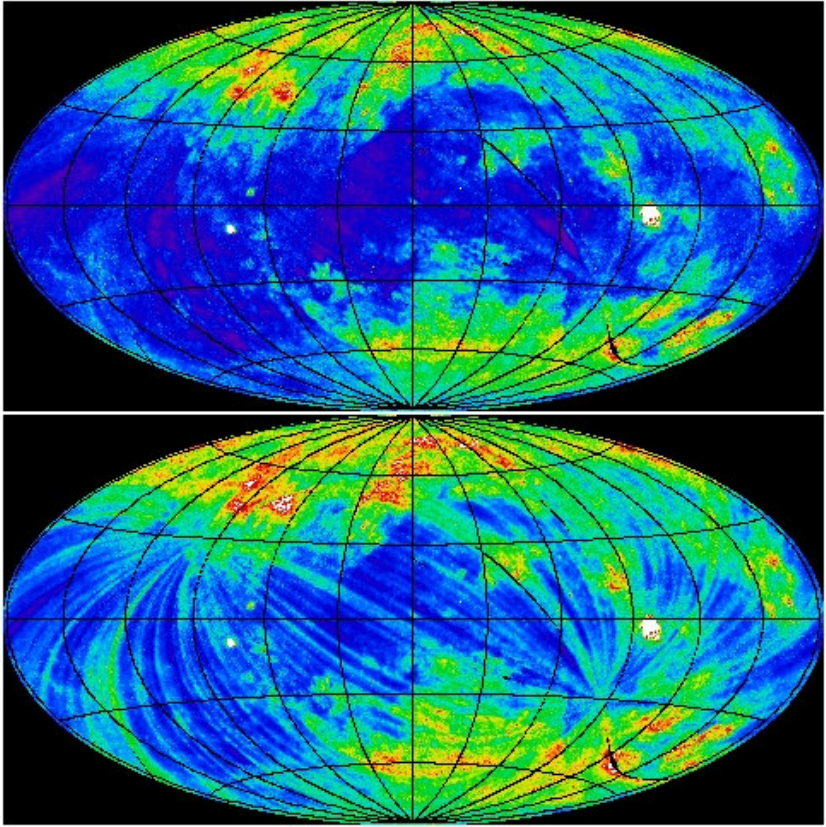

**Figure 47.** Upper panel) ROSAT All-Sky Survey map of the cosmic X-ray background at ¼ keV (Snowden et al. 1997). The data are displayed using an Aitoff projection in Galactic coordinates centered on the Galactic center with longitude increasing to the left and latitude increasing upwards. Low intensity is indicated by purple and blue while red indicates higher intensity. Lower panel) Same as above except the contaminating long-term enhancements (SWCX emission) were not removed. The striping is due to the survey geometry where great circles on the sky including the ecliptic poles were scanned precessing at ~1° per day.

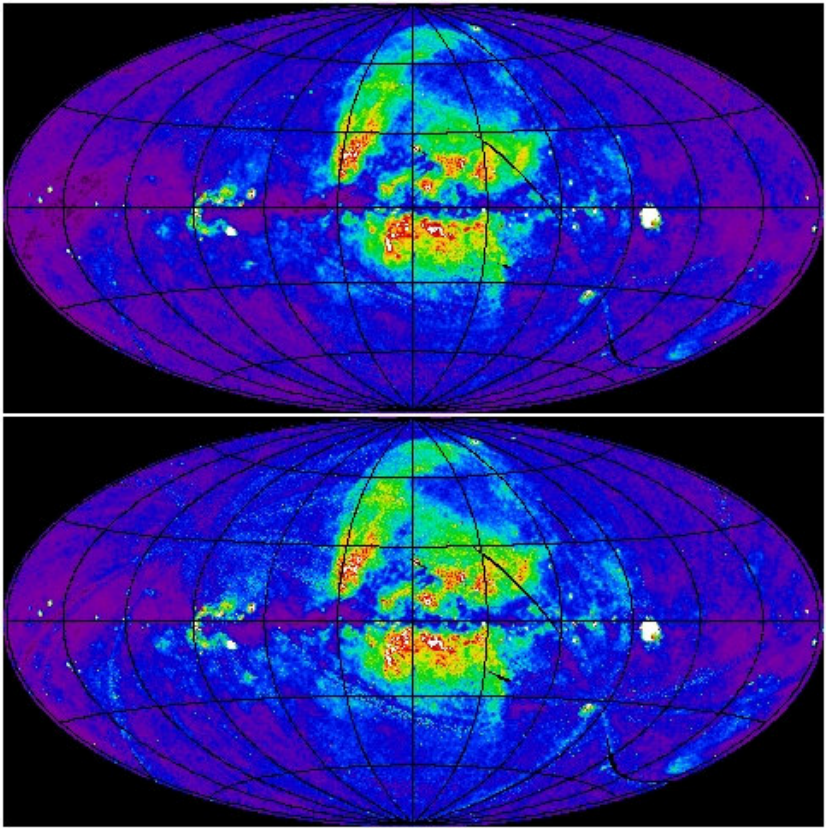

**Figure 48.** Same as Figure 47 except for the RASS ¾ keV band.

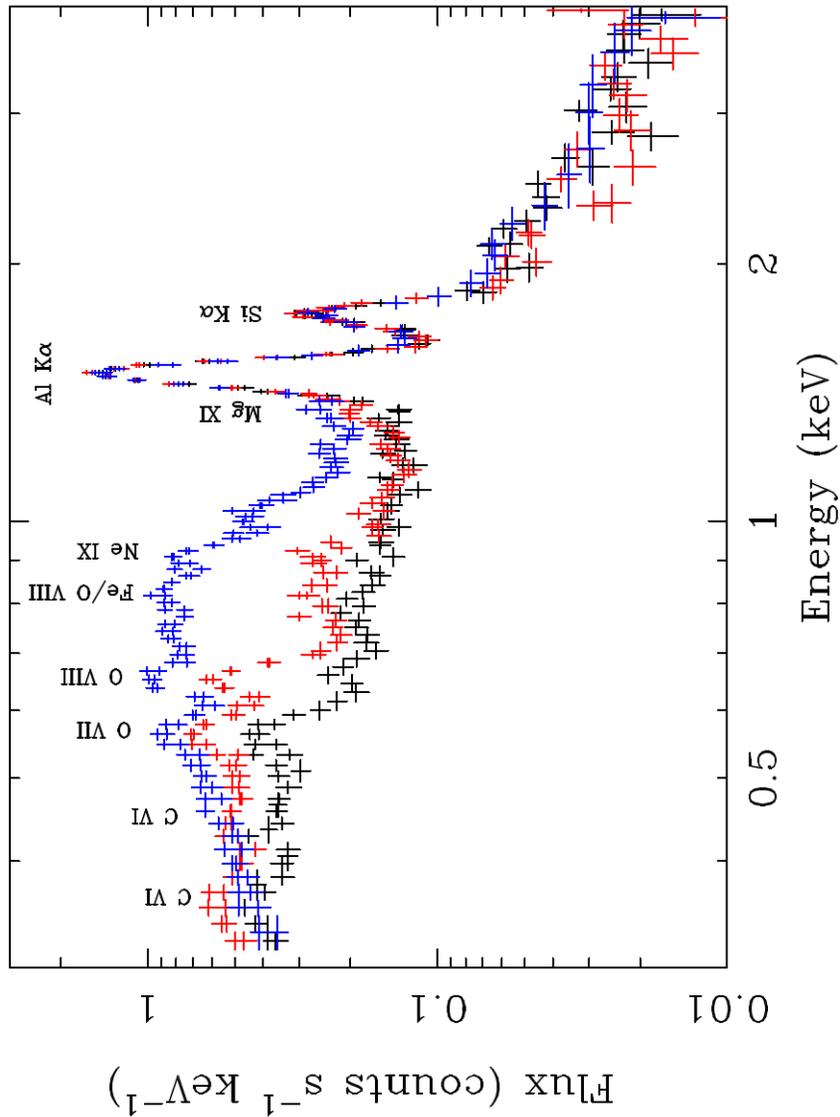

**Figure 49**: XMM-Newton MOS spectra from two directions on the sky are plotted. The blue spectrum is from a relatively bright region near the Galactic center (l,b~348.6,22.4) while the red and black spectra are from a relatively dim region at higher latitude (l,b~125.9,54.8). The excess of the red spectrum over the black is from SWCX. The Al Kalpha and Si Kalpha lines are instrumental but the other lines are astrophysical in origin.

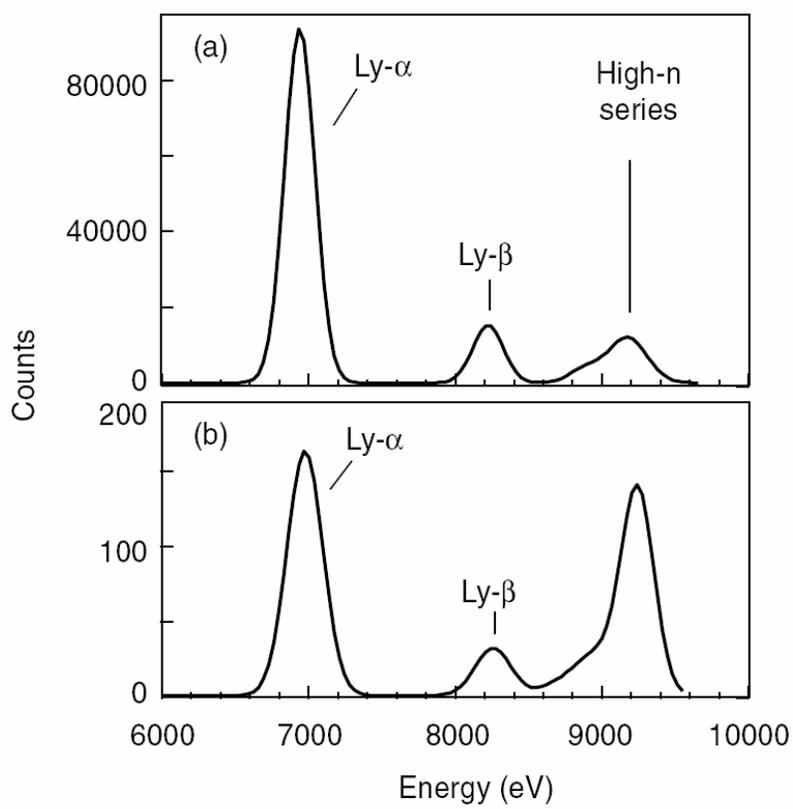

**Figure 50.** X-ray emission of $Fe^{25+}$ produced (a) by charge exchange in the reaction $Fe^{26+} + N_2$ and (b) by direct electron-impact excitation. Both spectra were measured on an electron beam ion trap at the Lawrence Livermore National Laboratory by Wargelin et al. (2005) using an intrinsic germanium detector. The Lyman lines are labeled in standard notation.

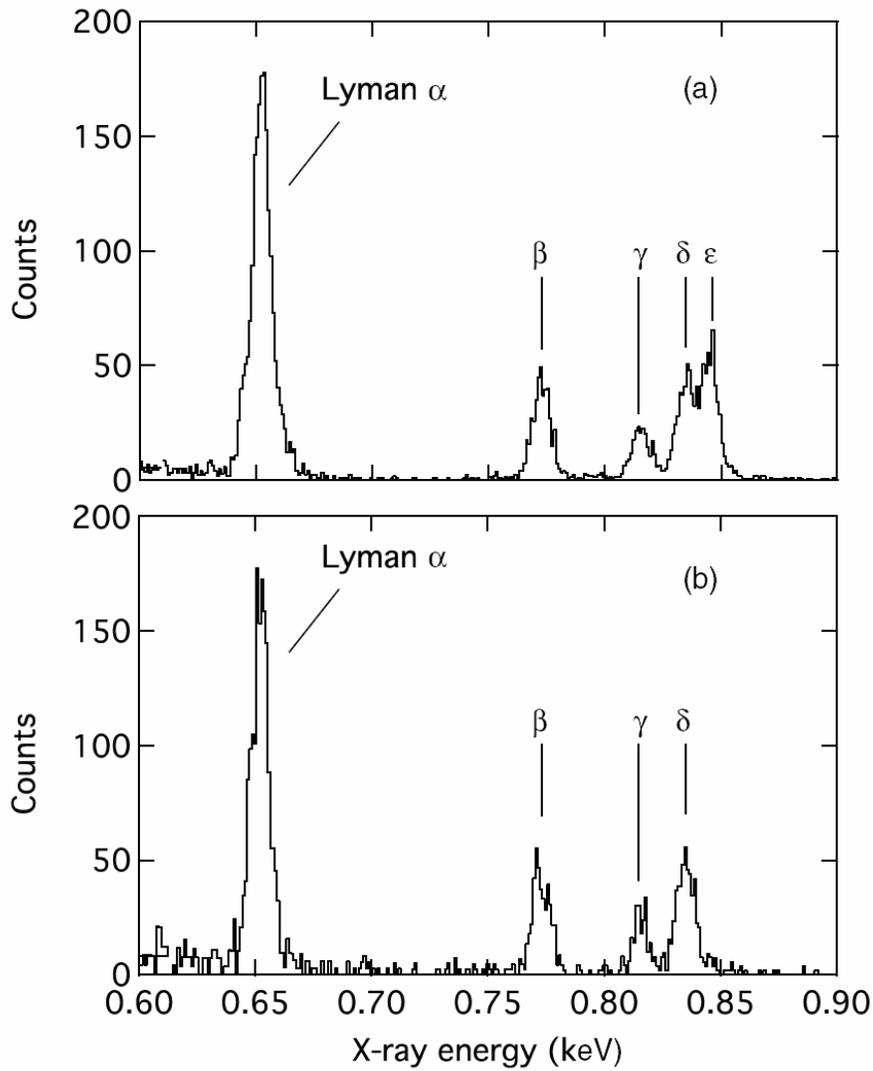

**Figure 51.** K-shell emission of H-like $O^{7+}$ produced by charge exchange with (a) $CH_4$ and (b) $N_2$. The spectra were measured on an electron beam ion trap at the Lawrence Livermore National Laboratory by Beiersdorfer et al. (2005b). Note that Lyman-$\varepsilon$ emanating from $n = 6$ is only seen in (a).

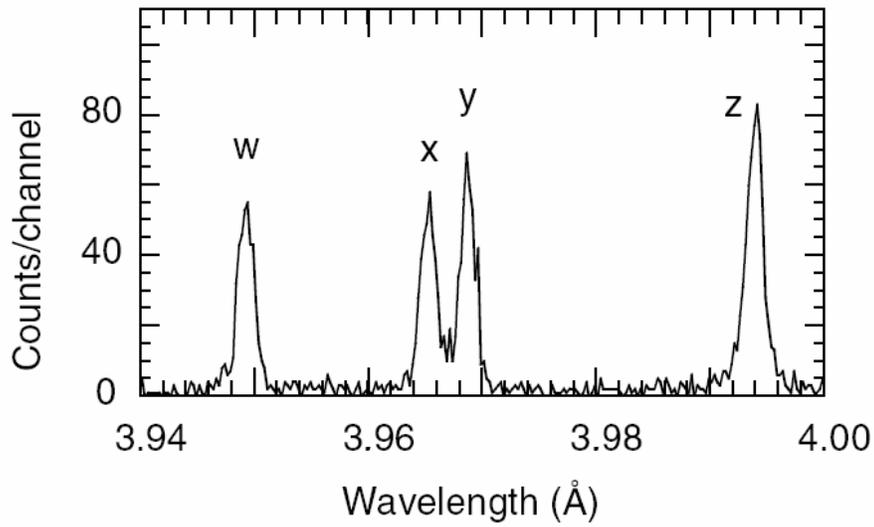

**Figure 52.** Spectrum of $Ar^{16+}$ excited in the charge exchange reaction between $Ar^{17+}$ ions and a 80 keV neutral deuterium beam. The spectrum was recorded on the NSTX tokamak by Beiersdorfer et al. (2005a). The labels *w, x, y* and *z* denote the resonance, intercombination, and forbidden transitions from upper levels $1s2p\ ^1P_1$, $1s2p\ ^3P_2$, $1s2p\ ^3P_1$, and $1s2s\ ^3S_1$, respectively, to the $1s^2\ ^1S_0$ ground state.

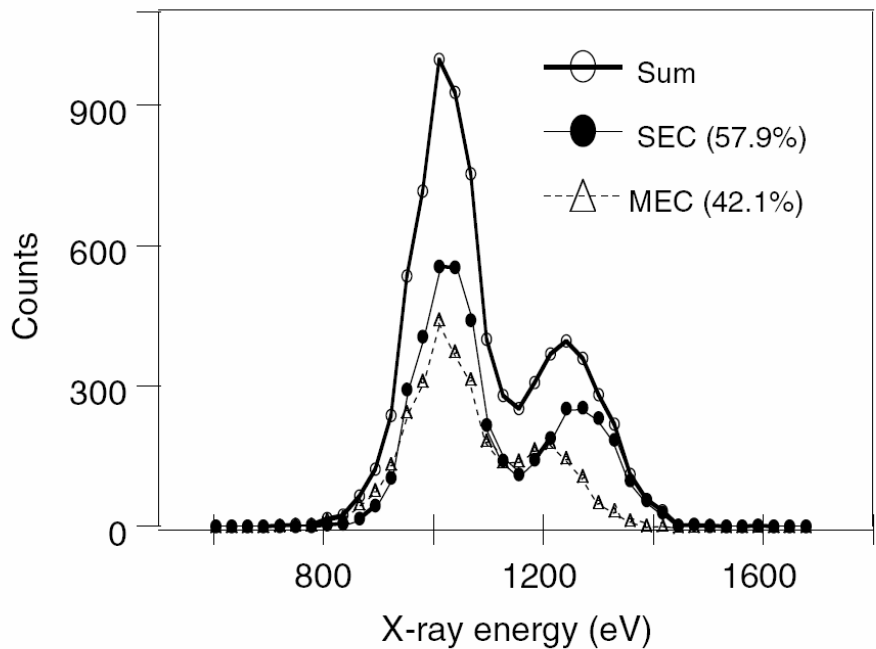

**Figure 53.** X-ray spectra formed in the collision of $Ne^{10+}$ ions with neutral neon. The measurements were performed using the ECR facility at the University of Nevada Reno by Ali et al. (2005) at a collision energy is about 100 keV. Emission from single electron capture (SEC) and emission from multiple electron capture (MEC) is shown separately.

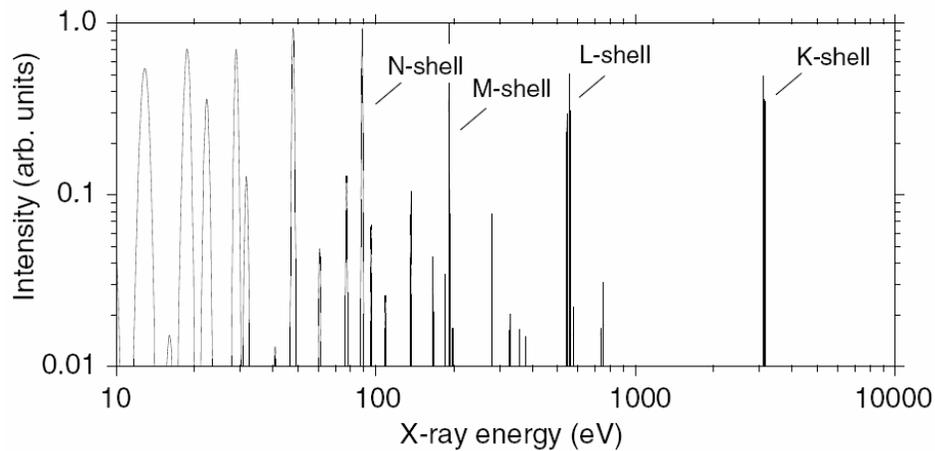

**Figure 54.** Predicted line emission from $Ar^{16+}$ following the charge exchange between $Ar^{17+}$ ions and atomic hydrogen at a collision energy of 40 keV. Lines are represented with a 1 eV line width using a collisonal-radiative model and charge exchange cross sections based on the classical trajectory Monte-Carlo method, as described by Beiersdorfer et al. (2005a).